%% file: dissertation_v2.tex
\documentclass[
        12pt,
]{ociamthesis}

\renewcommand{\baselinestretch}{1.25}

\usepackage{amssymb}
\usepackage{amsmath}
\usepackage{graphicx}
\usepackage{subcaption}
\usepackage{tensor}
\usepackage{hyperref}
\usepackage{pdfpages}

\usepackage[numbers,sort&compress]{natbib}
\setcitestyle{square,comma,numbers,sort&compress}

\newsavebox{\commentboxcontent}

\title{A Semi-Analytical Loss Cone Theory for Tidal Disruption Event Rates Around Kerr Black Holes}

\author{Wenkang Xin}
\college{}

\degree{Master of Mathematical and Theoretical Physics}
\degreedate{Trinity 2026}

\makeatletter
\def\@makechapterhead#1{
        \vspace*{0\p@}
        {\parindent \z@ \raggedright \normalfont
                \ifnum \c@secnumdepth >\m@ne
                        \huge\bfseries \@chapapp\space \thechapter
                        \par\nobreak
                        \vskip 20\p@
                \fi
                \interlinepenalty\@M
                \Huge \bfseries #1\par\nobreak
                \vskip 40\p@
        }}
\def\@makeschapterhead#1{
        \vspace*{0\p@}
        {\parindent \z@ \raggedright
                \normalfont
                \interlinepenalty\@M
                \Huge \bfseries  #1\par\nobreak
                \vskip 40\p@
        }}
\makeatother

\begin{document}

\setcounter{secnumdepth}{3}
\setcounter{tocdepth}{3}

\maketitle
\includepdf[pages=-]{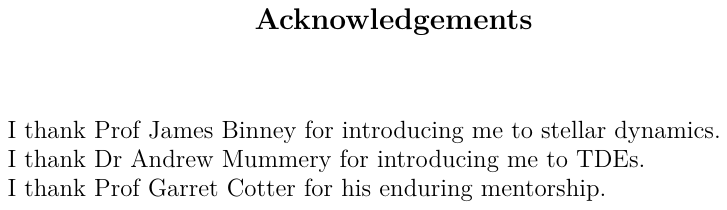}
\begin{abstract}
\input{abstract}
\end{abstract}

\begin{romanpages}
\tableofcontents
\end{romanpages}

\include{chapter0}
\include{chapter1}

\include{chapter2}

\include{chapter3}

\include{chapter4}

\appendix
\include{appendix1}
\include{appendix2_v2}
\include{appendix3_v2}

\addcontentsline{toc}{chapter}{References}
\renewcommand{\bibname}{References}
\begingroup
\footnotesize
\setlength{\bibsep}{1pt}
\renewcommand{\baselinestretch}{0.9}
\bibliography{tidal_rate}
\bibliographystyle{naturemag}
\endgroup

\end{document}

%% file: abstract.tex
A tidal disruption event (TDE) occurs when a star is scattered onto a near-radial orbit and is torn apart by a black hole (BH)'s tidal field. The angular momentum threshold for disruption is set by general relativistic tidal dynamics, while the supply of stars to the disruption zone is governed by Newtonian stellar dynamics. A spinning BH breaks the spherical symmetry of the disruption boundary, so a star's survival depends on both the magnitude and the orientation of its angular momentum. Existing treatments either assume a non-spinning BH or rely on numerical simulations of spinning BHs. We develop the first semi analytical framework that incorporates spin-dependent loss cone boundaries into TDE rate theory. Using a novel tidal tensor formalism, we compute inclination-dependent thresholds for tidal disruption and direct capture by the event horizon. We then revisit the classical one dimensional loss cone problem with nested disruption and capture boundaries, deriving a closed form capture fraction valid across all loss cone regimes. Finally, we formulate a two dimensional Fokker--Planck equation describing simultaneous diffusion in angular momentum magnitude and orientation. Through a perturbative treatment, we demonstrate that while the Kerr disruption boundary induces a first-order bias favouring the disruption of retrograde stars, the global TDE rate is remarkably insensitive to black hole spin. This approach offers a tractable route to including spin and orbital inclination in population-level TDE studies.

%% file: chapter0.tex
\chapter{Introduction and Review}

\section{Introduction} \label{sec:introduction}

In dense galactic nuclei, stochastic N-body interactions scatter stars onto near-radial orbits about the central black hole (hereafter BH) of the host galaxy. If the tidal force experienced by a star along its orbit overcomes its self-gravity, the star will be torn apart, producing a luminous flare that can outshine the entire host galaxy \cite{Rossi2021}. These so-called ``tidal disruption events'' (hereafter TDEs) provide unique windows into distant galactic nuclei and quiescent BHs that would otherwise remain hidden.

TDEs are rare; roughly 200 instances have been identified across the electromagnetic spectrum \cite{Gezari2021}, at a discovery rate that has increased from a few per year to several per month \cite{Velzen2021}. Facilities such as the Vera Rubin Observatory promise a dramatic increase in detections, projected to approach one per hour \cite{Bricman2020}. This will present rich opportunities to connect theory with observations, unlocking the potential of TDEs as a powerful probe of BH physics and galaxy evolution.

Significant progress has been made in understanding individual events. We now have a converging picture of the disruption process itself \cite{Rossi2021}, and subsequent accretion is increasingly well understood with relativistic disc accretion theories \cite{Mummery2025}, enabling inferences of BH mass from a single TDE observation \cite{Mummery2023}. However, the stellar-dynamical problem of how stars are supplied to the disruption zone has received much less attention \cite{Stone2020}.

In assessing the astrophysical role of TDE populations, the critical quantity is the event rate in a given galaxy. Sensitive to properties of both the BH and the stellar environment, TDE rates can be used to place statistical constraints on BH demographics \cite{Greene2020} and properties of galactic nuclei \cite{Stone2020}. Moreover, quantifying the role of TDEs in BH growth sheds light on the co-evolution of BHs and their host galaxies, bridging the gap between small-scale nuclear processes and large-scale galactic and cosmological evolution \cite{Wang2025}.

TDEs are strong-gravity phenomena where general relativity (GR) can play a significant role. One motivation is that observations of such events are biased towards high BH masses. This is because the light curve of a TDE is ``effectively Eddington-limited'' \cite{Velzen2020}, meaning that its peak luminosity scales linearly with the BH mass $M$. In a flux-limited survey, this implies that the detection distance scales as $M^{1/2}$, and the detection volume scales as $M^{3/2}$. This simple argument\footnote{The author would like to thank Andrew Mummery for pointing out this selection effect.} suggests that the observed TDE population will be dominated by events around more massive BHs, exactly where relativistic effects are the most important.

Newtonian approximations also fail to describe the crucial process of ``direct capture'' by the BH's event horizon. The horizon radius scales as $M$, whereas the tidal disruption radius typically scales as $M^{1/3}$. TDEs hence occur progressively closer to the horizon for more massive BHs, making relativistic effects more pronounced. For sufficiently massive BHs, the disruption radius can even fall within the horizon; stars are swallowed whole before they can be disrupted, producing no observable flares. This motivates the concept of a ``Hills mass'' above which no observable TDEs can occur \cite{Hills1975}. 

The picture is further complicated by the fact that astrophysical BHs are expected to be rapidly rotating \cite{Reynolds2021}. This rotation breaks spherical symmetry, rendering the dynamics dependent on the star's orbital inclination. Furthermore, the properties of the accretion flow and the resulting flare also depend on the orbital inclination \cite{Stone2019, Hayasaki2016}. Taken together, the observational bias towards massive BHs, the existence of the Hills mass, and the inclination dependence induced by spin make a relativistic framework necessary to accurately interpret the observed TDE population.

\section{Review of TDE Rate Theory} \label{sec:review}

The modern theory of TDE rates stands on two largely separate strands of work: relativistic calculations of how stars are disrupted in the strong-field region of a BH, and stellar-dynamical calculations of how they enter the zone of destruction.

\subsection{Newtonian loss cone theory}

On the stellar dynamics side, classical loss-cone literature established the basic language of the problem. Frank \& Rees \cite{Frank1976} first proposed the possibility of the disruption of stars by a central BH in a galactic nucleus, and raised the idea of a loss cone, a region in angular momentum space where stars come dangerously close to the BH and are removed on orbital timescales. Seminal works by Lightman \& Shapiro \cite{Lightman1977} and Cohn \& Kulsrud \cite{Cohn1978} placed this picture on an analytical footing by orbit-averaging the collisional Boltzmann equation, reducing the problem to a one-dimensional Fokker--Planck equation.

Subsequent work applied and refined the treatment, particularly in the intermediate regime where the timescales of loss cone depletion and refilling are comparable. In particular, Magorrian \& Tremaine \cite{Magorrian1999} and Wang \& Merritt \cite{Wang2004} provided early applications of the theory to realistic galaxy models; Merritt \cite{Merritt2013} presented a comprehensive review of loss cone theory and provided a fully analytical treatment of the intermediate regime. A large body of work (e.g.\ \cite{Yao2023, Chang2025, Stone2015}) has relied on this classical theory to estimate TDE rates, with the aim of using the rate to determine properties of the host galaxy and its central BH.

\subsection{Need for 2D loss cone theory}

Most applications of the classical loss cone theory rely on two assumptions of symmetry: the galaxy is spherically symmetric and the disruption threshold only depends on the total angular momentum $L$ of a star. The first assumption allows one to use $L$ as a phase space coordinate, which is the basis of the one-dimensional treatment that only considers diffusion in $L$ at fixed energies $E$ \cite{Lightman1977, Cohn1978}. The second allows the loss cone boundary to be treated as a single critical angular momentum magnitude $L_{\text{tde}}$. The spherical galaxy assumption has been challenged in several works (e.g.\ \cite{Magorrian1999, Vasiliev2013}), which have shown that non-spherical potentials can significantly enhance TDE rates by providing additional channels for stars to enter the loss cone.

On the other hand, very little work has challenged the orientation-independent loss cone assumption. Yet the spinning nature of astrophysical BHs suggests that the disruption threshold must depend on the orientation of the angular momentum. Kesden's numerical work \cite{Kesden2012} marked an important step forward by showing how BH spin can significantly affect the observed TDE rates. Two key effects are at play. First, the tidal disruption radius, and thus the critical disruption angular momentum $L_{\text{tde}}$, depends on the orientation of the star's angular momentum. Second, stars with low enough angular momentum plunge directly into the horizon without producing observable emission. This defines a critical capture angular momentum $L_{\text{cap}}$, which also depends on the orbital inclination.

The interplay between these two effects can lead to strong modifications of the TDE rate curves, the most notable effect being a significant spin-induced increase in the Hills mass, which is defined as the BH mass where $L_{\text{tde}} < L_{\text{cap}}$ for all orbital inclinations (see \cite{Mummery2023,Xin2025}). For solar-type stars disrupted by a non-spinning BH, the Hills mass is around $10^{8} M_{\odot}$; for a maximally spinning BH, the Hills mass can be as high as $10^{9} M_{\odot}$, well into the SMBH regime. By compiling a population of observed events, we can reconstruct the TDE rate as a function of mass by inferring the host BH mass for each event. The location of the high-mass cutoff in this population distribution is statistically sensitive to the underlying BH spin distribution, thereby providing a method to constrain BH spin demographics \cite{Mummery2023}. This is, of course, only possible if we have a robust theoretical framework for calculating these rates as a function of both BH mass and spin.

The results of Kesden \cite{Kesden2012} were based on the limiting assumption of extremely efficient diffusion processes, and inclination-dependent effects were inferred from Monte Carlo simulations of test orbits. Nevertheless, the work clearly demonstrated that to account for spinning BHs, the classical 1D theory must be extended from tracking just the magnitude of angular momentum to a 2D theory that also models its orientation. A more recent work by Singh \& Kesden \cite{Singh2024} attempted to understand the inclination dependence of TDE rates by reducing the 2D problem to the classical 1D framework, but ultimately found that a full 2D treatment is necessary to capture the relevant physics.

\subsection{Relativistic tidal dynamics}

The prerequisite for constructing any 2D loss-cone theory is a mathematical description of the absorbing boundaries. General relativistic calculations can be used to find the exact disruption and capture thresholds, provided that we have a good understanding of the dynamics around a Kerr BH and how much tidal force a star experiences along its orbit. Geodesics in Kerr spacetime are characterised by three conserved quantities that have a direct asymptotic mapping to the classical orbital parameters: energy $E$, angular momentum about the spin axis $L_{z}$, and the Carter constant $\mathcal{Q}$ which relates to the inclination of the orbit \cite{Carter1968, Bardeen1972, Wilkins1972}. This mapping is necessary to connect the relativistic calculations with Newtonian stellar dynamics.

The pioneering work by Marck \cite{Marck1983} provided the first solutions to the relativistic tidal tensor in Kerr spacetime, which was widely used in subsequent studies (e.g., \cite{Kesden2012, Singh2024}). However, the formalism relies on the construction of a so-called parallel-transported tetrad, which is a set of coordinates carefully moved and rotated along a star's trajectory to ensure that it describes the local frame of the star. This makes it difficult to incorporate the method into a loss cone calculation, which requires the evaluation of the tidal tensor at the pericentre to determine whether the star is disrupted.

A recent work \cite{Xin2025} developed a new tidal tensor formalism that relies on successive frame transformations instead of parallel transport, allowing one to directly compute the tidal tensor at any point along a geodesic by performing simple matrix multiplications. This new approach is more computationally efficient and easier to implement, making it ideal for incorporation into a loss cone theory.

Taken together, the existing literature forms a clear progression of two tracks of work converging towards the need for a 2D loss cone theory. Classical stellar-dynamical theory explains how stars are supplied to the loss cone; relativistic dynamics explains how strong gravity modifies disruption and capture thresholds. The natural next step is to combine these ingredients into a single framework that can treat TDE rates around spinning BHs in a self-consistent way. This is precisely the goal of this dissertation.

\section{Outline and Conventions}

Chapter 2 reviews Kerr geodesic and tidal dynamics and calculates the disruption and capture thresholds as explicit functions of orbital inclination. Chapter 3 develops a rigorous 1D loss-cone framework that incorporates these relativistic boundaries, including a transparent treatment of direct capture. Chapter 4 extends the 1D framework to formulate a genuinely 2D theory of loss cone dynamics. We summarise our findings and discuss future directions in Chapter 5.

In relativistic calculations, we use natural units $G = c = 1$ and only restore physical units when connecting to Newtonian stellar dynamics. Plain Greek indices such as $\mu, \nu, \rho$ are used for the distant observer frame; Roman indices in parentheses $(a), (b), (c)$ are for the zero angular momentum observer frame; indices with tildes $\tilde{\alpha}, \tilde{\beta}, \tilde{\gamma}$ are for the local freely falling frame. Throughout, we model the star as a test particle following a time-like geodesic.

In stellar-dynamical calculations, we assume that the stellar population consists of a single species of solar-type stars, that the galactic potential is spherical $\Phi(r)$, and that the galaxy is isotropic so that the distribution $f_{\star}(E)$ is a function of energy $E$ only.

%% file: chapter1.tex
\chapter{Relativistic Dynamics and Tidal Tensors}

Calculating TDE rates requires a mathematical description of the boundaries within which stars are either torn apart or swallowed by the horizon. These processes occur in the strong-field region near the BH, where the gravitational effect of the stellar cluster is entirely negligible and the dynamics are determined exclusively by the Kerr geometry. Establishing the relativistic loss cone boundaries requires a detailed analysis of the Kerr spacetime.

\section{Relativistic Dynamics} \label{sec:relativistic_dynamics}

In GR, a rotating BH is described by the Kerr metric, which represents the geometry of spacetime around a point mass with angular momentum. This metric is characterised by two parameters, the mass $M$ and the spin parameter $a$, and can be represented in an unconventional but convenient form as follows. Any stationary, axisymmetric spacetime in Boyer-Lindquist coordinates can be written as \cite{Bardeen1972}:
\begin{equation}
    \mathrm{d}s^{2} = -e^{2\nu} \mathrm{d}t^{2} + e^{2\psi} \left( \mathrm{d}\phi - \omega \mathrm{d}t \right)^{2} + e^{2\mu_{1}} \mathrm{d}r^{2} + e^{2\mu_{2}} \mathrm{d}\theta^{2} \, .
    \label{eq:general_metric}
\end{equation}
The Kerr metric is defined by the following metric functions:
\begin{equation}
    \begin{split}
        &e^{2\nu} = \frac{\Sigma \Delta}{\mathcal{A}} \, , \qquad e^{2\psi} = \sin^{2}{\theta} \frac{\mathcal{A}}{\Sigma} \, , \\
        &e^{2\mu_{1}} = \frac{\Sigma}{\Delta} \, , \qquad e^{2\mu_{2}} = \Sigma \, , \\
        &\omega = \frac{2Mra}{\mathcal{A}} \, ,
    \end{split}
\end{equation}
where:
\begin{equation}
    \begin{split}
        \Sigma &\equiv r^{2} + a^{2}\cos^{2}{\theta} \, , \\
        \Delta &\equiv r^{2} - 2Mr + a^{2} \, , \\
        \mathcal{A} &\equiv (r^{2} + a^{2})^{2} - \Delta a^{2} \sin^{2}{\theta} \, .
    \end{split}
\end{equation}
If $a^{2} < M^{2}$, the metric describes a BH with horizons at $r_{\pm} = M \pm \sqrt{M^{2} - a^{2}}$. We set $G = c = M = 1$ throughout this chapter. By doing so, we scale all physical quantities as follows:
\begin{equation}
\begin{split}
    [r] &= \frac{GM}{c^{2}} \approx 1.48 \times 10^6 \, \text{km} \left( \frac{M}{10^6 M_\odot} \right) \, , \\
    [t] &= \frac{GM}{c^{3}} \approx 4.93 \, \text{s} \left( \frac{M}{10^6 M_\odot} \right) \, , \\
    [L] &= \frac{GM}{c} \approx 4.43 \times 10^{11} \, \text{km}^{2} \, \text{s}^{-1} \left( \frac{M}{10^6 M_\odot} \right) \, ,
    \label{eq:units}
\end{split}
\end{equation}
where $[L]$ denotes the specific angular momentum of a test particle.

The motion of a test particle in Kerr spacetime is governed by the geodesic equations:
\begin{equation}
    \frac{\mathrm{d}^{2}x^{\mu}}{\mathrm{d}\tau^{2}} + \tensor{\Gamma}{^{\mu}_{\alpha \beta}} \frac{\mathrm{d}x^{\alpha}}{\mathrm{d}\tau} \frac{\mathrm{d}x^{\beta}}{\mathrm{d}\tau} = 0 \, ,
\end{equation}
where $x^{\mu} = (t, r, \theta, \phi)$ is the 4-position, $\tau$ is the proper time and $\tensor{\Gamma}{^{\mu}_{\alpha \beta}}$ are the Christoffel symbols associated with the Kerr metric.

The particle has four conserved quantities: rest mass, energy, axial angular momentum, and the Carter constant. The first three are associated with the spacetime's symmetries, whereas the last one is due to the fact that the Hamilton-Jacobi equation for geodesic motion in Kerr spacetime is separable \cite{Carter1968}. The conserved quantities are defined as follows:
\begin{equation}
\begin{split}
    m^{2} &= -g_{\mu \nu} p^{\mu} p^{\nu} \, , \\
    \mathcal{E} &\equiv -p_{t} \, , \\
    L_{z} &\equiv p_{\phi} \, , \\
    \mathcal{Q} &\equiv p_{\theta}^{2} + \cos^{2}{\theta} \left[ a^{2} (\mathcal{E}^{2} - m^{2}) + \frac{L_{z}^{2}}{\sin^{2}{\theta}} \right] \, ,
\end{split}
\end{equation}
where we have defined the Carter constant $\mathcal{Q}$ and the four-momentum $p^{\mu} = m \mathrm{d}x^{\mu}/\mathrm{d}\tau$.

The conserved quantities can be scaled by the rest mass $m$ such that $m = 1$. Because there are four constants of motion, the geodesic equations can be reduced to first integrals of motion, which are well-known in the literature (see e.g.\ \cite{Wilkins1972}):
\begin{equation}
    \begin{split}
        \Sigma \frac{\mathrm{d}t}{\mathrm{d}\tau} &= -a(a \mathcal{E} \sin^{2}{\theta} - L_{z}) + (r^{2} + a^{2}) \frac{T}{\Delta} \, , \\
        \Sigma^{2} \left( \frac{\mathrm{d}r}{\mathrm{d}\tau} \right)^{2} &= R(r) \, , \\
        \Sigma^{2} \left( \frac{\mathrm{d}\theta}{\mathrm{d}\tau} \right)^{2} &= \Theta(\theta) \, , \\
        \Sigma \frac{\mathrm{d}\phi}{\mathrm{d}\tau} &= -\left( a \mathcal{E} - \frac{L_{z}}{\sin^{2}{\theta}} \right) + a \frac{T(r)}{\Delta} \, .
    \end{split}
    \label{eq:kerr_first_integrals}
\end{equation}
The functions $T$, $R$, and $\Theta$ are defined as:
\begin{equation}
    \begin{split}
        T(r) &\equiv \mathcal{E} (r^{2} + a^{2}) - aL_{z} \, , \\
        R(r) &\equiv T^{2} - \Delta [r^{2} + \mathcal{Q} + (L_{z} - a\mathcal{E})^{2}] \, , \\
        \Theta(\theta) &\equiv \mathcal{Q} - \cos^{2}{\theta} \left[ a^{2} (1 - \mathcal{E}^{2}) + \frac{L_{z}^{2}}{\sin^{2}{\theta}} \right] \, .
    \end{split}
\end{equation}

Note that in the Kerr metric, a degeneracy exists in the choice of signs between $L_{z}$ and $a$, since the forms of the first integrals are invariant (except for a trivial sign change in $\dot{\phi}$) if both quantities switch signs. We restrict our study to $a \ge 0$ and allow $L_{z}$ to take any value. Orbits with $L_{z} > 0$ are co-rotating (prograde) with the BH, while orbits with $L_{z} < 0$ are counter-rotating (retrograde).

For the purpose of calculating TDE rates, we model the star as a test particle equipped with the above conserved quantities, moving on a geodesic defined by the first integrals of motion. The star can be treated as an infalling particle from spatial infinity since the typical scale of the stellar cluster (the radius of influence $r_{h} = GM/\sigma^{2}$) is roughly $c^{2}/\sigma^{2} \sim 10^{7}$ times larger than the horizon scale ($r_{g} = GM/c^{2}$) for a typical velocity dispersion of $\sigma \sim 100 \, \mathrm{km/s}$. It is also a very good approximation to set $\mathcal{E} = 1$ for any infalling stars \cite{Kesden2012} because their orbital binding energy is negligible compared to the rest-mass energy, deviating by only $\mathcal{O}(\sigma^{2}/c^{2}) \sim 10^{-7}$.

It turns out that in the limit of $\mathcal{E} = m = 1$, $\mathcal{Q}$ represents the squared angular momentum projected on the equatorial plane \cite{Wilkins1972, Kesden2012}:
\begin{equation}
    \mathcal{Q} = p_{\theta}^{2} + L_{z}^{2} \cot^{2}{\theta} = L_{x}^{2} + L_{y}^{2} \, .
    \label{eq:q_as_equatorial_angular_momentum}
\end{equation}
This motivates the definition of the total angular momentum $L$:
\begin{equation}
    L^{2} \equiv \mathcal{Q} + L_{z}^{2} \, ,
\end{equation}
and the conserved inclination parameter $x$ or $\cos{\iota}$:
\begin{equation}
    x = \cos{\iota} \equiv \frac{L_{z}}{L} \, .
\end{equation}
This implies that, at least asymptotically and in the limit of $\mathcal{E} = 1$, a one-to-one mapping exists between the conserved quantities $(L, x)$ in Kerr spacetime and the classical orbital parameters specified by the magnitude and orientation of the (Keplerian) angular momentum vector. Note that positive $x$ corresponds to prograde orbits, while negative $x$ corresponds to retrograde orbits; $x = \pm 1$ corresponds to equatorial orbits, while $x = 0$ corresponds to polar orbits.

\section{Innermost Bound Spherical Orbits} 

\begin{figure}[h]
    \centering
    \includegraphics[width=0.6\textwidth]{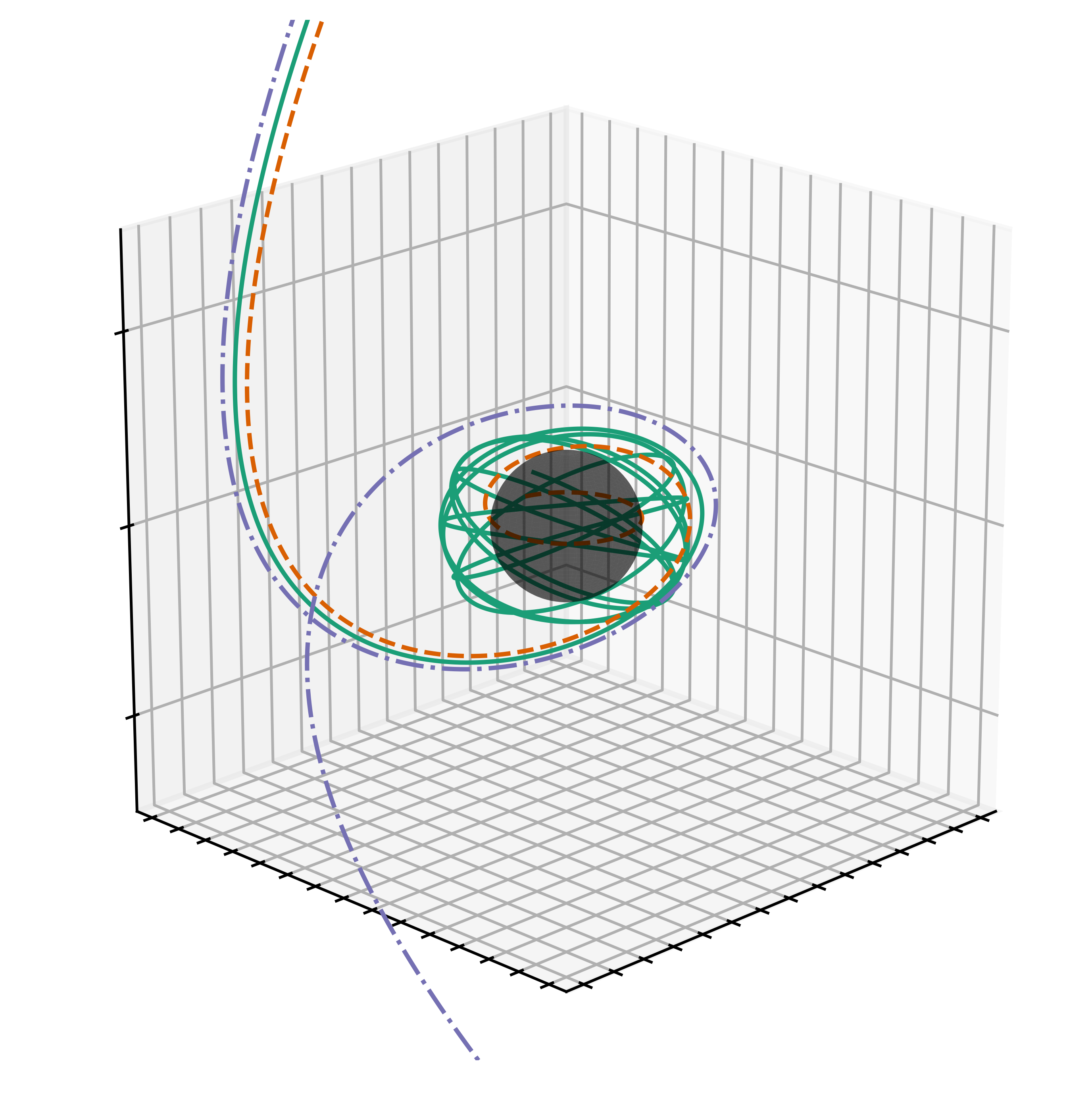}
    \caption{Trajectories showing the separatrix nature of the innermost bound spherical orbit (IBSO). The IBSO (green solid) approaches a constant radius asymptotically; an orbit with slightly lower angular momentum (orange dashed) terminates as it plunges into the horizon (dark grey sphere); an orbit with a slightly higher angular momentum (purple dash-dotted) returns to infinity after a parabolic flyby.}
    \label{fig:ibso}
\end{figure}

We now introduce a special class of orbits particularly relevant for the capture of stars by the horizon: the innermost bound spherical orbits (IBSOs). For stars starting at rest from spatial infinity, IBSOs represent the \textit{separatrix} between trajectories that plunge into the horizon and those that return to infinity after a parabolic flyby: a star with angular momentum below the IBSO threshold plunges, while one above it returns to infinity (Figure~\ref{fig:ibso}). For a plunging star, even if disruption occurs outside the horizon, the debris will rapidly fall into the BH without producing observable emission \cite{Mummery2023}. Therefore, the IBSO angular momentum sets the lower bound for observable TDEs.

IBSOs are defined by the following conditions:
\begin{equation}
\begin{split}
    \mathcal{E} = 1 \, , \quad R(r) = 0 \, , \quad R'(r) = 0 \, .
\end{split}
\end{equation}
The first unity energy condition corresponds to ``innermost bound'' while the second and third conditions correspond to ``spherical'' (constant radius) orbits. These conditions lead to the twin constraints on $r$ and $L$ for a given inclination $x$:
\begin{equation}
    \begin{split}
        0 &= (r - 1) L^{2} + 2axL - (a^{2} + 3r^{2}) \, , \\
        0 &= 9 r^{4} - (16 + 6 L^{2}) r^{3} + (6 a^{2} + 10 L^{2} + L^{4}) r^{2} \\
        &\quad- (2 a^{2} L^{2} + 2 L^{4}) r + a^{4} - 2 a^{2} L^{2} + L^{4} \, .
    \end{split}
    \label{eq:ibso_kerr}
\end{equation}
In general, given an inclination parameter $x$, these constraints can be numerically solved to find the radius and angular momentum of the IBSO. Stars with $L < L_{\text{cap}} \equiv L_{\text{IBSO}}$ are directly captured by the horizon, while those with $L > L_{\text{cap}}$ can produce observable TDEs if they are disrupted outside the horizon.

\section{Relativistic Tidal Tensor} \label{sec:tidal_tensor}

Having established the orbital dynamics of a test particle in Kerr spacetime, we wish to know the tidal forces experienced by the particle as it moves along its trajectory. Consider two nearby geodesics separated by an infinitesimal displacement vector $\delta x^{\mu}$. We write them as $X^{\mu}(\tau)$ and $X'^{\mu}(\tau) = X^{\mu}(\tau) + \delta x^{\mu}(\tau)$, where $\tau$ is the proper time. To first order in $\delta x^{\mu}$, the separation evolves according to the geodesic deviation equation (see e.g.\ \cite[Chapter 7]{Hobson2006}):
\begin{equation}
    \frac{\mathrm{D}^{2}}{\mathrm{D} \tau^{2}} \delta x^{\mu} = \tensor{R}{^{\mu}_{\alpha \nu \beta}} U^{\alpha} U^{\beta} \delta x^{\nu} \, .
    \label{eq:geodesic_deviation}
\end{equation}
$U \equiv (\dot{t}, \dot{r}, \dot{\theta}, \dot{\phi})$ is the 4-velocity of the particle, where the dot denotes differentiation with respect to proper time, and $\mathrm{D}/\mathrm{D}\tau$ is the corresponding covariant derivative. The geodesic deviation equation describes how nearby trajectories deviate due to spacetime curvature. That is, it describes linear tidal effects in a relativistic spacetime.

The tidal tensor is defined as:
\begin{equation}
    \tensor{C}{^{\mu}_{\nu}} \equiv -\tensor{R}{^{\mu}_{\alpha \nu \beta}} U^{\alpha} U^{\beta} \, ,
    \label{eq:tidal_tensor}
\end{equation}
so that we can write:
\begin{equation}
    \frac{\mathrm{D}^{2}}{\mathrm{D} \tau^{2}} \delta x^{\mu} = -\tensor{C}{^{\mu}_{\nu}} \delta x^{\nu} \, ,
    \label{eq:tidal_acceleration}
\end{equation}
in analogy with Newtonian tidal forces.

Equation~\ref{eq:geodesic_deviation} describes the deviation of geodesics as measured by the distant observer in Boyer-Lindquist coordinates. However, the relevant coordinate systems for computing \textit{physical} tidal forces are those that describe the local freely falling (LFF) frame of the test particle, as this is the tidal force the particle itself would measure \cite{Pirani2009}. Therefore, a procedure is required to transform the tidal tensor $\tensor{C}{^{\mu}_{\nu}}$ from the distant observer frame to the particle's LFF frame.

Historically, the transformation of $\tensor{C}{^{\mu}_{\nu}}$ to the LFF frame is done by constructing a so-called parallel-transported tetrad along the geodesic. This is a set of coordinates that fully describes the LFF frame of the test particle at all points, allowing the local tidal tensor to be directly evaluated \cite{Manasse1963}. However, constructing this tetrad is highly non-trivial and is known for only a few special cases, such as Kerr spacetime \cite{Marck1983}. Furthermore, one must track an angular variable used to rotate the tetrad along the trajectory to ensure parallel transport. Consequently, determining the local tidal tensor at any point requires integrating the entire history of the geodesic, rendering the procedure intractable for stellar-dynamical calculations.

A recent study \cite{Xin2025} provided a general framework for calculating the relativistic tidal tensor in any stationary, axisymmetric spacetime. Instead of attempting to fully describe the local frame via a parallel-transported tetrad, the framework exploits the fact that $\tensor{C}{^{\mu}_{\nu}}$ is a good rank-2 tensor that can be transformed between frames. This allows the tidal tensor to be moved from the distant observer frame to the LFF frame via a sequence of carefully designed transformations. Effectively, this changes a \textit{calculus} problem (rotating a tetrad along a geodesic) into a \textit{matrix algebra} problem (transforming a tensor between frames). Crucially, because the ZAMO tetrad can always be defined locally, the tidal tensor at any point can be evaluated without knowing the particle's history.

\section{ZAMO Transformation} \label{sec:zamo}

A so-called ``zero angular momentum observer'' (ZAMO) is characterised by the 4-position $r = \text{const}$, $\theta = \text{const}$, and $\phi = \omega t + \text{const}$. The orthonormal tetrad carried by this observer is given by \cite{Bardeen1972}:
\begin{equation}
    \begin{split}
        &\tensor{\mathbf{e}}{^\mu_{(t)}} = e^{-\nu} \left( \frac{\partial }{\partial t} + \omega \frac{\partial }{\partial \phi} \right) \, , \quad \tensor{\mathbf{e}}{^\mu_{(r)}} = e^{-\mu_{1}} \frac{\partial }{\partial r} \, , \\
        &\tensor{\mathbf{e}}{^\mu_{(\theta)}} = e^{-\mu_{2}} \frac{\partial }{\partial \theta} \, , \quad \tensor{\mathbf{e}}{^\mu_{(\phi)}} = e^{-\psi} \frac{\partial }{\partial \phi} \, .
    \end{split}
\end{equation}
The corresponding dual basis (covariant basis) is:
\begin{equation}
    \begin{split}
        &\tensor{\mathbf{e}}{_{\mu}^{(t)}} = e^{\nu} \mathrm{d}t \, , \quad \tensor{\mathbf{e}}{_{\mu}^{(r)}} = e^{\mu_{1}} \mathrm{d}r \, , \\
        &\tensor{\mathbf{e}}{_{\mu}^{(\theta)}} = e^{\mu_{2}} \mathrm{d}\theta \, , \quad \tensor{\mathbf{e}}{_{\mu}^{(\phi)}} = e^{\psi} \left( -\omega\mathrm{d}t + \mathrm{d}\phi \right) \, .
    \end{split}
\end{equation}
This tetrad can be used to project any tensor into the ZAMO frame, a transformation that corresponds to describing physical quantities by their components as measured by the local ZAMO observer. Crucially, it is simple to verify by explicit computation that:
\begin{equation}
    \tensor{g}{_{(a)(b)}} = \tensor{\mathbf{e}}{^\mu_{(a)}} \tensor{\mathbf{e}}{^\nu_{(b)}} g_{\mu\nu} = \eta_{(a)(b)} = \mathrm{diag}(-1,1,1,1) \, ,
\end{equation}
meaning that the local transformation laws of special relativity apply in this frame since the ZAMO observer infers a flat spacetime locally. 

With the ZAMO tetrad defined, a test particle's 4-velocity $U^{\mu}$ in the distant observer frame can be projected onto the ZAMO basis:
\begin{equation}
    U^{(a)} = U^{\mu} \tensor{\mathbf{e}}{_{\mu}^{(a)}} \, ,
\end{equation}
where $U^{(a)}$ is the 4-velocity of the particle measured in the ZAMO frame.

The Riemann tensor can also be transformed to the ZAMO frame:
\begin{equation}
    \tensor{R}{^{(a)}_{(b)(c)(d)}} = \tensor{\mathbf{e}}{_{\mu}^{(a)}} \tensor{\mathbf{e}}{^{\nu}_{(b)}} \tensor{\mathbf{e}}{^{\rho}_{(c)}} \tensor{\mathbf{e}}{^{\sigma}_{(d)}} \tensor{R}{^{\mu}_{\nu \rho \sigma}} \, .
\end{equation}
Since we place the ZAMO at the test particle's position and spacetime is locally Minkowski in the ZAMO frame, a simple Lorentz boost yields the Riemann tensor in the LFF frame (i.e.\ the physically relevant frame):
\begin{equation}
    \tensor{R}{^{\tilde{\mu}}_{\tilde{\alpha} \tilde{\nu} \tilde{\beta}}} = \tensor{\varLambda}{_{(a)}^{\tilde{\mu}}} \tensor{\varLambda}{^{(b)}_{\tilde{\alpha}}} \tensor{\varLambda}{^{(c)}_{\tilde{\nu}}} \tensor{\varLambda}{^{(d)}_{\tilde{\beta}}} \tensor{R}{^{(a)}_{(b)(c)(d)}} \, ,
\end{equation}
where the Lorentz boost matrix can be explicitly written as:
\begin{equation}
    \begin{split}
        &\tensor{\varLambda}{^{\mu}_{\nu}} \equiv \mathbb{I}_{4} + \begin{bmatrix}
            \gamma - 1            & -\gamma \beta^{r}                                      & -\gamma\beta^{\theta}                                     & -\gamma\beta^{\phi}                                       \\
            -\gamma\beta^{r}      & (\gamma - 1) \frac{(\beta^{r})^{2}}{\beta^{2}}         & (\gamma - 1) \frac{\beta^{r}\beta^{\theta}}{\beta^{2}}    & (\gamma - 1) \frac{\beta^{r}\beta^{\phi}}{\beta^{2}}      \\
            -\gamma\beta^{\theta} & (\gamma - 1) \frac{\beta^{\theta}\beta^{r}}{\beta^{2}} & (\gamma - 1) \frac{(\beta^{\theta})^{2}}{\beta^{2}}       & (\gamma - 1) \frac{\beta^{\theta}\beta^{\phi}}{\beta^{2}} \\
            -\gamma\beta^{\phi}   & (\gamma - 1) \frac{\beta^{\phi}\beta^{r}}{\beta^{2}}   & (\gamma - 1) \frac{\beta^{\phi}\beta^{\theta}}{\beta^{2}} & (\gamma - 1) \frac{(\beta^{\phi})^{2}}{\beta^{2}}
        \end{bmatrix} \, ,
    \end{split}
\end{equation}
where $\mathbb{I}_{4}$ is the $4 \times 4$ identity matrix $\mathbb{I}_{4}={\rm diag}(1, 1, 1, 1)$ and:
\begin{equation}
    \begin{split}
        \gamma &\equiv \frac{1}{\sqrt{1 - \beta^{2}}} \, , \\
        \beta &\equiv \left[ {(\beta^{r})^{2} + (\beta^{\theta})^{2} + (\beta^{\phi})^{2}} \right]^{1/2} \, , \\
        \beta^{i} &\equiv U^{(i)} / U^{(t)} \quad \text{for} \quad i = r, \theta, \phi \, .
    \end{split}
\end{equation}
With the Riemann tensor now known in the LFF frame, the physical local tidal tensor can be constructed:
\begin{equation}
    \tensor{C}{^{\tilde{\mu}}_{\tilde{\nu}}} = -\tensor{R}{^{\tilde{\mu}}_{\tilde{\alpha} \tilde{\nu} \tilde{\beta}}} U^{\tilde{\alpha}} U^{\tilde{\beta}} \, ,
    \label{eq:local_tidal_tensor}
\end{equation}
where $U^{\tilde{\alpha}} = (1, 0, 0, 0)$ is the 4-velocity of the particle in its own LFF frame.

The above steps provide a general algorithm for deriving the local tidal tensor using relativistic frame transformations. Given a stationary, axisymmetric metric (e.g.\ Kerr spacetime), one can always compute the Riemann tensor in the distant observer frame $(t, r, \theta, \phi)$, transform it to the ZAMO frame, project the particle's 4-velocity, and then Lorentz boost to the LFF frame to construct $\tensor{C}{^{\tilde{\mu}}_{\tilde{\nu}}}$.

Returning to Kerr spacetime, the local tidal tensor $\tensor{C}{^{\tilde{\mu}}_{\tilde{\nu}}}$ can be derived as a function of the 4-velocity $U^{\mu}$ and 4-position $x^{\mu}$ of the particle in the distant observer frame. Since Kerr geodesics are equipped with the first integrals of motion, the 4-velocity can be reduced to the conserved quantities and the 4-position, such that:
\begin{equation}
    \tensor{C}{^{\tilde{\mu}}_{\tilde{\nu}}} = \tensor{C}{^{\tilde{\mu}}_{\tilde{\nu}}}(t, r, \theta, \phi, \mathcal{E}, L, x) \, ,
\end{equation}
where we have used the conserved quantity pair $(L, x = L_{z}/L)$.

The tidal tensor is a $4 \times 4$ matrix encoding the physical tidal forces experienced by the star. The eigenvalues of the tidal tensor represent forces along the corresponding eigenvectors. As can be inferred from Equation~\ref{eq:tidal_acceleration}, a negative (positive) eigenvalue corresponds to a stretching (compressing) force along the corresponding eigenvector. In general, $\tensor{C}{^{\tilde{\mu}}_{\tilde{\nu}}}$ has one zero eigenvalue (time direction) and three non-zero spatial eigenvalues, among which one is negative (stretching) and two are positive (compressing). The sum of the eigenvalues vanishes because in vacuum the Ricci tensor is identically zero ($R_{\mu\nu} = 0$), which forces $\tensor{C}{^{\tilde{\mu}}_{\tilde{\nu}}}$ to be traceless \cite{Xin2025}.

In general, the tidal eigenvalues depend strongly on the radial coordinate $r$, while the polar angle $\theta$ plays a secondary role; there is no dependence on $t$ or $\phi$ due to the stationarity and axisymmetry of the spacetime. The tidal forces are therefore strongest at the orbital pericentre $r_{\text{min}}$:
\begin{equation}
    \tensor{C}{^{\tilde{\mu}}_{\tilde{\nu}}}(r = r_{\text{min}}, \theta = \pi/2, \mathcal{E} = 1, L, x) = \tensor{C}{^{\tilde{\mu}}_{\tilde{\nu}}}_{\text{max}}(L, x) \, .
    \label{eq:max_tidal_tensor}
\end{equation}
Here, we choose to evaluate the tidal tensor at the equatorial plane ($\theta = \pi/2$) at pericentre. This is a necessary concession because the exact polar angle $\theta_{\text{peri}}$ at pericentre cannot be determined from the conserved quantities alone, as stars begin their trajectories at arbitrary initial radii. Nevertheless, this choice is justified because the tidal eigenvalues oscillate at the sub-$10\%$ level at fixed $r$ as $\theta$ varies \cite{Xin2025}. Since the tidal force is the strongest on the equatorial plane, this approximation slightly overestimates the tidal effect. Alternatively, one could evaluate the tensor at the minimum polar angle $\theta_{\text{min}}$ (the turning point of the latitudinal motion), which is determined by setting $p_{\theta} = 0$ in Equation~\ref{eq:q_as_equatorial_angular_momentum}:
\begin{equation}
    \sin{\theta_{\text{min}}} = \sqrt{\frac{L_{z}^{2}}{\mathcal{Q} + L_{z}^{2}}} = \left\lvert x \right\rvert \, .
\end{equation}
Since $\theta_{\text{min}}$ represents the maximum possible deviation from the equatorial plane, evaluating the tidal tensor at $\theta_{\text{min}}$ sets a lower bound that slightly underestimates the tidal forces.

The tidal tensor has eigenvalues of the dimension $T^{-2}$ measured in units of $c^{6}G^{-2}M^{-2}$ (see Equation~\ref{eq:units}). A tidal eigenvalue represents an acceleration per unit length along the corresponding eigenvector; a star of characteristic length $R_{*}$ experiences a physical tidal acceleration of the size:
\begin{equation}
    a_{\text{tidal}} = \left\lvert \lambda \right\rvert \frac{c^{6}}{G^{2}M^{2}} R_{*} \, ,
\end{equation}
where $\left\lvert \lambda \right\rvert$ is the (dimensionless) magnitude of the stretching tidal eigenvalue.

The stretching tidal acceleration must overcome the star's attracting gravitational self-acceleration, which can be estimated as:
\begin{equation}
    a_{\text{self}} = \eta \frac{GM_{*}}{R_{*}^{2}} \, ,
\end{equation}
where $M_{*}$ is the mass of the star.

The dimensionless parameter $\eta$ is a calibration factor that depends on the stellar structure (e.g.\ polytropic index \cite{Tejeda2017}) and the precise disruption criterion (e.g.\ partial disruption \cite{Bortolas2023}). For the purpose of this dissertation, we set $\eta = 1$ throughout. For tidal disruption to occur, $a_{\text{tidal}} \gtrsim a_{\text{self}}$ or equivalently:
\begin{equation}
    \left\lvert \lambda \right\rvert \gtrsim \left\lvert \lambda \right\rvert_{\text{crit}} \equiv \frac{G^{3} M_{*} M^{2}}{c^{6} R_{*}^{3}} \, .
    \label{eq:tidal_condition}
\end{equation}

\section{Anisotropic Loss Cone Boundaries} \label{sec:loss_cone}

Equation~\ref{eq:tidal_condition} defines the critical value $\left\lvert \lambda \right\rvert_{\text{crit}}$ required for tidal disruption. For each geodesic specified by the pair $(L, x)$, the maximum local tidal tensor $\tensor{C}{^{\tilde{\mu}}_{\tilde{\nu}}}_{\text{max}}$ can be evaluated according to Equation~\ref{eq:max_tidal_tensor} to find the stretching eigenvalue $\left\lvert \lambda \right\rvert_{\text{max}}$. If $\left\lvert \lambda \right\rvert_{\text{max}} > \left\lvert \lambda \right\rvert_{\text{crit}}$, the star is disrupted.

\begin{figure}[h]
    \centering
    \begin{subfigure}{0.49\linewidth}
        \centering
        \includegraphics[width=\linewidth,keepaspectratio]{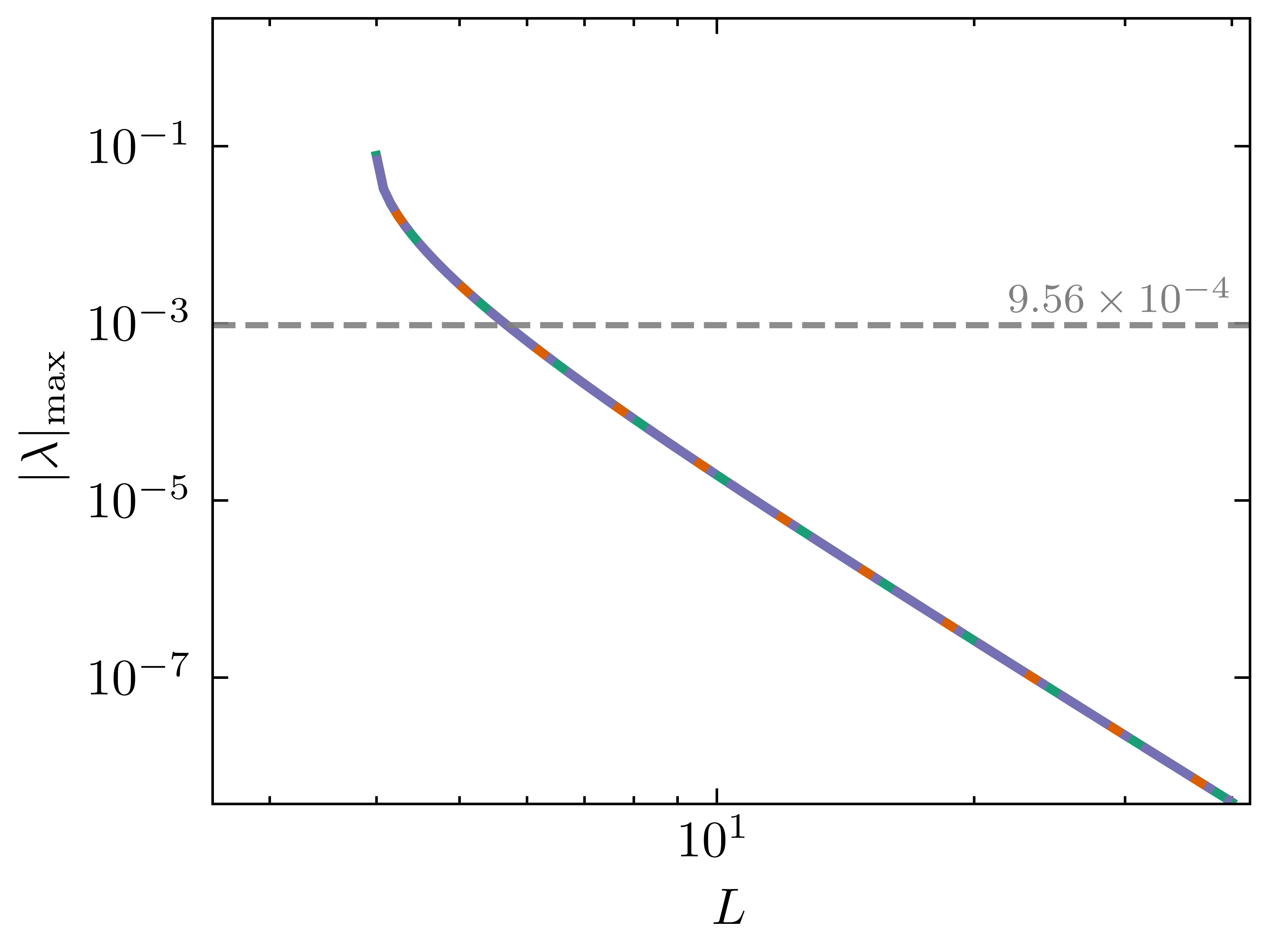}
        \caption{$a = 0.0$}
    \end{subfigure}
    \hfill
    \begin{subfigure}{0.49\linewidth}
        \centering
        \includegraphics[width=\linewidth,keepaspectratio]{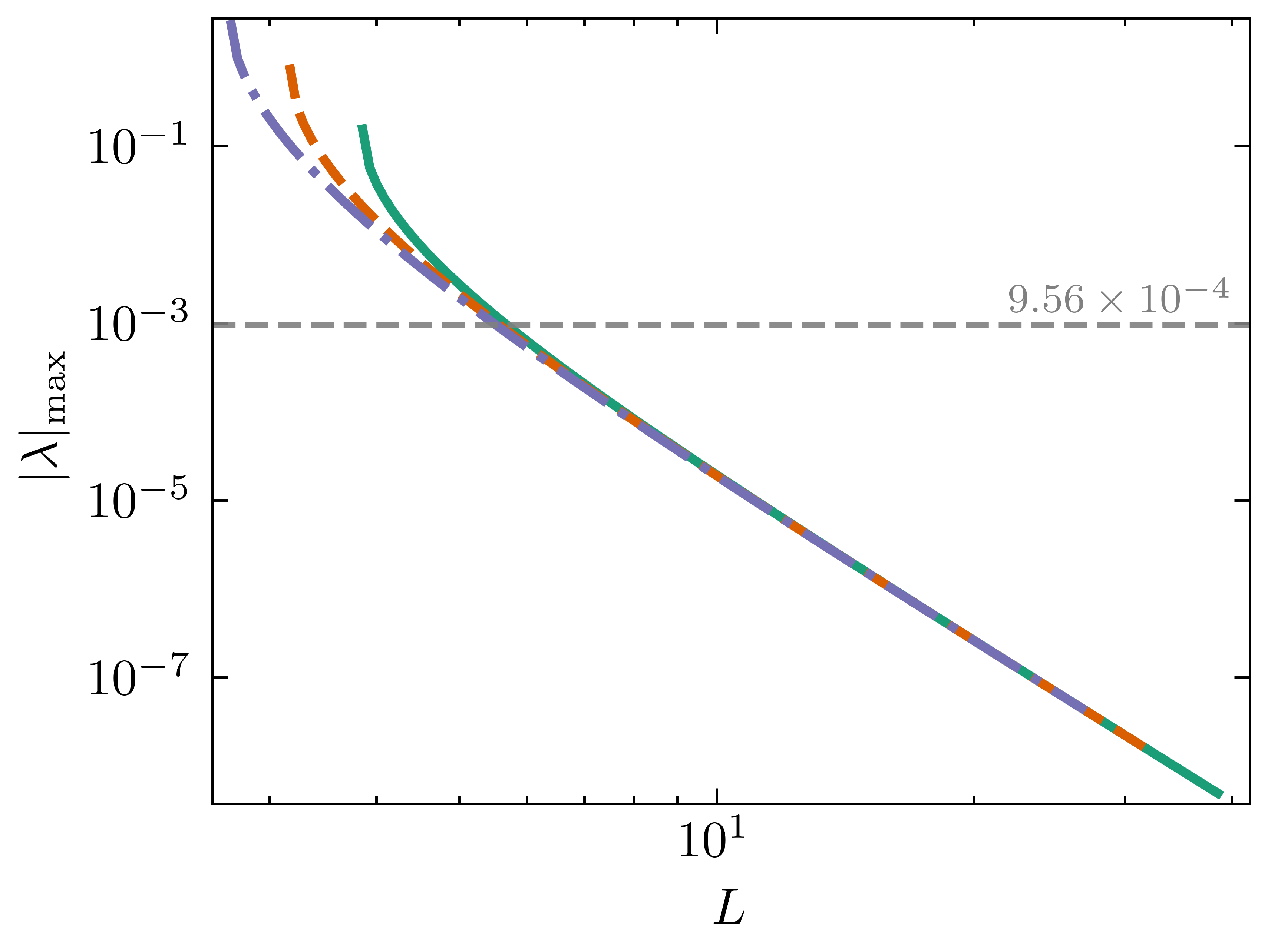}
        \caption{$a = 0.99$}
    \end{subfigure}
    \caption{The magnitude of the largest eigenvalue of the tidal tensor as a function of $L$ for different values of $x = L_{z}/L$ with $x = 0.0$ (green solid), $x = 0.5$ (orange dashed) and $x = 0.9$ (purple dash-dotted). Grey horizontal dashed line indicates the critical tidal eigenvalue $\left\lvert \lambda \right\rvert_{\text{crit}}$ for $M = 10^{7} M_{\odot}$, $M_{*} = 1 M_{\odot}$, $R_{*} = 1 R_{\odot}$.}
    \label{fig:lambda_vs_l}
\end{figure}

In Figure~\ref{fig:lambda_vs_l}, we plot $\left\lvert \lambda \right\rvert_{\text{max}}$ as a function of $L$ for different values of $x$. Note that in the GR convention, angular momenta are measured in units of $GMc^{-1}$ (see Equation~\ref{eq:units}). The eigenvalue increases with decreasing $L$ as a power law until the low-$L$ limit, where the curve turns as the pericentre approaches the IBSO radius and relativistic effects dominate. The intersection of the $\left\lvert \lambda \right\rvert_{\text{max}}$ curve with the horizontal $\left\lvert \lambda \right\rvert_{\text{crit}}$ line defines the critical angular momentum $L_{\text{tde}}$ for disruption.

\begin{figure}[h]
    \centering
    \begin{subfigure}{0.49\linewidth}
        \centering
        \includegraphics[width=\linewidth,keepaspectratio]{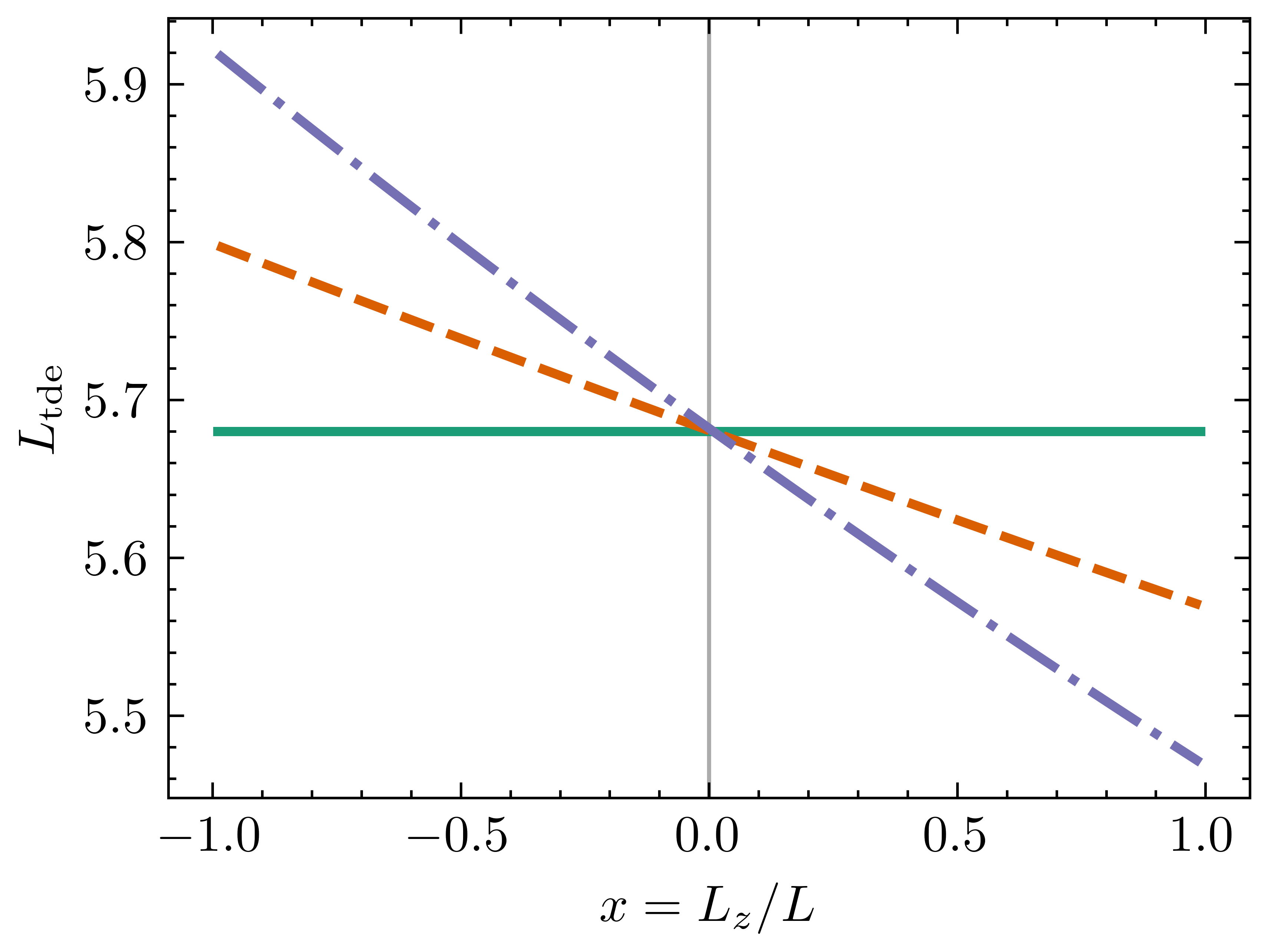}
        \caption{Tidal disruption boundary $L_{\text{tde}}(x)$}
        \label{fig:l_tde_vs_chi}
    \end{subfigure}
    \hfill
    \begin{subfigure}{0.49\linewidth}
        \centering
        \includegraphics[width=\linewidth,keepaspectratio]{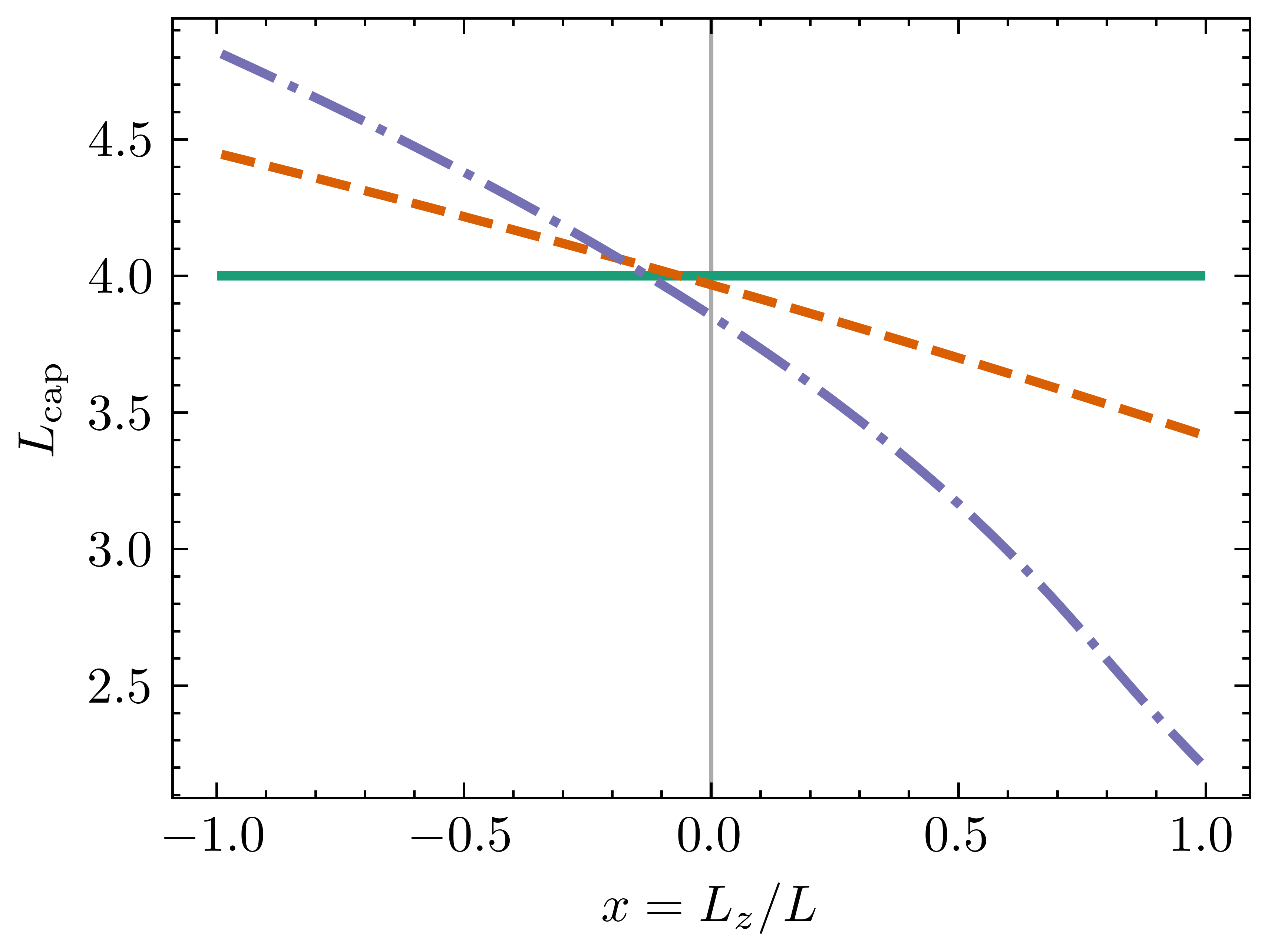}
        \caption{Direct capture boundary $L_{\text{cap}}(x)$}
        \label{fig:l_cap_vs_chi}
    \end{subfigure}
    \caption{Loss cone boundaries as a function of inclination $x = L_{z}/L$ for different BH spins $a$ with $a = 0.0$ (green solid), $a = 0.5$ (orange dashed) and $a = 0.99$ (purple dash-dotted). $L_{\text{tde}}$ is evaluated with $M = 10^{7} M_{\odot}$, $M_{*} = M_{\odot}$, $R_{*} = R_{\odot}$.}
    \label{fig:l_tde_lcap_vs_x}
\end{figure}

We search for $L_{\text{tde}}$ by first fixing the stellar parameters to $M_{*} = M_{\odot}$ and $R_{*} = R_{\odot}$ (representing a solar-type star) and choosing the BH mass $M$ and spin $a$. For each value of $x$, we determine the threshold $L_{\text{tde}}$ at which the stretching eigenvalue of $\tensor{C}{^{\tilde{\mu}}_{\tilde{\nu}}}_{\text{max}}$ equals $\left\lvert \lambda \right\rvert_{\text{crit}}$. The searching algorithm is available in the supporting code repository\footnote{\url{https://github.com/Walker-Xin/tidal_rate}}. The results are plotted in Figure~\ref{fig:l_tde_vs_chi} for three different BH spins.

For non-zero spin, $L_{\text{tde}}$ is lower for higher values of $x$. Since positive $x$ corresponds to prograde motion, this implies that prograde stars are \textit{less} likely to be tidally disrupted than retrograde ones. This is somewhat counterintuitive, as one might expect co-rotating orbits to be more susceptible to disruption. The reason is that for a given $L$, a positive $x$ ``pushes'' the first root of $R(r) = 0$ to a larger radius; prograde orbits have a larger pericentre than retrograde ones at fixed $L$. Since TDE occurrence depends primarily on pericentre depth, prograde stars are less likely to be disrupted than retrograde ones for the same $L$.

Observe that $L_{\text{tde}}$ in Figure~\ref{fig:l_tde_vs_chi} is almost linear in $x$, with a gradient much smaller than the value at $x = 0$. This means that one may write:
\begin{equation}
    L_{\text{tde}}(x) \approx L_{\text{tde}}^{(0)} (1 + \epsilon x) \, ,
    \label{eq:l_tde_linear_approx}
\end{equation}
where $\epsilon$ is a small, negative parameter that depends on $M$ and $a$.

\begin{figure}[h]
    \centering
    \includegraphics[width=0.6\linewidth,keepaspectratio]{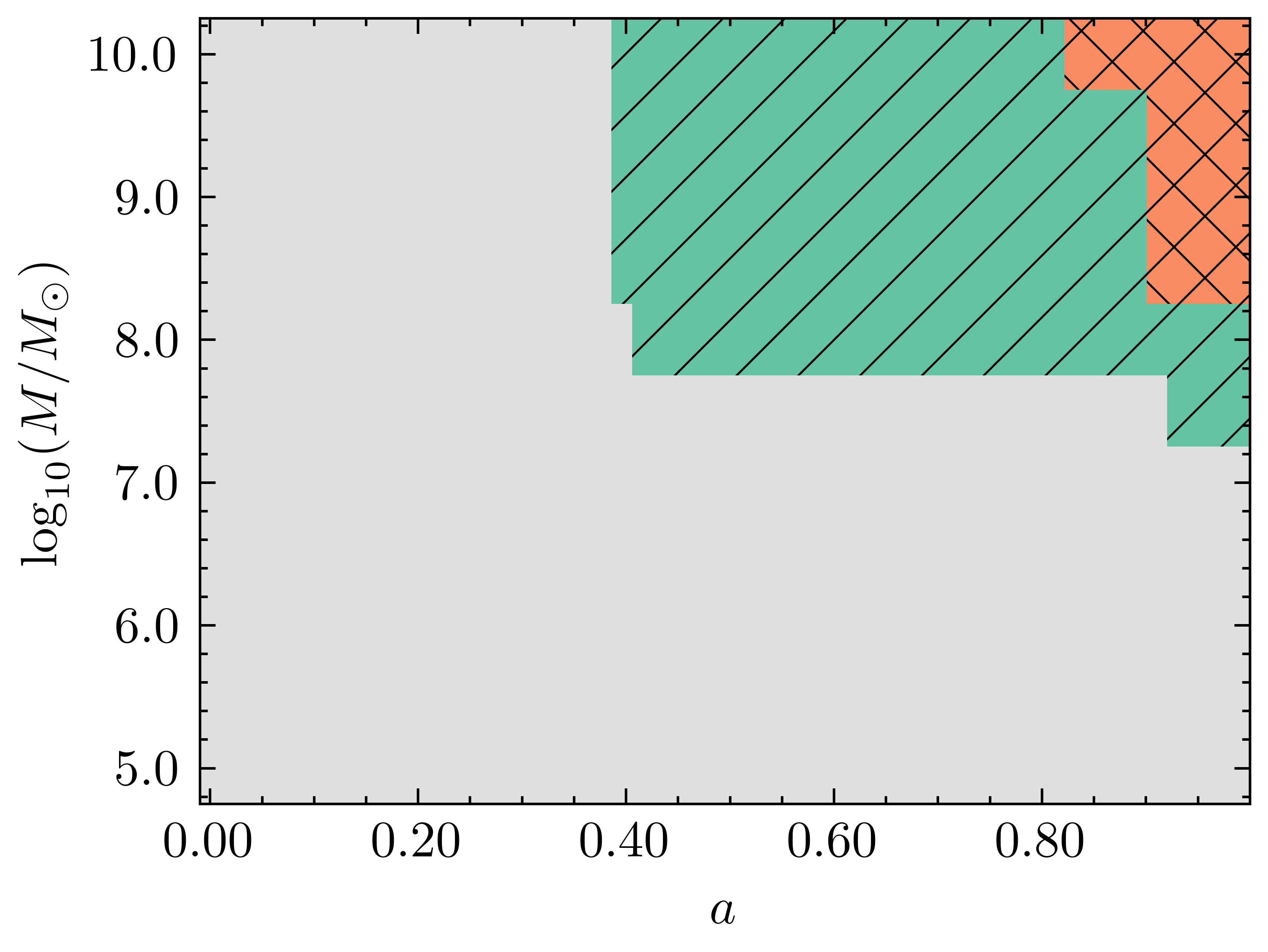}
    \caption{Heatmap showing failures of $L_{\text{tde}}(x) \approx L_{\text{tde}}^{(0)} (1 + \epsilon x)$ across the $(M,a)$ parameter grid, with both linear-fit and small-slope conditions satisfied (grey solid), small-slope violated (green diagonal-hatching), and both conditions violated (orange cross-hatching).}
    \label{fig:breakdown_class}
\end{figure}

We evaluate the validity of this linear approximation by calculating $L_{\text{tde}}(x)$ over a grid of $M$ and $a$. For each $(M, a)$ pair, we perform a linear regression on $L_{\text{tde}}(x)$ to extract the coefficient of determination $R^{2}$ and the slope parameter $\epsilon$. We define the linear fit as robust if $R^{2} > 0.99$, and the small-slope condition as satisfied if $|\epsilon| < 0.1$. Figure~\ref{fig:breakdown_class} demonstrates that the assumption of linearity is highly robust, failing only at the extreme high-mass, high-spin limit; the small-slope condition is more easily violated, though failure remains confined to the high-mass, high-spin regime. 

As will be shown in Chapter 4, the approximate linearity of the boundary ($R^{2} > 0.99$) is more structurally important than the strict small-slope condition ($|\epsilon| < 0.1$). Our analysis relies on a perturbative expansion in $\epsilon$, and therefore remains qualitatively well behaved even when the small-slope condition is marginally violated, provided that the boundary remains approximately linear.

A similar analysis can be performed for the direct capture boundary. Recall that $L_{\text{cap}}$ is defined by the IBSO conditions (Equations~\ref{eq:ibso_kerr}), which can be numerically solved for a given $x$ and $a$. Figure~\ref{fig:l_cap_vs_chi} plots $L_{\text{cap}}(x)$ for different spins. The capture threshold is much more sensitive to $x$, exhibiting strong non-linearities for moderate to high spins. This sensitivity arises because the capture boundary is defined by IBSOs that come very close to the horizon, where relativistic effects are most pronounced.

Figure~\ref{fig:l_tde_lcap_vs_x} defines the region of valid angular momentum for tidal disruption. For a given $x$, observable disruption only happens if $L_{\text{cap}}(x) \leq L \leq L_{\text{tde}}(x)$. Modulo the star's properties, there generally exists a critical value of BH mass $M = M_{\text{Hills}}$ such that $L_{\text{tde}} \leq L_{\text{cap}}$ for all $x$. The BH can no longer disrupt stars before they plunge into the horizon. This defines the Hills mass above which no TDEs can occur, and it turns out to be a strong function of the BH spin \cite{Xin2025}, as will be shown in the next chapter.

%% file: chapter2.tex
\chapter{One Dimensional Loss Cone Theory}

In this chapter, we turn to the stellar-dynamical problem of supplying stars to the black hole. We employ Newtonian stellar dynamics to model how relaxation processes drive stars into the relativistic loss cone determined in the previous chapter.

\section{Stellar Dynamics of Tidal Disruption Events} \label{sec:stellar_dynamics}

In dense galactic nuclei, stars are continually perturbed by relaxation processes that may scatter them onto near-radial orbits about the central BH. Based on the discussion in the previous chapter, any star with angular momentum $L < L_{\text{tde}}$ will be either tidally disrupted or directly captured by the BH within one orbital period. This defines the loss cone (LC) in the phase space of the stellar distribution, within which stars are removed on dynamical timescales. Meanwhile, relaxation processes cause stars to change their orbital parameters and repopulate the LC. The interplay between depletion and repopulation determines the rate of TDEs in a galaxy.

Although both correspond to the \textit{same} underlying stellar population, it is crucial to distinguish two groups of stars: the macroscopic field stars and the TDE-inducing test stars. Let $f_{\star}$ be the distribution function (DF) of the field stars representing the bulk of the galaxy, which behaves as an effectively collisionless, quasi-steady fluid. The field DF $f_{\star}$ is governed by Jeans' theorem and depends only on the integrals of motion (see e.g.\ \cite[Chapter 4]{Binney2008}).

On the other hand, let $f$ be the test star DF, which represents the subset of stars that can potentially be disrupted. The test DF $f$ cannot be described solely in terms of integrals of motion because its value changes drastically along an orbit: $f$ is depleted to zero right after pericentre passage, and is gradually replenished along the orbit. This means that $f$ must be a function of orbital phase, and a purely orbit-averaged Fokker--Planck treatment is insufficient.

The loss cone is depleted on the dynamical timescale $t_{\text{dyn}}$, which is roughly the orbital period of stars near the LC boundary. The repopulation timescale can be estimated as:
\begin{equation}
    t_{\text{rep}} \sim \frac{L_{\text{tde}}^{2}}{D} \, ,
\end{equation}
where the coefficient $D$ characterises the ``rate of change'' of the squared angular momentum of stars due to relaxation processes.

We define a dimensionless parameter $q$ to compare the depletion and repopulation timescales:
\begin{equation}
    q = \frac{t_{\text{dyn}}}{t_{\text{rep}}} \sim \frac{D P}{L_{\text{tde}}^{2}} \, ,
    \label{eq:q_parameter_heuristic}
\end{equation}
where the (radial) orbital period $P$ is used as an estimate of the dynamical timescale.

Two limiting regimes emerge. If repopulation is extremely efficient such that $t_{\text{rep}} \ll t_{\text{dyn}}$ and $q \gg 1$, any removed test star will be rapidly replaced by another with the same orbital parameters. This is the full LC regime, where the test DF $f$ becomes the same as the field DF $f_{\star}$ throughout phase space. The disruption rate is then directly proportional to the LC's volume.

Conversely, if the repopulation is inefficient such that $t_{\text{rep}} \gg t_{\text{dyn}}$ and $q \ll 1$, stars are removed much faster than they are replaced. This is the empty LC regime, where $f$ is approximately zero inside the boundary and only approaches $f_{\star}$ well outside it. The disruption rate is determined by the flux generated by $f$.

The most dynamically complex regime is the transition region where $q \sim 1$. In this case, $f$ must be split into two parts. The part interior to the LC develops a dependence on the orbital phase, while the exterior part is independent of the orbital phase but only approaches $f_{\star}$ asymptotically. To find the flux of stars, one needs to know the value of $f$ at the boundary, which can only be determined by solving for $f$ throughout the transition regime.

The consideration of $q$ highlights an important point sometimes overlooked in the literature: the ``fullness'' of the loss cone is an energy-dependent property as the period $P(E)$ varies strongly with energy. For weakly bound stars (i.e.\ long orbital periods), the loss cone is relatively full because $q$ is large. Conversely, highly bound stars (i.e.\ short orbital periods) occupy the empty regime. An understanding of all diffusion regimes is required to determine TDE rates in a given galaxy.

The following sections present a derivation of the classical 1D loss-cone theory, focusing primarily on diffusion in angular momentum magnitude due to two-body relaxation. Here, ``1D'' refers to diffusion in the angular momentum magnitude alone, while the ``2D'' theory developed in Chapter 4 also incorporates diffusion in the angular momentum orientation. Furthermore, although resonant relaxation can be highly efficient near the loss-cone boundary, the formalism can be adapted to include it by modifying the diffusion coefficients without structural changes. We discuss this in Chapter 5.

\section{1D Full Loss Cone Solution} \label{sec:full_loss_cone_rates}

In the full LC regime, replenishment occurs so rapidly that the test DF matches the prescribed field DF everywhere. Therefore, the TDE rate is simply a phase-space integral of $f_{\star}$ over the loss cone. The number of stars in a differential phase-space volume is given by (see e.g.\ \cite[Chapter 4]{Binney2008}):
\begin{equation}
    N \, \mathrm{d}^{3}\boldsymbol{\theta} \, \mathrm{d}^{3}\mathbf{J} = f_{\star}(\mathbf{J}) \, \mathrm{d}^{3}\boldsymbol{\theta} \, \mathrm{d}^{3}\mathbf{J} \, ,
\end{equation}
where we invoke action-angle coordinates $(\boldsymbol{\theta}, \mathbf{J})$ and Jeans' theorem in writing $f_{\star}$ as a function of actions $\mathbf{J}$ only.

Action-angle coordinates are canonical coordinates related to the usual position--velocity coordinates $(\mathbf{r}, \mathbf{v})$ via a canonical transformation (see e.g.\ \cite[Chapter 3]{Binney2008}). While infinitely many such canonical coordinate systems exist, for a set to qualify as action-angle variables, the actions must be constants of motion, ensuring that their conjugate angles evolve linearly in time.

Since we assume that the galaxy is isotropic, there are three Newtonian constants of motion: energy $E$, total angular momentum $L$, and axial component of the angular momentum $L_{z}$. Action variables can be constructed from these constants, with a common choice being $(L_{z}, L, J_{r})$, where the radial action $J_{r}$ is defined as:
\begin{equation}
    J_{r} = \frac{1}{\pi} \int_{r_{-}}^{r_{+}} \sqrt{2E - 2\Phi(r) - \frac{L^{2}}{r^{2}}} \, \mathrm{d}r = \frac{1}{\pi} \int_{r_{-}}^{r_{+}} v_{r} \, \mathrm{d}r \, ,
    \label{eq:radial_action}
\end{equation}
where $r_{-}$ and $r_{+}$ are the pericentre and apocentre of the orbit, respectively, $\Phi(r)$ is the gravitational potential of the galaxy, and $v_{r}$ is the radial velocity.

We can integrate over angles since the angle variables are periodic and the field DF is independent of angles. This gives:
\begin{equation}
    N \, \mathrm{d}^{3}\mathbf{J} = (2\pi)^{3} f_{\star}(\mathbf{J}) \, \mathrm{d}^{3}\mathbf{J} \, .
\end{equation}
Starting with the standard choice of action variables $(L_{z}, L, J_{r})$:
\begin{equation}
    N \, \mathrm{d}L_{z} \, \mathrm{d}L \, \mathrm{d}J_{r} = (2\pi)^{3} f_{\star}(L_{z}, L, J_{r}) \, \mathrm{d}L_{z} \, \mathrm{d}L \, \mathrm{d}J_{r} \, .
\end{equation}
The radial action $J_{r}$ can be changed to energy $E$ using the relation:
\begin{equation}
    \frac{\partial J_{r}}{\partial E} = \frac{1}{\pi} \int_{r_{-}}^{r_{+} } \frac{\mathrm{d}r}{v_{r}} = \frac{P(E, L)}{2\pi} \, ,
\end{equation}
where we define the radial orbital period:
\begin{equation}
    P(E, L) = 2 \int_{r_{-}}^{r_{+}} \frac{\mathrm{d}r}{v_{r}} \, .
    \label{eq:orbital_period}
\end{equation}
Replacing $L_{z}$ with the dimensionless inclination variable $x = L_{z}/L$ via the relation $\mathrm{d}L_{z} \, \mathrm{d}L = L \, \mathrm{d}x \, \mathrm{d}L$, we obtain:
\begin{equation}
    N \, \mathrm{d}x \, \mathrm{d}L \, \mathrm{d}E = (2\pi)^{2} P(E, L) f_{\star}(x, L, E) L \, \mathrm{d}x \, \mathrm{d}L \, \mathrm{d}E \, .
    \label{eq:phase_space_number_element}
\end{equation}
Note that the above steps are essentially a more explicit calculation of the Jacobian of the transformation from $(L_{z}, L, J_{r})$ to $(E, L, x)$.

\subsection{Integrating over the loss cone}

Because the background galaxy is isotropic, the field DF depends only on energy, i.e.\ $f_{\star} = f_{\star}(E)$. Furthermore, since the orbital period (Equation~\ref{eq:orbital_period}) depends only weakly on $L$, and TDE-inducing orbits are highly eccentric with $L \ll L_{c}$, we adopt the approximation $P(E, L) \approx P(E)$.

To evaluate the TDE rate, we integrate over the region of phase space that corresponds to observable TDEs, defined by $L_{\text{cap}} \leq L \leq L_{\text{tde}}$. Following the results of the previous chapter, both $L_{\text{cap}}$ and $L_{\text{tde}}$ are treated as functions of the inclination $x$ only with no dependence on energy.

The number of stars in the loss cone with energy $E$ and inclination $x$ is obtained by integrating over Equation~\ref{eq:phase_space_number_element} over $L$:
\begin{equation}
\begin{split}
        N \, \mathrm{d}x \, \mathrm{d}E &= (2\pi)^{2} P(E) f_{\star}(E) \, \mathrm{d}x \, \mathrm{d}E \int_{L_{\text{cap}}(x)}^{L_{\text{tde}}(x)} L \, \mathrm{d}L \\
        &= 2 \pi^{2} P(E) f_{\star}(E) \left[ L_{\text{tde}}^{2}(x) - L_{\text{cap}}^{2}(x) \right] \, \mathrm{d}x \, \mathrm{d}E \, .
\end{split}
\label{eq:number_element}
\end{equation}
To obtain the total rate, we divide the phase-space density by the orbital period $P(E)$. This yields a differential flux of:
\begin{equation}
    F \, \mathrm{d}x \, \mathrm{d}E = 2 \pi^{2} f_{\star}(E) \left[ L_{\text{tde}}^{2}(x) - L_{\text{cap}}^{2}(x) \right] \, \mathrm{d}x \, \mathrm{d}E \, ,
    \label{eq:flux_element}
\end{equation}
so the total TDE rate is given by the integral:
\begin{equation}
    \Gamma = 4 \pi^{2} \int_{-1}^{1} \frac{L_{\text{tde}}^{2}(x) - L_{\text{cap}}^{2}(x)}{2} \, \mathrm{d}x \int_{E} f_{\star}(E) \, \mathrm{d}E \, ,
    \label{eq:full_loss_cone_rate}
\end{equation}
where the integration over inclination represents an averaging over $x$.

This equation emphasises an important point: in the full LC regime, rate calculation is straightforward once the boundaries $L_{\text{tde}}$ and $L_{\text{cap}}$ are known. Because rapid repopulation maintains $f \approx f_{\star}$ everywhere, the rate calculation reduces to a purely geometric phase-space integral. Consequently, the spin-induced inclination dependence manifests as an averaging over the variable $x$. This explains how anisotropic boundaries can be already incorporated within the classical 1D framework.

\subsection{Comparing with previous numerical results}

To conclude the discussion of the full LC regime, we follow the calculation of Kesden \cite{Kesden2012} to obtain TDE rates for different BH masses and spins. Start by assuming a Maxwellian distribution function:
\begin{equation}
    f_{\star}(E) = \frac{n}{(2\pi \sigma^{2})^{3/2}} \exp{\left( -\frac{E}{\sigma^{2}} \right)} \, ,
\end{equation}
where $n$ is the number density of stars and $\sigma$ is the (one-dimensional) velocity dispersion in the hypothetical galactic nucleus.

Let us adopt the density profile of a singular isothermal sphere:
\begin{equation}
    \rho(r) = \frac{\sigma^{2}}{2\pi G r^{2}} \, ,
\end{equation}
with a cutoff at the radius of influence $r_{h} = GM/\sigma^{2}$.

\begin{figure}[h]
    \centering
    \begin{subfigure}[b]{0.49\linewidth}
        \includegraphics[width=\textwidth,keepaspectratio]{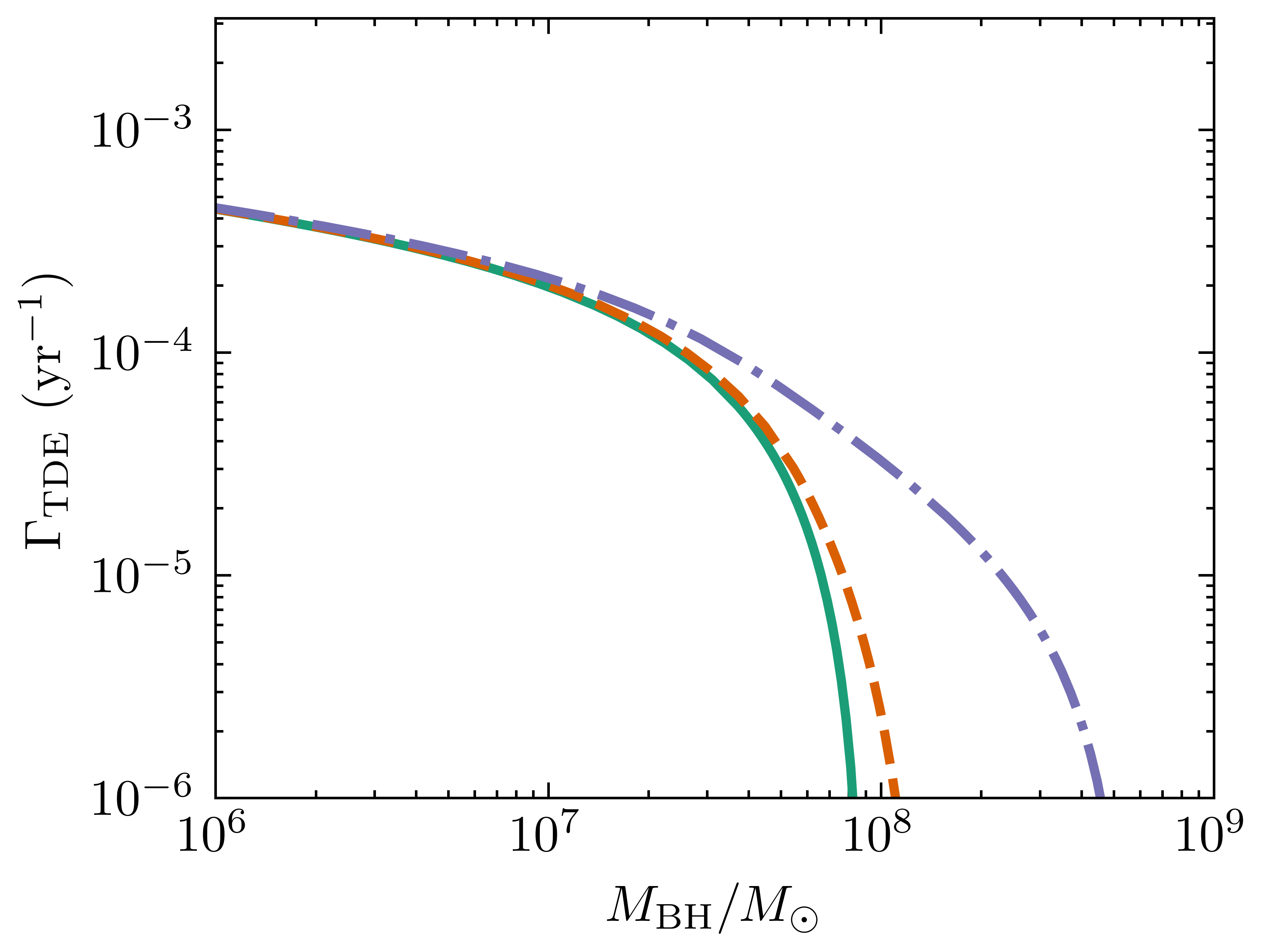}
        \caption{$a = 0$ (green solid), $a = 0.5$ (orange dashed), $a = 0.99$ (purple dash-dotted).}
    \end{subfigure}
    \hfill
    \begin{subfigure}[b]{0.49\linewidth}
        \includegraphics[width=\textwidth,keepaspectratio]{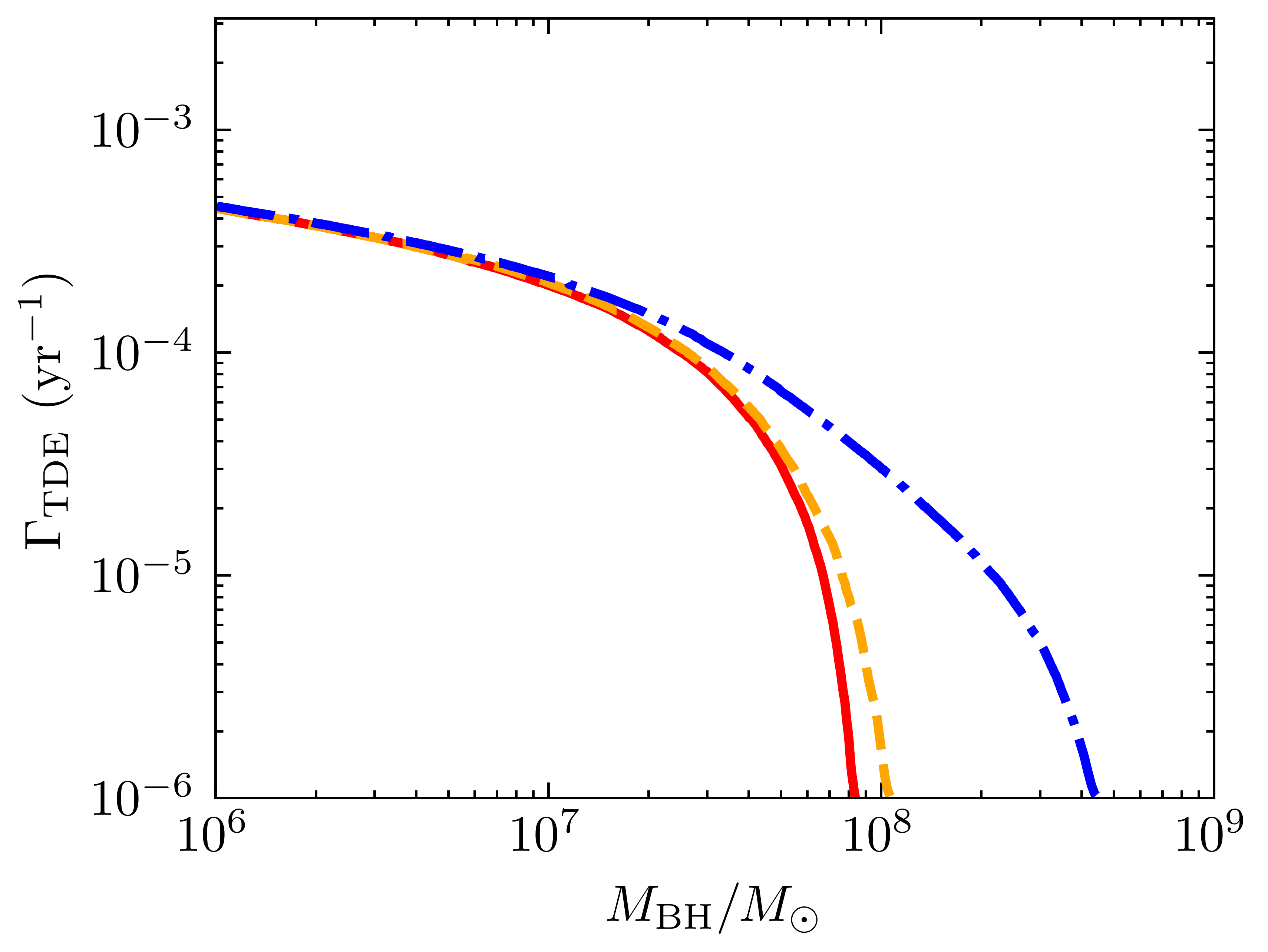}
        \caption{$a = 0$ (red solid), $a = 0.5$ (orange dashed), $a = 0.99$ (blue dash-dotted).}
    \end{subfigure}
    \caption{TDE rates as a function of BH mass for different BH spins. Left: Our calculations. Right: Reference results adapted from Kesden 2012 \cite{Kesden2012}.}
    \label{fig:tde_rate_vs_bh_mass}
\end{figure}

The energy integral evaluates to:
\begin{equation}
    \int_{0}^{\infty} f_{\star}(E) \, \mathrm{d}E = \frac{n}{(2\pi \sigma^{2})^{3/2}} \int_{0}^{\infty} \exp{\left( -\frac{E}{\sigma^{2}} \right)} \, \mathrm{d}E = \frac{n}{(2\pi)^{3/2} \sigma} \, .
\end{equation}
Substituting $n = \rho(r = r_{h})/M_{*}$ into Equation~\ref{eq:full_loss_cone_rate} yields:
\begin{equation}
    \Gamma = \frac{\sigma^{5}}{(2\pi)^{1/2} G^{3} M^{2} M_{*}} \int_{-1}^{1} \frac{L_{\text{tde}}^{2}(x) - L_{\text{cap}}^{2}(x)}{2} \, \mathrm{d}x \, .
\end{equation}
At this point, Kesden \cite{Kesden2012} resorted to Monte Carlo simulations of geodesics to determine disruption rates, as it was not possible to calculate $L_{\text{tde}}(x)$ analytically. With the semi-analytical framework for $L_{\text{tde}}(x)$ and $L_{\text{cap}}(x)$ established in Chapter 2, the TDE rate can now be computed directly for arbitrary BH parameters.
Following Kesden, we adopt the $M-\sigma$ relation from Schulze \& Gebhardt \cite{Schulze2011}:
\begin{equation}
    \frac{M}{10^{6} M_{\odot}} = 7.58 \left( \frac{\sigma}{100 \, \mathrm{km/s}} \right)^{4.32} \, .
    \label{eq:m_sigma_relation}
\end{equation}
Figure~\ref{fig:tde_rate_vs_bh_mass} plots the TDE rates as a function of BH mass, comparing our semi-analytical calculations with Kesden's numerical results \cite{Kesden2012}. We find excellent agreement in the overall shape of the curves. The minor discrepancies at the high-mass cutoffs reflect methodological differences: our approach computes the Hills mass limit semi-analytically, whereas the Monte Carlo geodesic simulations of Kesden are subject to finite sampling resolution near the cutoff.

\section{1D Diffusion Equation} \label{sec:1d_diffusion_equation}

While the full loss cone rates are straightforward to evaluate, they are only valid when replenishment is sufficiently rapid. As discussed in Section~\ref{sec:stellar_dynamics}, a complete description of TDE rates requires modelling the intermediate and empty regimes. This requires solving for the test DF $f$, which is depleted by the BH and replenished on orbital timescales. Below, we reproduce the derivation of the classical loss cone formalism, following Merritt \cite[Chapter 6]{Merritt2013a}.

Consider the collisional Boltzmann equation governing the test DF, written in action-angle coordinates:
\begin{equation}
    \frac{\partial f}{\partial t} + \dot{\boldsymbol{\theta}} \cdot \frac{\partial f}{\partial \boldsymbol{\theta}} + \dot{\mathbf{J}} \cdot \frac{\partial f}{\partial \mathbf{J}} = \left( \frac{\partial f}{\partial t} \right)_{c} \, .
    \label{eq:boltzmann_equation}
\end{equation}
Because the system is in steady state ($\partial f / \partial t = 0$), and the actions are constants of motion along unperturbed orbits ($\dot{\mathbf{J}} = 0$), the first and third terms vanish. Since the background potential is spherically symmetric, $f$ depends only on the radial angle $\theta_{r}$. The left-hand side therefore simplifies to:
\begin{equation}
    \dot{\theta_{r}} \frac{\partial f}{\partial \theta_{r}} = \left( \frac{\mathrm{d}\theta_{r}}{\mathrm{d}r} \frac{\mathrm{d}r}{\mathrm{d}t} \right) \left( \frac{\partial f}{\partial r} \frac{\mathrm{d}r}{\mathrm{d}\theta_{r}} \right) = v_{r} \frac{\partial f}{\partial r} \, ,
\end{equation}
where we identify the radial velocity $v_{r} = \mathrm{d}r/\mathrm{d}t$.

Following standard loss-cone theory, we neglect diffusion in energy $E$, leaving only diffusion in angular momentum $L$. The reason is twofold. First, the LC boundary is nearly independent of energy, so diffusion in $E$ contributes minimally to replenishing the loss cone. Second, diffusion in $L$ is much more rapid than in $E$, thereby dominating the replenishment process \cite{Lightman1977}. Consequently, one may treat energy as a fixed parameter and restrict the evolution of $f$ to $L$ at each $E$. Under these assumptions, the collisional term can be written as \cite[Chapter 5]{Merritt2013a}:
\begin{equation}
    \left( \frac{\partial f}{\partial t} \right)_{c} = - \frac{\partial }{\partial R}(f \langle \Delta R \rangle) + \frac{1}{2} \frac{\partial^{2}}{\partial R^{2}}(f \langle (\Delta R)^{2} \rangle) \, ,
    \label{eq:collisional_term}
\end{equation}
where a dimensionless variable is introduced:
\begin{equation}
    R \equiv \frac{L^{2}}{L_{c}^{2}(E)} \, ,
\end{equation}
where $L_{c}(E)$ is the angular momentum of a circular orbit with energy $E$.

The dimensionless variable $R$ is used because its domain is bounded between $0$ and $1$. TDE-inducing orbits are highly eccentric, corresponding to $R \ll 1$. In this limit, the local diffusion coefficients in $R$ are given by (see Appendices~\ref{app:diffusion_derivation} and~\ref{app:diffusion_derivation_coordinate_transformation}):
\begin{equation}
\begin{split}
    \langle \Delta R \rangle &= \frac{r^{2} \langle (\Delta v_{\perp})^{2} \rangle}{L_{c}^{2}} + \mathcal{O}(R) \, , \\
    \langle (\Delta R)^{2} \rangle &= \frac{2 r^{2} \langle (\Delta v_{\perp})^{2} \rangle}{L_{c}^{2}} R + \mathcal{O}(R^{2}) \, ,
\end{split}
\end{equation}
so that to the lowest order:
\begin{equation}
    \langle \Delta R \rangle = \frac{1}{2} \frac{\partial }{\partial R} \langle (\Delta R)^{2} \rangle \, .
    \label{eq:R_differential_relation}
\end{equation}
Here, $\langle (\Delta v_{\perp})^{2} \rangle$ (also $\langle \Delta v_{\parallel} \rangle$ and $\langle (\Delta v_{\parallel})^{2} \rangle$) represent the local diffusion coefficients in ordinary velocity space. They can be calculated from first principles by considering the cumulative effects of two-body scattering (see e.g.\ \cite[Appendix L]{Binney2008}). The details are omitted here, as their precise forms are irrelevant for our discussion; instead, we treat them as known quantities once a field DF is specified.

The collisional term can be rewritten in a more compact form:
\begin{equation}
    \left( \frac{\partial f}{\partial t} \right)_{c} = \frac{1}{2} \frac{\partial }{\partial R} \left[ \langle (\Delta R)^{2} \rangle \frac{\partial f}{\partial R} \right] \, ,
    \label{eq:collisional_term_compact}
\end{equation}
so that the Boltzmann equation becomes:
\begin{equation}
    v_{r} \frac{\partial f}{\partial r} = \frac{\langle (\Delta R)^{2} \rangle}{2R} \frac{\partial }{\partial R} \left( R \frac{\partial f}{\partial R} \right) = D \frac{\partial }{\partial R} \left( R \frac{\partial f}{\partial R} \right) \, ,
    \label{eq:boltzmann_equation_simplified}
\end{equation}
where the factor $ \langle (\Delta R)^{2} \rangle / 2R$ is pulled out because it is independent of $R$ to the lowest order:
\begin{equation}
    D \equiv \lim_{R \to 0} \frac{\langle (\Delta R)^{2} \rangle}{2R} = \frac{r^{2} \langle (\Delta v_{\perp})^{2} \rangle}{L_{c}^{2}} \, ,
    \label{eq:diffusion_rate}
\end{equation}
where we note that $D$ can be viewed as a ``diffusion rate'' in $R$-space.

Equations~\ref{eq:R_differential_relation} and~\ref{eq:collisional_term_compact} have a deeper physical significance beyond mathematical convenience. In a general stellar dynamical Fokker--Planck formulation, the drift coefficient $\langle \Delta R \rangle$ consists of two distinct components: a gradient of the diffusion tensor $\partial_{R} \langle (\Delta R)^{2} \rangle/2$ driven by random fluctuations, and a dynamical friction term representing the ``back reaction'' of the field stars responding to the test star's wake \cite{Binney1988}. The fact that Equation~\ref{eq:R_differential_relation} holds means that this dynamical friction vanishes for highly eccentric orbits, and the total drift is entirely determined by the diffusion gradient. This explains why the collisional term collapses into a flux-conserving divergence form $\sim \partial_{R}\left[ R (\partial_{R} f) \right]$. We discuss this in Chapter 5.

Following Cohn \& Kulsrud \cite{Cohn1978}, we define a new variable $\tau$:
\begin{equation}
\begin{split}
    \tau(E, R) &\equiv \int_{r_{-}}^{r} \frac{\mathrm{d}r}{v_{r}} \frac{\langle (\Delta R)^{2} \rangle}{2R} \Bigg/ \oint \frac{\mathrm{d}r}{v_{r}} \frac{\langle (\Delta R)^{2} \rangle}{2R} \, , \\
    &= \frac{1}{\bar{D}(E)P(E)} \int_{r_{-}}^{r} \frac{\mathrm{d}r}{v_{r}} D \, ,
\end{split}
\label{eq:tau_definition}
\end{equation}
where $P(E)$ is the orbital period and $\bar{D}(E)$ is the orbit-averaged diffusion rate:
\begin{equation}
    \bar{D}(E) = \frac{1}{P(E)} \oint \frac{\mathrm{d}r}{v_{r}} D \, .
    \label{eq:orbit_averaged_diffusion_rate}
\end{equation}
The variable $\tau$ is unrelated to physical time, serving instead as a dimensionless, normalised coordinate for the orbital phase. As $\tau$ increases from $0$ to $1$, the star moves from pericentre $r_{-}$ to apocentre $r_{+}$, and back to pericentre. In terms of $\tau$, the Boltzmann equation (Equation~\ref{eq:boltzmann_equation_simplified}) becomes:
\begin{equation}
    \frac{\partial f}{\partial \tau} = P(E) \bar{D}(E) \frac{\partial }{\partial R} \left( R \frac{\partial f}{\partial R} \right) \, ,
    \label{eq:boltzmann_intermediate}
\end{equation}
which has a similar form to the heat equation in cylindrical coordinates.

To make this resemblance explicit, introduce a new variable $y$:
\begin{equation}
    y = \frac{R}{P(E) \bar{D}(E)} \, ,
\end{equation}
resulting in a diffusion equation for $f(\tau, y)$:
\begin{equation}
    \frac{\partial f}{\partial \tau} = \frac{\partial }{\partial y} \left( y \frac{\partial f}{\partial y} \right) \, .
    \label{eq:boltzmann_cylindrical}
\end{equation}

Now consider the boundary conditions. Outside the LC ($y > y_{\text{lc}}$), two-body relaxation operates on timescales much longer than the orbital period, so $f$ is independent of orbital phase. Inside the LC ($y < y_{\text{lc}}$), $f$ develops a dependence on $\tau$. Since no stars survive passage through the disruption zone, we require:
\begin{equation}
    f(0^{+}, y) = 0 \, ,
\end{equation}
which states that the distribution is depleted immediately after successive pericentre passages but can be replenished along the orbit.

Finally, a smoothness condition must also be imposed at $y = 0$:
\begin{equation}
    \frac{\partial f}{\partial L} \propto \frac{\partial f}{\partial \sqrt{y}} = 0 \, ,
\end{equation}
so that the flux vanishes at $L = 0$ as negative angular momentum is unphysical.

\section{Boundary Layer Problem} \label{sec:boundary_layer_problem}

Solving and matching Equation~\ref{eq:boltzmann_cylindrical} is termed a ``boundary layer'' problem \cite{Merritt2013a} because the loss cone represents an extremely narrow region in angular momentum space where the test DF $f$ exhibits steep gradients. The equation can be solved inside and outside the LC separately. Mathematically, the interior solution is analogous to a heat-conduction problem of a cylinder that starts at zero temperature and is warmed by a steady temperature $f(y_{\text{lc}})$ at its outer edge. Solving for the unknown edge value $f(y_{\text{lc}})$ requires matching the fluxes of the interior and exterior solutions at the boundary.

Inspecting the form of Equation~\ref{eq:boltzmann_intermediate}, the flux of stars is given by:
\begin{equation}
    F(E) = 4\pi^{2} P(E) \bar{D}(E) L_{c}^{2}(E) \left( R \frac{\partial f}{\partial R} \right) \, ,
\end{equation}
where the factor $4\pi^{2} L_{c}^{2}$ can be inferred from Equation~\ref{eq:flux_element}.

We now determine the interior and exterior solutions and match their fluxes to solve for $f(y_{\text{lc}})$.

\subsection{Interior solution}

Equation~\ref{eq:boltzmann_cylindrical} can be solved inside the LC using separation of variables. This standard but tedious calculation is postponed to Appendix~\ref{app:diffusion_solution}. The solution is:
\begin{equation}
    f(\tau, y) = f(y_{\text{lc}}) \left[ 1 - \frac{2}{\sqrt{y_{\text{lc}}}} \sum_{m = 1}^{\infty} \frac{\exp{( -\beta_{m}^{2}\tau/4 )}}{\beta_{m}} \frac{J_{0}(\beta_{m} \sqrt{y})}{J_{1}(\beta_{m} \sqrt{y_{\text{lc}}})} \right] \, ,
    \label{eq:inside_loss_cone_solution}
\end{equation}
where $J_{0}$ and $J_{1}$ are Bessel functions of the first kind of order 0 and 1, respectively, and $\beta_{m} \sqrt{y_{\text{lc}}}$ are consecutive zeros of $J_{0}$.

The flux of stars at the boundary is:
\begin{equation}
    F(E) = 4\pi^{2} P \bar{D} L_{c}^{2} \int_{0}^{1} \left( y \frac{\partial f}{\partial y} \right)_{y = y_{\text{lc}}} \, \mathrm{d}\tau \, ,
\end{equation}
where integration over $\tau$ is necessary because the flux varies along the orbit.

Instead of carrying out the differentiation, we can be eliminate the $y$-derivative by integrating both sides of Equation~\ref{eq:boltzmann_cylindrical} from $0$ to $y_{\text{lc}}$:
\begin{equation}
    \int_{0}^{y_{\text{lc}}} \frac{\partial f}{\partial \tau} \, \mathrm{d}y = \int_{0}^{y_{\text{lc}}} \frac{\partial }{\partial y} \left( y \frac{\partial f}{\partial y} \right) \, \mathrm{d}y = \left( y \frac{\partial f}{\partial y} \right)_{y = y_{\text{lc}}} \, .
\end{equation}
This yields an expression for the flux in agreement with Merritt \cite[Equation 6.57]{Merritt2013a}:
\begin{equation}
\begin{split}
        F(E) &= 4\pi^{2} P(E) \bar{D}(E) L_{c}^{2}(E) \int_{0}^{1} \int_{0}^{y_{\text{lc}}} \frac{\partial f}{\partial \tau} \, \mathrm{d}y \, \mathrm{d}\tau \\
        &= 4\pi^{2} P(E) \bar{D}(E) L_{c}^{2}(E) \int_{0}^{y_{\text{lc}}} f(1, y) \, \mathrm{d}y \, ,
\end{split}
\label{eq:flux_inside_loss_cone_integral}
\end{equation}
where $f(0^{+}, y) = 0$ is used to eliminate the lower limit of the integral over $\tau$.

Using the properties of Bessel functions, this integral evaluates to:
\begin{equation}
\begin{split}
    \int_{0}^{y_{\text{lc}}} f(1, y) \, \mathrm{d}y &= f(y_{\text{lc}}) \left[ y_{\text{lc}} - \frac{2}{\sqrt{y_{\text{lc}}}} \sum_{m = 1}^{\infty} \frac{\exp{( -\beta_{m}^{2}/4 )}}{\beta_{m} J_{1}(\beta_{m}\sqrt{y_{\text{lc}}})} \int_{0}^{y_{\text{lc}}} J_{0}(\beta_{m} \sqrt{y}) \, \mathrm{d}y \right] \\
    &= f(y_{\text{lc}}) y_{\text{lc}} \left[ 1 - 4 \sum_{m = 1}^{\infty} \frac{\exp{(-\alpha_{m}^{2}q/4)}}{\alpha_{m}^{2}} \right] \\
    &= f(y_{\text{lc}}) y_{\text{lc}} \xi(q) \, ,
\end{split}
\label{eq:lc_flux_difficult_integral}
\end{equation}
where $\alpha_{m}$ are the zeros of $J_{0}$, the fullness parameter $q$ is defined as:
\begin{equation}
    q(E) \equiv \frac{P(E) \bar{D}(E)}{R_{\text{lc}}} = \frac{1}{y_{\text{lc}}(E)} \, ,
    \label{eq:fullness_parameter}
\end{equation}
and we define a special function $\xi(q)$:
\begin{equation}
    \xi(q) \equiv 1 - 4 \sum_{m = 1}^{\infty} \frac{\exp{(-\alpha_{m}^{2}q/4)}}{\alpha_{m}^{2}} \, .
    \label{eq:xi_function}
\end{equation}
Changing back to $R$-space, the flux at the LC boundary can be written as:
\begin{equation}
    F(E) = 4\pi^{2} L_{c}^{2}(E) R_{\text{lc}}(E) f(R_{\text{lc}}) \xi(q) \, .
\end{equation}

\subsection{Exterior solution}

Outside the loss cone, the test DF is independent of $\tau$. Setting $\partial f/\partial \tau = 0$, Equation~\ref{eq:boltzmann_intermediate} can be integrated to give:
\begin{equation}
    f(R) = A \ln{(R/R_{0})} \, .
\end{equation}

We demand that $f(R = 1) = f_{\star}(E, L_{c})$, where $f_{\star}(E, L_{c})$ is the field DF fitted to the cluster without regard to TDEs. This yields the exterior solution:
\begin{equation}
    f(R) = f_{\star}(E, L_{c}) \frac{\ln{(R/R_{0})}}{\ln{(1/R_{0})}} \, .
    \label{eq:outside_loss_cone_solution}
\end{equation}
The flux demanded by the exterior solution is:
\begin{equation}
\begin{split}
    F(E) &= 4\pi^{2} P(E) \bar{D}(E) L_{c}^{2}(E) \left( R \frac{\partial f}{\partial R} \right)_{R = R_{\text{lc}}} \\
    &= 4\pi^{2} P(E) \bar{D}(E) L_{c}^{2}(E) \frac{f_{\star}(E, L_{c})}{\ln{(1/R_{0})}} \, .
    \label{eq:flux_outside_loss_cone}
\end{split}
\end{equation}
Note that the exterior flux is independent of where the derivative is evaluated, which is expected because flux is globally conserved in a steady state system.

\subsection{Matching the solutions}

Because the fluxes demanded by the interior and exterior solutions must match at the boundary, we can solve self-consistently for the unknown value $f(R_{\text{lc}})$:
\begin{equation}
    f(R_{\text{lc}}) = \frac{PD}{R_{\text{lc}} \xi} \frac{f_{\star}(E, L_{c})}{\ln{(1/R_{0})}} = \frac{q}{\xi} \frac{f_{\star}(E, L_{c})}{\ln{(1/R_{0})}} \, .
\end{equation}
Substituting this back into the exterior solution (Equation~\ref{eq:outside_loss_cone_solution}) yields a relation between $R_{0}$ and $R_{\text{lc}}$:
\begin{equation}
    \ln{(R_{\text{lc}}/R_{0})} = \frac{q}{\xi(q)} \implies R_{0}(E) = R_{\text{lc}}(E) \exp{[-q/\xi(q)]} \, .
    \label{eq:R0_Rlc_relation}
\end{equation}
Finally, substituting these relations back into the flux expressions yields two equivalent forms of the stellar flux into the loss cone:
\begin{equation}
    \begin{split}
        F(E) &= 4\pi^{2} L_{c}^{2} R_{\text{lc}} \frac{f_{\star}(E, L_{c})}{\xi^{-1} + q^{-1} \ln{(1/R_{\text{lc}})}} \\
        &= 4\pi^{2} P(E) \bar{D}(E) L_{c}^{2} \frac{f_{\star}(E, L_{c})}{\ln{(1/R_{0})}} \, .
    \end{split}
    \label{eq:final_flux}
\end{equation}

$R_{0}$ is the value of $R$ where the exterior solution would extrapolate to zero if it were (naively) extended into the loss cone. Equation~\ref{eq:R0_Rlc_relation} shows that $R_{0} < R_{\text{lc}}$ for any finite $q$, and that $R_{0}$ approaches the boundary $R_{\text{lc}}$ only in the empty loss-cone limit $q \to 0$.

\subsection{Different loss cone regimes}

Let us look at the loss-cone fullness parameter $q$ in more detail:
\begin{equation}
    q(E) = \frac{P(E)}{R_{\text{lc}}/\bar{D}(E)} \, .
\end{equation}
The numerator is the orbital period, which quantifies the dynamical timescale on which the LC is depleted; the denominator is the ratio of the LC size to the rate of change in $R$, which quantifies the diffusion timescale. This formal result matches the heuristic version introduced in Equation~\ref{eq:q_parameter_heuristic}. The fullness parameter $q$ emerges naturally from the boundary layer analysis, and it controls the behaviour of the test DF, thereby determining the disruption rate in different loss cone regimes.

\textbf{Empty regime ($q \ll 1$):} In this limit, stars are highly bound and close to the BH. The approximation $R_{0} \approx R_{\text{lc}}$ holds because the loss cone is nearly empty, and the test DF drops to zero at the boundary. The TDE rate is bottlenecked by the diffusion flux across the boundary. Substituting $R_{0} \approx R_{\text{lc}}$ into the second expression of Equation~\ref{eq:final_flux} yields:
\begin{equation}
    F_{\text{empty}}(E) = 4\pi^{2} P(E) \bar{D}(E) L_{c}^{2}(E) \frac{f_{\star}(E, L_{c})}{\ln{(1/R_{\text{lc}})}} \, .
    \label{eq:final_empty_loss_cone_flux}
\end{equation}

\textbf{Full regime ($q \gg 1$):} In this limit, stars are weakly bound and far from the BH. $R_{0} \to 0$ and the test DF remains nearly equal to the field DF throughout most of the loss cone, dropping to zero only in a narrow region near $R = 0$. The disruption rate is determined by the instantaneous phase-space density inside the loss cone. Setting $\xi \to 1$ (see Equation~\ref{eq:xi_function}) in the first expression of Equation~\ref{eq:final_flux} yields:
\begin{equation}
    F_{\text{full}}(E) = 4\pi^{2} L_{c}^{2}(E) R_{\text{lc}}(E) f_{\star}(E, L_{c}) \, ,
    \label{eq:final_full_loss_cone_flux}
\end{equation}
which agrees with Equation~\ref{eq:flux_element} if direct capture and inclination dependence of the boundaries are ignored.

\textbf{Intermediate regime ($q \sim 1$):} Both the diffusion flux at the boundary and the phase-space density inside the LC contribute to the rate. In realistic galactic nuclei, this regime must be modelled accurately because a significant fraction of TDEs originate from energies near this transition (see e.g.\ \cite{Magorrian1999,Wang2004}). The total rate is obtained by integrating Equation~\ref{eq:final_flux} over energy:
\begin{equation}
    \Gamma = 4\pi^{2} \int L_{c}^{2} R_{\text{lc}} \frac{f_{\star}(E, L_{c})}{\xi^{-1} + q^{-1} \ln{(1/R_{\text{lc}})}} \, \mathrm{d}E \, .
    \label{eq:final_tde_rate}
\end{equation}
This equation completes the classical one-dimensional loss-cone theory, providing a solution for the TDE rate across all loss cone regimes.

\subsection{A reciprocal-sum approximation} \label{sec:bypassing_boundary_layer_analysis}

A useful approximation follows from noting that:
\begin{equation}
    \frac{1}{F_{\text{full}}(E)} + \frac{1}{F_{\text{empty}}(E)} = \frac{1}{4\pi^{2} L_{c}^{2} f_{\star} R_{\text{lc}}} \left[ 1 + \frac{R_{\text{lc}}}{P \bar{D}} \ln{(1/R_{\text{lc}})} \right] \approx \frac{1}{F(E)} \, ,
\end{equation}
where the last step follows from setting $\xi(q) \approx 1$ in Equation~\ref{eq:final_flux}, which is valid as long as $q$ is not too small \cite[Chapter 6]{Merritt2013a}.

The reciprocal sum of the full and empty fluxes provides a good approximation to the actual flux. This structure is physically transparent, yet to our knowledge has not been identified explicitly in the literature. We will use it to motivate an approximate treatment of the intermediate regime in the two-dimensional loss-cone problem.

\section{Direct Capture Rates} \label{sec:direct_capture_rates}

While the 1D boundary-layer analysis has been widely discussed in the literature (e.g., \cite{Cohn1978, Merritt2013, Chang2025}), many studies accounted for direct capture only in the empty and full LC limits. Servin \& Kesden \cite{Servin2017} recognised the importance of direct capture in the intermediate regime. They determined the capture rate by computing the flux of stars into the capture region $R \le R_{\text{cap}}$, where $R_{\text{cap}} = L_{\text{cap}}^{2}/L_{c}^{2}$. The interior solution (Equation~\ref{eq:inside_loss_cone_solution}) can be integrated over this narrower interval:
\begin{equation}
    F_{\text{cap}}(E) = 4\pi^{2} P(E) \bar{D}(E) L_{c}^{2}(E) \int_{0}^{y_{\text{cap}}} f(1, y) \, \mathrm{d}y \, .
    \label{eq:capture_flux_integral}
\end{equation}
Servin \& Kesden then proceeded to evaluate the observable TDE rate by subtracting the capture flux integral from the total flux integral (Equation~\ref{eq:flux_inside_loss_cone_integral}). We show that the subtraction can be reduced to a transparent analytical expression.

Let the ratio of the capture angular momentum to the tidal disruption angular momentum be defined as $\varphi_{c} \equiv L_{\text{cap}}/L_{\text{tde}} = \sqrt{y_{\text{cap}}/y_{\text{lc}}}$. Evaluating the above integral following Equation~\ref{eq:lc_flux_difficult_integral} yields a similar expression:
\begin{equation}
\begin{split}
    \int_{0}^{y_{\text{cap}}} f(1, y) \, \mathrm{d}y &= f(y_{\text{lc}}) \left[ y_{\text{cap}} - \frac{2}{\sqrt{y_{\text{lc}}}} \sum_{m = 1}^{\infty} \frac{\exp{(-\beta_{m}^{2}/4)}}{\beta_{m} J_{1}(\beta_{m}\sqrt{y_{\text{lc}}})} \int_{0}^{y_{\text{cap}}} J_{0}(\beta_{m} \sqrt{y}) \, \mathrm{d}y \right] \\
    &= f(y_{\text{lc}}) y_{\text{lc}} \left[ \varphi_{c}^{2} - 4 \varphi_{c} \sum_{m = 1}^{\infty} \frac{\exp{(-\alpha_{m}^{2}q/4)}}{\alpha_{m}^{2}} \frac{J_{1}(\alpha_{m} \varphi_{c})}{J_{1}(\alpha_{m})} \right] \\
    &\equiv f(y_{\text{lc}}) y_{\text{lc}} \xi_{\text{cap}}(q, \varphi_{c}) \, .
\end{split}
\label{eq:capture_flux_difficult_integral}
\end{equation}
We conclude that the capture flux is a scaled-down version of the total flux:
\begin{equation}
    F_{\text{cap}}(E) = F(E) \mathcal{C}_{\text{cap}} \, ,
\end{equation}
where the capture fraction $\mathcal{C}_{\text{cap}}$ is the ratio between the two special functions:
\begin{equation}
    \mathcal{C}_{\text{cap}} \equiv \frac{\xi_{\text{cap}}(q, \varphi_{c})}{\xi(q)} = \frac{\varphi_{c}^{2} - 4 \varphi_{c} \sum_{m = 1}^{\infty} \frac{\exp{(-\alpha_{m}^{2}q/4)}}{\alpha_{m}^{2}} \frac{J_{1}(\alpha_{m} \varphi_{c})}{J_{1}(\alpha_{m})}}{1 - 4 \sum_{m = 1}^{\infty} \frac{\exp{(-\alpha_{m}^{2}q/4)}}{\alpha_{m}^{2}}} \, .
    \label{eq:capture_fraction}
\end{equation}
Equation~\ref{eq:capture_fraction} isolates direct capture into a dimensionless quantity $\mathcal{C}_{\text{cap}}$ depending only on the loss-cone fullness $q$ and boundary ratio $\varphi_{c}$. This factorisation makes the asymptotic behaviour of the problem much more transparent.

\begin{figure}[ht]
    \centering
    \includegraphics[width=0.6\linewidth]{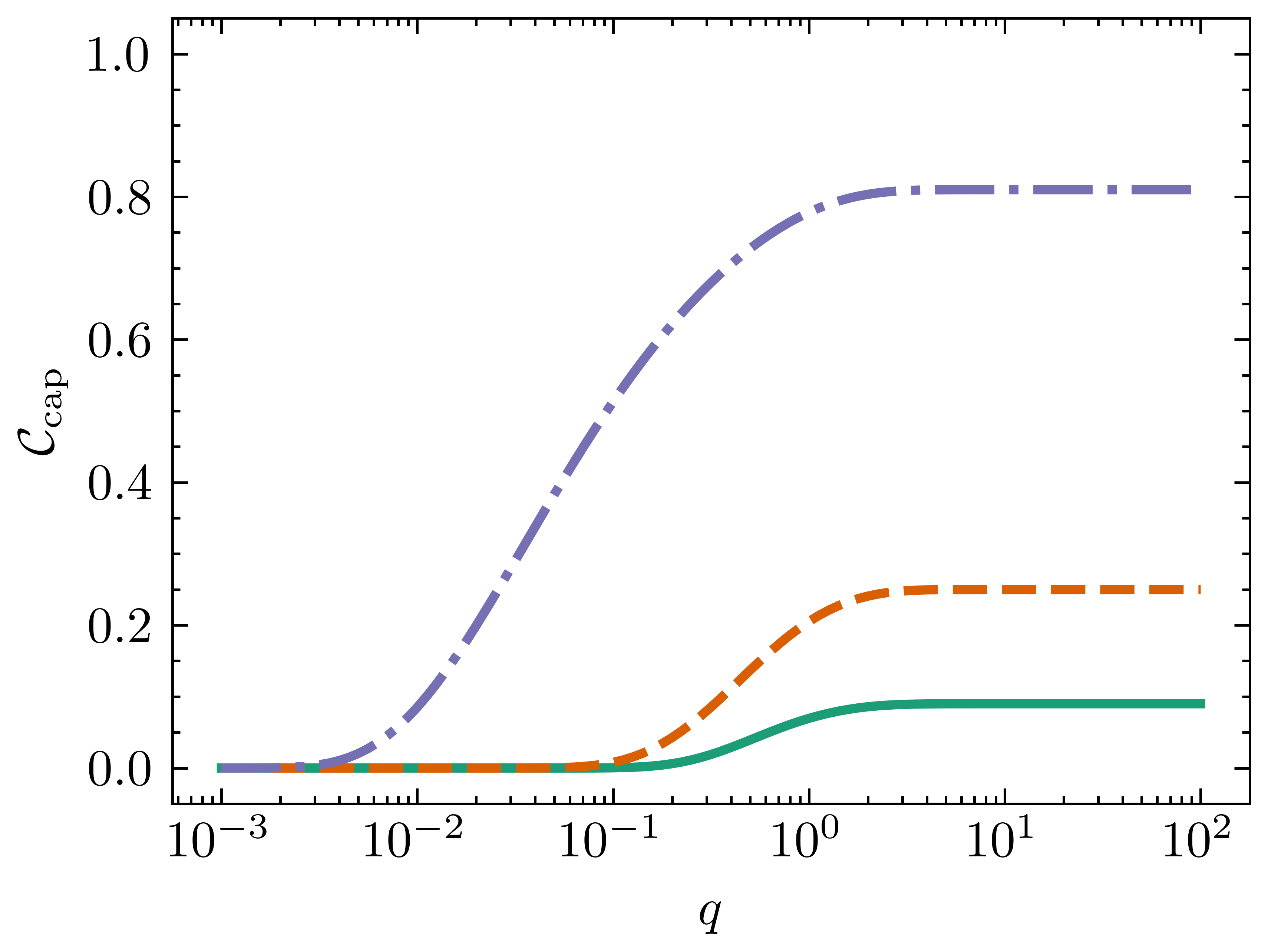}
    \caption{Capture fraction $\mathcal{C}_{\text{cap}}$ as a function of $q$ for different values of $\varphi_{c}$, with $\varphi_{c} = 0.3$ (green solid), $\varphi_{c} = 0.5$ (orange dashed), $\varphi_{c} = 0.9$ (purple dash-dotted).}
    \label{fig:capture_fraction_vs_q_curves}
\end{figure}

Figure~\ref{fig:capture_fraction_vs_q_curves} shows the capture fraction $\mathcal{C}_{\text{cap}}$ as a function of $q$ for various values of $\varphi_{c}$. In the full regime ($q \gg 1$), $\mathcal{C}_{\text{cap}}$ approaches a constant $\varphi_{c}^{2}$, whereas in the empty regime ($q \ll 1$), it approaches zero. In the intermediate regime, the capture fraction varies smoothly, with the transition region expanding as $\varphi_{c}$ increases. These observations align with physical intuition.

\textbf{Full regime ($q \gg 1$):} The diffusion step size $q \sim \bar{D}$ is large and the orbital period is long, allowing the star's angular momentum to randomise during its trajectory. By the time the star arrives at pericentre, the probability of direct capture rather than disruption is simply the ratio $\mathcal{C}_{\text{cap}} \approx \varphi_{c}^{2}$ of the areas of the respective loss cones. Analytically, as $q \gg 1$, $\xi(q) \approx 1$ and $\xi_{\text{cap}}(q, \varphi_{c}) \approx \varphi_{c}^{2}$, so the capture fraction reduces to:
\begin{equation}
    \mathcal{C}_{\text{cap}} = \frac{\xi_{\text{cap}}(q, L_{\text{cap}}/L_{\text{tde}})}{\xi(q)} \approx \left( \frac{L_{\text{cap}}}{L_{\text{tde}}} \right)^{2} \, ,
\end{equation}
which agrees with Equation~\ref{eq:flux_element} if inclination dependence is ignored.

\textbf{Empty regime ($q \ll 1$):} The diffusion step size is small and the orbital period is short. It is difficult for a star to accumulate enough change in $R$ along its orbit; the star is destroyed almost immediately after its angular momentum falls below the TDE threshold, and it is very unlikely to diffuse all the way to the capture threshold. Hence, the capture fraction is close to zero.

\textbf{Hills mass limit ($\varphi_{c} \to 1$):} The capture region becomes identical to the disruption region in that limit, and $\xi_{\text{cap}} \to \xi$ as $\varphi_{c} \to 1$. Therefore, $\mathcal{C}_{\text{cap}} \to 1$ and the observable TDE rate goes to zero. This is the Hills mass limit, where all stars that enter the loss cone plunge into the horizon without producing observable TDEs.

For supermassive BHs with $M_\bullet \gtrsim 10^{7} M_{\odot}$, the capture and disruption boundaries become comparable ($\varphi_{c} \to 1$). Consequently, the transition region where $\mathcal{C}_{\text{cap}}$ varies rapidly expands significantly, and its $q$-dependence becomes even more pronounced. Because the dominant TDE contributions in typical galactic nuclei arise from the intermediate regime ($q \sim 1$) \cite{Magorrian1999}, properly accounting for a smooth $\mathcal{C}_{\text{cap}}$ is essential for any realistic model of observable TDE demographics.

We also note that the capture fraction is ``model-agnostic'' in the sense that it can be applied to any suitable galaxy model as long as $q(E)$ and $\varphi_{c}$ are established. $\mathcal{C}_{\text{cap}}$ can be multiplied to any existing 1D rate calculation to account for direct capture as a post-processing correction, without the need to re-derive the full solution over the nested boundaries. We finally write down the general expression for the observable TDE flux, accounting for direct capture across all regimes (see Equation \ref{eq:final_flux}):
\begin{equation}
    F_{\text{obs}}(E) = [1 - \mathcal{C}_{\text{cap}}(q, L_{\text{cap}}/L_{\text{tde}})] 4\pi^{2} L_{c}^{2} R_{\text{lc}} \frac{f_{\star}(E, L_{c})}{\xi^{-1} + q^{-1} \ln{(1/R_{\text{lc}})}} \, .
\end{equation}
This equation completes our analysis of the one-dimensional loss-cone problem.

%% file: chapter3.tex
\chapter{Two Dimensional Loss Cone Theory}

\section{Towards an Inclination-Dependent Theory} \label{sec:towards_an_inclination_dependent_theory}

This chapter presents a loss cone theory that accounts for inclination-dependent loss cone (LC) boundaries. We reiterate the assumptions made throughout this work: an isotropic, steady-state galaxy hosting a spinning BH. This creates a hybrid geometry where the global system is symmetric but hosts an anisotropic absorbing boundary. Without the global symmetry of the galaxy, we do not have the luxury of both the magnitude and orientation of angular momentum being conserved; without the anisotropic boundary, there would be little motivation to consider the 2D problem since the system is entirely symmetrical.

Recall that we have been carefully distinguishing between the test star distribution function (DF) $f$ and the field star DF $f_{\star}$. When the boundary becomes anisotropic, the test DF naturally develops a dependence on the inclination $x$. The field DF remains isotropic: the loss cone occupies a negligible fraction of the total angular momentum space, so its depletion has little measurable effect on the bulk stellar distribution.

As seen in Section~\ref{sec:full_loss_cone_rates}, anisotropy can be easily accounted for in the full LC regime by integrating the field DF over the LC. However, the problem becomes complicated in the empty LC regime where TDEs are driven by the flux at the boundary. Stars may wander into the zone of destruction by either lowering the magnitude of their angular momentum or changing the orientation. To construct a 2D loss-cone theory, we must understand how diffusions in both variables happen at the same time.

\section{Coordinate Choice and Diffusion Coefficients} \label{sec:angular_diffusion_coefficients}

Before undertaking any calculations, one has to decide whether to use $(L, L_{z})$ or $(L, x = L_{z}/L)$ as phase space coordinates. The latter set is vastly superior because it prevents cross-terms in the diffusion coefficients. Intuitively, $x$ is the orientation of the angular momentum vector so it is orthogonal to the magnitude $L$; contrast this to using $L_{z}$, which is a projection of the vector and has a non-trivial correlation with $L$. In the subsequent treatment, we will stick with $R = L^{2}/L_{c}^{2}$ to represent the magnitude of the angular momentum, as is standard in loss-cone literature.

As argued in Section~\ref{sec:1d_diffusion_equation}, diffusion in energy can be ignored and we compute the diffusion coefficients in $R$ and $x$. This calculation is done via two complementary methods in Appendices~\ref{app:diffusion_derivation} and~\ref{app:diffusion_derivation_coordinate_transformation}. The latter follows the general coordinate transformation outlined in \cite[Chapter 5]{Merritt2013a}, while the former employs a vector perturbation method, offering a transparent geometric interpretation of the diffusion structure. Both approaches yield the same result:
\begin{equation}
\begin{split}
    \langle \Delta R \rangle &= \frac{r^{2} \langle (\Delta v_{\perp})^{2} \rangle}{L_{c}^{2}} + \mathcal{O}(R) \, , \\
    \langle \Delta x \rangle &= - \frac{r^{2} \langle (\Delta v_{\perp})^{2} \rangle}{4 L_{c}^{2}} \frac{x}{R} \, , \\
    \langle (\Delta R)^{2} \rangle &= \frac{2 r^{2} \langle (\Delta v_{\perp})^{2} \rangle}{L_{c}^{2}} R + \mathcal{O}(R^{2}) \, , \\
    \langle \Delta R \Delta x \rangle &= 0 \, , \\
    \langle (\Delta x)^{2} \rangle &= \frac{r^{2} \langle (\Delta v_{\perp})^{2} \rangle}{2 L_{c}^{2}} \frac{-x^{2} + \sin^{2}{\theta_{r}}}{R} \, ,
\end{split}
\label{eq:diffusion_coefficients}
\end{equation}
where as expected, no cross-terms between $R$ and $x$ appear.

We use $\theta_{r}$ and $\phi_{r}$ to denote the polar and azimuthal angles of the position vector of a star. Because the two-body relaxation timescale is much longer than the orbital period, a star completes many orbits before its conserved quantities get altered. Consequently, Equations~\ref{eq:diffusion_coefficients} (specifically the $\sin^{2}{\theta_{r}}$ term) need to be averaged over an orbit. To do this, notice that the star is travelling on an orbit of inclination $\iota = \cos^{-1}{(L_{z}/L)}$ such that its polar angle $\theta_{r}$ satisfies (see e.g.\ \cite[Chapter 3]{Binney2008}):
\begin{equation}
    \cos{\theta_{r}} = \sin{\iota} \cos{\phi_{r}} = \sin{\iota} \cos{(\varpi + \psi)} \, ,
\end{equation}
where $\psi$ is the true anomaly and $\varpi$ is an angle that describes the position of the pericentre in the orbital plane (argument of periapsis).

Due to efficient relativistic precession near a BH, $\varpi$ can be treated as uniformly distributed between $0$ and $2\pi$ \cite{Merritt2013}. Therefore:
\begin{equation}
    \left\langle \cos^{2}{\phi_{r}} \right\rangle = \left\langle \frac{1 + \cos{(2\varpi + 2\psi)}}{2} \right\rangle_{\varpi} = \frac{1}{2} \, ,
\end{equation}
where the average over the cosine term vanishes because $\varpi$ is uniformly distributed.

Computing $\langle \sin^{2}{\theta_{r}} \rangle$ then proves the desired result:
\begin{equation}
    \langle \sin^{2}{\theta_{r}} \rangle = 1 - \langle \cos^{2}{\theta_{r}} \rangle = 1 - \sin^{2}{\iota} \langle \cos^{2}{\phi_{r}} \rangle = \frac{1 + x^{2}}{2} \, ,
\label{eq:sin2_theta_average}
\end{equation}
where the definition $x = \cos{\iota}$ is used in the last step.

This averaging is necessary because it gives the following satisfying result:
\begin{equation}
    \langle (\Delta x)^{2} \rangle = \frac{r^{2} \langle (\Delta v_{\perp})^{2} \rangle}{2 L_{c}^{2}} \frac{-x^{2} + (1 + x^{2})/2}{R} = \frac{r^{2} \langle (\Delta v_{\perp})^{2} \rangle}{4 L_{c}^{2}} \frac{1 - x^{2}}{R} \, ,
\label{eq:diffusion_x_after_average}
\end{equation}
implying:
\begin{equation}
    \langle \Delta x \rangle = \frac{1}{2} \frac{\partial }{\partial x} \langle (\Delta x)^{2} \rangle \, ,
    \label{eq:x_differential_relation}
\end{equation}
which is the expected relation between the drift and diffusion coefficients, similar to the relation between $\langle \Delta R \rangle$ and $\langle (\Delta R)^{2} \rangle$ (see Equation~\ref{eq:R_differential_relation}).

Notice that Equation~\ref{eq:diffusion_x_after_average} vanishes at $x = \pm 1$. Physically, this is necessary to prevent the distribution from ``leaking out'' of the physical domain $-1 \leq x \leq 1$. This imperative property is only enabled by the averaging of $\sin^{2}{\theta_{r}}$ over the orbit, serving as a consistency check of our result.

The collisional term can be written down using Equations~\ref{eq:R_differential_relation} and~\ref{eq:x_differential_relation}:
\begin{equation}
    \begin{split}
        \left( \frac{\partial f}{\partial t} \right)_{c} &= - \frac{\partial }{\partial R} (f \langle \Delta R \rangle) + \frac{1}{2} \frac{\partial^{2}}{\partial R^{2}} (f \langle (\Delta R)^{2} \rangle) - \frac{\partial }{\partial x} (f \langle \Delta x \rangle) + \frac{1}{2} \frac{\partial^{2}}{\partial x^{2}} (f \langle (\Delta x)^{2} \rangle) \\
        &= D(E) \left[ \frac{\partial }{\partial R} \left( R \frac{\partial f}{\partial R} \right) +  \frac{\partial }{\partial x} \left( \frac{1 - x^{2}}{8 R} \frac{\partial f}{\partial x} \right) \right] \, ,
    \end{split}
\end{equation}
where $D(E)$ is the same ``diffusion rate'' as defined in the 1D theory (Equation~\ref{eq:diffusion_rate}):
\begin{equation}
    D(E) = \frac{r^{2} \langle (\Delta v_{\perp})^{2} \rangle}{L_{c}^{2}} \, .
\end{equation}

\section{2D Diffusion Equation} \label{sec:2d_diffusion_equation}

We can now write down the Boltzmann equation (see Equation~\ref{eq:boltzmann_equation_simplified}) for $f$:
\begin{equation}
        v_{r} \frac{\partial f}{\partial r} = D(E) \left[ \frac{\partial }{\partial R} \left( R \frac{\partial f}{\partial R} \right) +  \frac{\partial }{\partial x} \left( \frac{1 - x^{2}}{8 R} \frac{\partial f}{\partial x} \right) \right] \, .
\end{equation}
Introducing the same orbital phase variable $\tau$ as in the 1D case (see Equation~\ref{eq:tau_definition}), we arrive at a 2D diffusion equation:
\begin{equation}
    \boxed{
    \frac{\partial f}{\partial \tau} = P(E) \bar{D}(E) \left[ \frac{\partial }{\partial R} \left( R \frac{\partial f}{\partial R} \right) +  \frac{\partial }{\partial x} \left( \frac{1 - x^{2}}{8 R} \frac{\partial f}{\partial x} \right) \right] \, .
    }
    \label{eq:enhanced_diffusion_equation}
\end{equation}

This is an enhanced version of Equation~\ref{eq:boltzmann_intermediate} examined earlier, with an additional diffusion term in the $x$ direction. Equation~\ref{eq:enhanced_diffusion_equation} describes how an orbit, characterised by $R$ and $x$, is depleted at the pericentre $r_{-}$ (represented by $\tau = 0$) and replenished by diffusion in both $R$ and $x$ directions as the trajectory proceeds.

Remarkably, the 2D equation also collapses into a pure divergence form similar to the 1D case (Equation~\ref{eq:collisional_term_compact}). Just as dynamical friction exerts no net back reaction on $R$ in the highly eccentric limit, it also exerts no torque to tilt the orbital plane; the drift in $x$ originates entirely from the gradient of random diffusion (Equation~\ref{eq:x_differential_relation}). Note that this conclusion does not rely on $R \ll 1$ since there are no approximations made in the $x$-coefficients (see Equations~\ref{eq:diffusion_coefficients}). The emergence of a flux-conserving, pure divergence form holds significant implications when TDE rates are calculated. We also discuss the interpretation of this result in more detail in Chapter 5.

Similar to the 1D case, Equation~\ref{eq:enhanced_diffusion_equation} can be formally solved via separation of variables, yielding a solution in terms of special functions. The procedure is detailed in Appendix~\ref{app:diffusion_solution}. The interior $\tau$-dependent solution for the test DF is given by:
\begin{equation}
    f(\tau, R, x) = \sum_{l = 0}^{\infty} \sum_{m = 1}^{\infty} A_{lm} J_{\nu}(2 \sqrt{\lambda_{lm} R}) P_{l}(x) \exp{(-\lambda_{lm} P \bar{D} \tau)} \, ,
    \label{eq:inside_loss_cone_solution_2d}
\end{equation}
where $\nu = \sqrt{l(l + 1)/2}$ and eigenvalues $\lambda_{lm}$ and amplitudes $A_{lm}$ are determined by the boundary conditions.

Outside the LC, $\partial f/\partial \tau = 0$ so the equation becomes Laplace-type:
\begin{equation}
    \frac{\partial}{\partial R} \left( R \frac{\partial f}{\partial R} \right) + \frac{\partial}{\partial x} \left[ \frac{1 - x^{2}}{8R} \frac{\partial f}{\partial x} \right] = 0 \, .
\end{equation}
This is more easily solved by employing $s = \ln{R}$:
\begin{equation}
    \frac{\partial^{2} f}{\partial s^{2}} + \frac{\partial}{\partial x} \left[ \frac{1 - x^{2}}{8} \frac{\partial f}{\partial x} \right] = 0 \, .
\end{equation}
Assuming a separable solution $f(s, x) = \mathcal{S}(s) \mathcal{X}(x)$, the angular part is again solved by Legendre polynomials $P_{l}(x)$ where $k = l(l + 1)/8$. The radial part is:
\begin{equation}
    \frac{1}{\mathcal{S}} \frac{\mathrm{d}^{2} \mathcal{S}}{\mathrm{d} s^{2}} = k \, ,
\end{equation}
which is solved by exponentials:
\begin{equation}
    \mathcal{S}(s) = C \exp{(\sqrt{k} s)} + D \exp{(-\sqrt{k} s)} \, ,
\end{equation}
unless $k = 0$, in which case $\mathcal{S}(s) = C + D s$.

The exterior solution is thus given by:
\begin{equation}
    f(R, x) = C_{0} + D_{0} \ln{R} + \sum_{l = 1}^{\infty} \left[ C_{l} R^{\sqrt{l(l + 1)/8}} + D_{l} R^{-\sqrt{l(l + 1)/8}} \right] P_{l}(x) \, .
    \label{eq:outside_loss_cone_solution_2d}
\end{equation}
Clearly, the $l = 0$ mode is the isotropic solution (Equation~\ref{eq:outside_loss_cone_solution}), while the $l \ge 1$ modes are the angular corrections to the 1D solution. The interior and exterior solutions must agree at the LC boundary $R = R_{\text{lc}}(x)$ where $f = f(R_{\text{lc}}(x), x)$, resulting in a complicated convolution equation for the coefficients $A_{lm}$, $C_{l}$ and $D_{l}$. It is not immediately clear how one can solve for these coefficients analytically, as can be done in the 1D case.

Instead, we will bypass this problem and attempt solutions at the two limiting regimes. In the empty LC limit, we can set $f(R_{\text{lc}}(x), x) = 0$ and focus on the exterior solution since the interior is assumed to be negligible; in the full LC limit, it has been shown in Section~\ref{sec:full_loss_cone_rates} that the flux is simply an integration of the field DF. In both cases, it is not necessary to understand the interior solution in detail. Furthermore, we need not worry about direct capture in the two limits, because capture is either negligible (empty LC) or constitutes a fixed fraction of the total flux (full LC).

\section{2D Empty Loss Cone Solution}

To derive the exterior solution under the empty loss cone approximation, we make extensive use of the fact that the LC boundary is linear in $x$ with a small slope (see Figure~\ref{fig:l_tde_vs_chi}), permitting a perturbative treatment of the problem.

\subsection{Perturbative expansion}

Impose the empty LC approximation and the outer boundary condition:
\begin{equation}
\begin{split}
    f(R_{\text{lc}}(x), x) &= 0 \, , \\
    f(1, x) &= f_{1} \, ,
\end{split}
\end{equation}
where $f_{1} = f(R = 1) = f_{\star}(E, L_{c})$ is the value of the test DF at the outer boundary, which can be fitted to the DF of the field stars.

The second condition implies $D_{l} = - C_{l}$ for $l \ge 1$ and $C_{0} = f_{1}$ in Equation~\ref{eq:outside_loss_cone_solution_2d}. The general solution then takes the form:
\begin{equation}
    f(R, x) = f_{1} + D_{0} \ln{R} + \sum_{l=1}^{\infty} C_{l} B_{l}(R) P_{l}(x) \, ,
\end{equation}
where the functions $B_{l}(R) \equiv R^{\Gamma_{l}} - R^{-\Gamma_{l}}$ and $\Gamma_{l} = \sqrt{l(l+1)/8}$ have been defined.

The LC boundary is linear in $x$ with a gentle slope $\epsilon$:
\begin{equation}
    R_{\text{lc}}(x) = R_{\text{lc}}^{(0)}(1 + \epsilon x) \, ,
\end{equation}
where $R_{\text{lc}}^{(0)}$ is the LC size at $x=0$ and $\epsilon$ is a small, negative parameter.

Note that strictly speaking, the slope parameter here is not the same as $\epsilon$ defined in Equation~\ref{eq:l_tde_linear_approx}. Since $R_{\text{lc}} \propto L_{\text{tde}}^{2}$, the slope of $R_{\text{lc}}$ is indeed $2\epsilon$. However, we will still use $\epsilon$ to denote the slope of $R_{\text{lc}}$ for notational simplicity.

The mode coefficients can be expanded in a power series in $\epsilon$:
\begin{equation}
        D_{0} = \sum_{n=0}^{\infty} D_{0}^{(n)} \epsilon^{n} \, , \quad C_{l} = \sum_{n=0}^{\infty} C_{l}^{(n)} \epsilon^{n} \, ,
\end{equation}
and the radial functions evaluated at the boundary are also expanded:
\begin{equation}
\ln{[R_{\text{lc}}^{(0)} (1 + \epsilon x)]} = \sum_{k=0}^{\infty} L_{k} x^{k} \epsilon^{k} \, , \quad B_{l}[R_{\text{lc}}^{(0)} (1 + \epsilon x)] = \sum_{k=0}^{\infty} \Phi_{l}^{(k)} x^{k} \epsilon^{k} \, .
\end{equation}
The Taylor coefficients for the logarithm are simply $L_{0} = \ln{R_{\text{lc}}^{(0)}}$, $L_{1} = 1$, $L_{2} = -1/2$, $L_{3} = 1/3$, and $L_{4} = -1/4$. For the basis functions $B_{l}$, we have:
\begin{equation}
\begin{split}
    \Phi_{l}^{(0)} &= \left(R_{\text{lc}}^{(0)}\right)^{\Gamma_{l}} - \left(R_{\text{lc}}^{(0)}\right)^{-\Gamma_{l}} \, , \\
    \Phi_{l}^{(1)} &= \Gamma_{l} \left(R_{\text{lc}}^{(0)}\right)^{\Gamma_{l}} + \Gamma_{l} \left(R_{\text{lc}}^{(0)}\right)^{-\Gamma_{l}} \, , \\
    \Phi_{l}^{(2)} &= \frac{1}{2}\Gamma_{l}^{2} \Phi_{l}^{(0)} - \frac{1}{2}\Phi_{l}^{(1)} \, , \\
    \Phi_{l}^{(3)} &= \frac{1}{6}\Gamma_{l}^{2} \Phi_{l}^{(1)} - \Phi_{l}^{(2)} - \frac{1}{6}\Phi_{l}^{(1)} \, .
\end{split}
\label{eq:Phi_coefficients}
\end{equation}
Substituting these series into the boundary condition $f(R_{\text{lc}}(x), x) = 0$ yields a master convolution equation:
\begin{equation}
    0 = f_{1} \delta_{n,0} + \sum_{k=0}^{n} D_{0}^{(n-k)} L_{k} x^{k} + \sum_{l=1}^{\infty} \sum_{k=0}^{n} C_{l}^{(n-k)} \Phi_{l}^{(k)} x^{k} P_{l}(x) \, ,
    \label{eq:master_convolution}
\end{equation}
which must hold independently for every order $\epsilon^{n}$.

By projecting this equation onto the Legendre polynomials $P_{m}(x)$, the coefficients can be solved order by order. Define the projection integrals:
\begin{equation}
    I_{k, l, m} = \frac{2m+1}{2} \int_{-1}^{1} x^{k} P_{l}(x) P_{m}(x) \, \mathrm{d}x \, .
\end{equation}
This integral is identical to the quantum matrix element $\langle l, 0 | x^{k} | m, 0 \rangle$, where $|l, 0\rangle$ is the spherical harmonic state with angular momentum quantum number $l$ and magnetic quantum number $m=0$ \cite{Binney2015}. $x^{k}$ serves as a ladder operator that couples different angular momentum states. From this analogy, the selection rules for the non-zero elements $I_{k, l, m}$ can be immediately established:
\begin{equation}
    I_{k, l, m} \neq 0 \quad \text{only if} \quad k + l + m \text{ is even and } |l - m| \le k \, .
\end{equation}
The former condition arises from the parity of the integrand, while the latter condition arises from ``conservation of angular momentum'' in the ladder operator picture \cite{Sakurai2021}. Two important properties of the coefficients $D_{0}^{(n)}$ and $C_{l}^{(n)}$ can be inferred.

\textbf{Mode hierarchy:} Because the process starts with an isotropic zeroth-order background ($l=0$), exciting the $m$-th angular mode requires a ladder operator of at least degree $m$. Since $x^{m}$ is always associated with $\epsilon^{m}$, the $m$-th mode cannot be excited at any order lower than $\epsilon^{m}$. Therefore, $C_{m}^{(n)} = 0$ for all $n < m$. This means that the angular modes follow a perturbative hierarchy:
\begin{equation}
    C_{l} = \mathcal{O}(\epsilon^{l}) \, .
\end{equation}

\textbf{Global parity:} To solve for the coefficients $D_{0}^{(n)}$, we project the master equation onto $P_{0} = 1$. Every term involves an integral $I_{k,l,0}$, which requires $k+l$ to be even by the first selection rule. From the mode hierarchy, any non-zero $C_{l}^{(n-k)}$ (including $D_{0}^{(n-k)}$) requires $n-k$ and $l$ to share the same parity, meaning $(n-k)+l$ is even. Summing these two even quantities gives $(k+l) + (n-k+l) = n + 2l$, which must be even. Therefore, $n$ must be an even number, implying that all odd-order corrections to $D_{0}$ vanish:
\begin{equation}
    D_{0}^{(1)} = D_{0}^{(3)} = D_{0}^{(5)} = \dots = 0 \, .
\end{equation}
Physically, this enforces the condition that the global flux must be invariant under the transformation $x \to -x$. As will be shown, the empty LC, global TDE flux is simply $2D_{0}$, with $D_{0}$ being the sum of all the $D_{0}^{(n)}$ coefficients. Since the global flux should not change if the sign of $x$ is flipped, which is equivalent to $\epsilon \to -\epsilon$ (the BH flipping its spin direction), $D_{0}^{(n)} = 0$ must hold for all odd $n$.

We now explicitly derive the first two non-zero terms in $D_{0}$. At the zeroth order $\mathcal{O}(\epsilon^{0})$, setting $n=0$ in Equation~\ref{eq:master_convolution} and projecting onto $m=0$ yields:
\begin{equation}
    0 = f_{1} + D_{0}^{(0)} L_{0} \implies D_{0}^{(0)} = \frac{f_{1}}{\ln{(1/R_{\text{lc}}^{(0)})}} \, .
\end{equation}
By the mode hierarchy rule, $C_{l}^{(0)} = 0$ for all $l \ge 1$. This perfectly recovers the unperturbed isotropic solution Equation~\ref{eq:outside_loss_cone_solution}.

At the second order $\mathcal{O}(\epsilon^{2})$, setting $n=2$ and projecting onto $m=0$ produces the first non-zero correction to $D_{0}$:
\begin{equation}
    0 = D_{0}^{(2)} L_{0} + D_{0}^{(0)} L_{2} I_{2,0,0} + C_{1}^{(1)} \Phi_{1}^{(1)} I_{1,1,0} \, .
\end{equation}
Using $L_{2} = -1/2$, $I_{2,0,0} = 1/3$, and $I_{1,1,0} = 1/3$, we find:
\begin{equation}
    D_{0}^{(2)} \ln R_{\text{lc}}^{(0)} = \frac{1}{6} D_{0}^{(0)} - \frac{1}{3} C_{1}^{(1)} \Phi_{1}^{(1)} = \frac{1}{6} D_{0}^{(0)} \left( 1 + 2 \frac{\Phi_{1}^{(1)}}{\Phi_{1}^{(0)}} \right) \, .
    \label{eq:D0_second_order}
\end{equation}
We can now write down the perturbative expansion of $D_{0}$ up to $\mathcal{O}(\epsilon^{4})$:
\begin{equation}
\begin{split}
    D_{0} &\approx D_{0}^{(0)} + \epsilon^{2} D_{0}^{(2)} + \epsilon^{4} D_{0}^{(4)} \\
    &\approx D_{0}^{(0)} \left[ 1 + \frac{R_{\text{lc}}^{(0)}}{3\ln{(1/R_{\text{lc}}^{(0)})}} \epsilon^{2} + \mathcal{O}(\epsilon^{4}) \right] \, .
\end{split}
\label{eq:D0_perturbative_expansion}
\end{equation}

To verify the mode hierarchy, we truncate the boundary condition equations at $l_{\text{max}} = 4$ and solve numerically for $\epsilon \in [10^{-4}, 10^{-1}]$. Table~\ref{tab:cl_scaling_fits} confirms the predicted scaling $C_{l} = \mathcal{O}(\epsilon^{l})$ to machine precision.

\begin{table}[htbp]
    \centering
    \caption{Linear-fit diagnostics of $\log{C_{l}}$ versus $\log{\epsilon}$.}
    \label{tab:cl_scaling_fits}
    \begin{tabular}{c c c c}
        \hline
        Mode $l$ & Measured slope & Standard error & $R^{2}$ \\
        \hline
        1  & 1.0000  & $7.93 \times 10^{-8}$ & 1.00000 \\
        2  & 2.0000  & $8.64 \times 10^{-6}$ & 1.00000 \\
        3  & 3.0000  & $1.2 \times 10^{-6}$  & 1.00000 \\
        4  & 4.0000  & $7.01 \times 10^{-7}$ & 1.00000 \\
        \hline
    \end{tabular}
\end{table}

\subsection{Conservation of flux} \label{sec:flux_conservation}

We now consider the flux at the LC boundary. Since the test distribution function is in steady state, the global flux is conserved and the integral of the flux vector on a closed surface must be zero. Following Equation~\ref{eq:enhanced_diffusion_equation}, the flux vector $\mathbf{F} = (F_{R}, F_{x})$ has two components:
\begin{equation}
    F_{R} = 2\pi^{2} P D L_{c}^{2} \left( R \frac{\partial f}{\partial R} \right) \, , \quad F_{x} = 2\pi^{2} P D L_{c}^{2} \left( \frac{1 - x^{2}}{8 R} \frac{\partial f}{\partial x} \right) \, .
    \label{eq:flux_vector}
\end{equation}
For subsequent calculations, we will drop the prefactor and only calculate the dimensionless fluxes, which can be written down in terms of the mode expansions:
\begin{equation}
\begin{split}
    F_{R} &= D_{0} + \sum_{l = 1}^{\infty} \sqrt{\frac{l(l + 1)}{8}} C_{l} \left[ R^{\sqrt{l(l + 1)/8}} + R^{-\sqrt{l(l + 1)/8}} \right] P_{l}(x) \, , \\
    \quad F_{x} &= \sum_{l = 1}^{\infty} \frac{1 - x^{2}}{8 R} C_{l} \left[ R^{\sqrt{l(l + 1)/8}} - R^{-\sqrt{l(l + 1)/8}} \right] P_{l}'(x) \, .
\end{split}
\label{eq:flux_components}
\end{equation}
Consider the exterior domain defined by $x \in [-1, 1]$ and $R \in [R_{\text{lc}}(x), 1]$. At $x = \pm 1$, there is no ``leakage'' of flux since $F_{x} = 0$ at $x = \pm 1$. At the outer boundary $R = 1$, the $x$-integrated flux is denoted as $F_{\text{out}}$, while at the inner boundary $R = R_{\text{lc}}(x)$, the $x$-integrated flux is denoted as $F_{\text{in}}$. Conservation of flux requires $F_{\text{out}} = F_{\text{in}}$.

It is immediately clear that at $R = 1$, the angular flux $F_{x}$ vanishes exactly and the radial flux $F_{R}$ can be integrated over $x$ to give:
\begin{equation}
    F_{\text{out}} = \int_{-1}^{1} \mathrm{d}x \, \left[ D_{0} + \sum_{l = 1}^{\infty} \sqrt{\frac{l(l + 1)}{8}} C_{l} \left( 1 + 1 \right) P_{l}(x) \right] = 2 D_{0} \, ,
\label{eq:empty_flux_out}
\end{equation}
where the orthogonality of Legendre polynomials eliminates all $l \ge 1$ modes.

This demonstrates that the global flux into the empty LC is simply $2D_{0}$, which follows the same form as the isotropic case (Equation~\ref{eq:final_empty_loss_cone_flux}). However, the value of $D_{0}$ itself suffers a geometrical correction due to the slope of the LC boundary, which is quantified by Equation~\ref{eq:D0_perturbative_expansion}:
\begin{equation}
    D_{0} \approx D_{0}^{\text{iso}} \left[ 1 + \frac{R_{\text{lc}}^{(0)}}{3\ln{(1/R_{\text{lc}}^{(0)})}} \epsilon^{2} + \mathcal{O}(\epsilon^{4}) \right] \, ,
    \label{eq:D0_correction}
\end{equation}
where we identify $D_{0}^{\text{iso}} = D_{0}^{(0)} = f_{1}/\ln{(1/R_{\text{lc}}^{(0)})}$ as the isotropic flux coefficient.

Substituting this into the flux calculation and restoring the prefactor:
\begin{equation}
    F_{\text{empty}}(E) \approx F_{\text{empty}}^{\text{iso}}(E) \left[ 1 + \frac{R_{\text{lc}}^{(0)}}{3\ln{(1/R_{\text{lc}}^{(0)})}} \epsilon^{2} + \mathcal{O}(\epsilon^{4}) \right] \, ,
    \label{eq:flux_correction}
\end{equation}
where we identify $F_{\text{empty}}^{\text{iso}}$ as the isotropic flux given by Equation~\ref{eq:final_empty_loss_cone_flux}.

The 2D, $x$-integrated empty LC flux therefore differs from its isotropic counterpart only at $\mathcal{O}(\epsilon^2)$. The effect is vanishingly weak since $\epsilon$ is a small parameter and $R_{\text{lc}}^{(0)}$ is also an extremely small number for typical galaxies. The correction is not due to how stars collectively diffuse into the LC since the \textit{form} of the total flux is unchanged. It is instead caused by the \textit{geometric effect} of the sloped LC boundary, which changes the value of $D_{0}$ and thus the total flux.

Equation~\ref{eq:flux_correction} provides a mapping between the 1D and 2D empty LC fluxes. For every boundary profile $R_{\text{lc}}(x)$, determine the middle point $R_{\text{lc}}^{(0)}$ and the slope $\epsilon$; use the middle point to calculate the isotropic flux $F^{\text{iso}}$ based on Equation~\ref{eq:flux_outside_loss_cone}; then apply the correction factor in Equation~\ref{eq:flux_correction} to get the final flux. The fact that there exists such a simple relation between the 1D and 2D problems is hitherto not obvious.

\subsection{Preferential disruption}

Although the total global flux is only enhanced by a negligible second-order factor, the local flux at specific values of $x$ undergoes significant first-order redistribution. To find the local disruption rate, this flux vector $\mathbf{F} = (F_{R}, F_{x})$ must be projected onto the normal vector $\hat{\mathbf{n}}$ of the boundary. The differential flux crossing an infinitesimal segment of the LC is:
\begin{equation}
        \mathrm{d}F = \mathbf{F} \cdot \hat{\mathbf{n}} \, \mathrm{d}l = F_{R} \mathrm{d}x - F_{x} \frac{\mathrm{d}R_{\text{lc}}}{\mathrm{d}x} \mathrm{d}x \, .
\end{equation}
This calculation can be vastly simplified by exploiting the empty LC condition, which imposes $f(R_{\text{lc}}(x), x) = 0$. Since the DF is identically zero everywhere along the boundary, its total derivative with respect to $x$ must also vanish:
\begin{equation}
        \frac{\mathrm{d}}{\mathrm{d}x} f(R_{\text{lc}}(x), x) = \frac{\partial f}{\partial x} + \frac{\partial f}{\partial R} \frac{\mathrm{d}R_{\text{lc}}}{\mathrm{d}x} = 0 \, .
\end{equation}
Substituting this into the definition of $F_{x}$ (see Equations~\ref{eq:flux_vector}):
\begin{equation}
\begin{split}
    F_{x} &= \frac{1 - x^{2}}{8 R} \frac{\partial f}{\partial x} \\
    &= -\frac{1 - x^{2}}{8 R} \frac{\partial f}{\partial R} \frac{\mathrm{d}R_{\text{lc}}}{\mathrm{d}x} \\
    &= -\frac{1 - x^{2}}{8 R^{2}} \frac{\mathrm{d}R_{\text{lc}}}{\mathrm{d}x} F_{R} \, .
\end{split}
\end{equation}
Putting this back into the expression for $\mathrm{d}F$:
\begin{equation}
\begin{split}
    \mathrm{d}F &= F_{R} \mathrm{d}x - F_{x} \frac{\mathrm{d}R_{\text{lc}}}{\mathrm{d}x} \mathrm{d}x \\
    &= F_{R} \left[ 1 + \frac{1 - x^{2}}{8 R^{2}} \left( \frac{\mathrm{d}R_{\text{lc}}}{\mathrm{d}x} \right)^{2} \right] \mathrm{d}x \, .
\end{split}
\end{equation}
So far no approximations have been made, so this result is exact. We can now substitute the LC boundary $R_{\text{lc}}(x) = R_{\text{lc}}^{(0)} (1 + \epsilon x)$ to get:
\begin{equation}
    \frac{\mathrm{d}F}{\mathrm{d}x} = F_{R}(x) \left[ 1 + \epsilon^{2} \frac{1-x^{2}}{8 (1 + \epsilon x)^{2}} \right] \, .
    \label{eq:exact_local_flux}
\end{equation}

This is a clean result showing that the angular flux only contributes at $\mathcal{O}(\epsilon^{2})$. Because of the geometric $(1-x^{2})$ factor, the angular flux contribution vanishes at the poles ($x = \pm 1$). To calculate the preferential disruption at the extremes, we only need to evaluate $F_{R}(\pm 1)$ at the LC boundary. The radial flux evaluated at the boundary is easily expanded to $\mathcal{O}(\epsilon)$ (see Equation~\ref{eq:flux_components}):
\begin{equation}
\begin{split}
    F_R(x) &\approx D_0^{(0)} + \epsilon C_1^{(1)} \Gamma_1 \left[ \left(R_{\text{lc}}^{(0)}\right)^{\Gamma_1} + \left(R_{\text{lc}}^{(0)}\right)^{-\Gamma_1} \right] x \\
    &= D_{0}^{(0)} + \epsilon C_1^{(1)} \Phi_1^{(1)} x \\
    &\approx D_{0}^{(0)} \left( 1 - \epsilon \frac{\Phi_1^{(1)}}{\Phi_1^{(0)}} x \right) \, .
\end{split}
\end{equation}
For small $R_{\text{lc}}^{(0)}$, the ratio of the $\Phi$ constants is $\Phi_1^{(1)}/\Phi_1^{(0)} \approx -1/2$. Thus, the local boundary-crossing flux simplifies to:
\begin{equation}
    \frac{\mathrm{d}F}{\mathrm{d}x} \approx D_0^{(0)} \left( 1 + \frac{\epsilon}{2} x \right) \, .
    \label{eq:empty_lc_local_flux}
\end{equation}
Evaluating this at the extremes $x=\pm1$, we find:
\begin{equation}
    \frac{\mathrm{d}F/\mathrm{d}x \lvert_{x = 1}}{\mathrm{d}F/\mathrm{d}x \lvert_{x = -1}} \approx \frac{1 + \epsilon/2}{1 - \epsilon/2} \approx 1 + \epsilon \, .
\end{equation}

Since $\epsilon$ is a small, negative slope as shown in Figure~\ref{fig:l_tde_vs_chi}, we conclude that stars on equatorial retrograde orbits ($x = -1$) are preferentially disrupted compared to stars on equatorial prograde orbits ($x = 1$) at a first-order level. This is a direct consequence of the fact that the LC boundary is larger for retrograde orbits than for prograde ones, allowing more stars to diffuse into the LC at retrograde inclinations.

\section{2D Full Loss Cone Solution}

The full LC solution follows directly from the phase-space integral derived in Section~\ref{sec:full_loss_cone_rates}. Since $f \approx f_{\star}$ throughout the loss cone in the full LC limit, the TDE rate is simply the integral of $f_{\star}$ over the loss-cone volume (Equations~\ref{eq:number_element} and~\ref{eq:flux_element}). Making use of $R_{\text{lc}}(x) = R_{\text{lc}}^{(0)} (1 + \epsilon x)$, the total flux can be calculated as:
\begin{equation}
\begin{split}
        F_{\text{full}}(E) &= 4\pi^{2} f_{\star}(E) L_{c}^{2} \int_{-1}^{1} \frac{R_{\text{lc}}(x)}{2} \, \mathrm{d}x \\
        &= 4\pi^{2} f_{\star}(E) L_{c}^{2} R_{\text{lc}}^{(0)} \, ,
\end{split}
\label{eq:full_lc_flux}
\end{equation}
which is exactly the same as the isotropic full LC flux (Equation~\ref{eq:final_full_loss_cone_flux}).

We therefore conclude that the global, $x$-integrated flux in both the empty and full LC limits has the same form as their 1D counterparts. On the other hand, the full LC local flux also has a first-order preferential disruption effect, which can be calculated by considering $\mathrm{d}F/\mathrm{d}x$ at the poles $x = \pm 1$:
\begin{equation}
    \frac{\mathrm{d}F/\mathrm{d}x \lvert_{x = 1}}{\mathrm{d}F/\mathrm{d}x \lvert_{x = -1}} = \frac{R_{\text{lc}}(1)}{R_{\text{lc}}(-1)} \approx \frac{1 + \epsilon}{1 - \epsilon} \approx 1 + 2\epsilon \, ,
    \label{eq:full_lc_local_flux}
\end{equation}
which is twice the preferential disruption effect in the empty LC case.

The global flux invariance holds even if the LC boundary is nonlinear in $x$. The empty global flux calculation in Equation~\ref{eq:empty_flux_out} does not require $R_{\text{lc}}(x)$ to be linear in $x$; the full calculation above in Equation~\ref{eq:full_lc_flux} remains valid as long as $R_{\text{lc}}(x)$ can be written as an expansion in Legendre polynomials, i.e.\ $R_{\text{lc}}(x) = R_{\text{lc}}^{(0)} + \sum_{l=1} a_{l} P_{l}(x)$. These observations are a structural consequence of conservation of flux (see Section~\ref{sec:flux_conservation}). The vanishing of angular fluxes at $x = \pm1$ means that instead of evaluating the global flux at the curved inner LC boundary, we are allowed to ``zoom out'' and move to the flat outer boundary $R = 1$, where the flux is determined solely by $F_{R}$. All angular modes $l \ge 1$ are suppressed when integrated over $x$ due to the orthogonality of Legendre polynomials, so the global flux is determined exclusively by the monopole coefficient, which only receives geometric corrections to its magnitude.

We have thus shown that the global flux invariance is structurally enforced by the symmetry of the 2D problem. Anisotropy only modifies the geometric prefactors, provided the boundary admits a sufficiently regular expansion.

\section{Understanding the 2D Intermediate Regime}

The resemblance of the conclusions drawn in the empty and full LC limits provides useful guidance for understanding the intermediate regime. Since the $x$-integrated flux in both limits reduces to the isotropic result, it is reasonable to expect that the global flux remains close to the 1D prediction throughout the intermediate regime. Physically, this robustness is intuitive: because the background galaxy is isotropic, the total TDE rate must be invariant under an inversion of the black hole spin ($\epsilon \to -\epsilon$), which is equivalent to $x \to -x$. Therefore, the leading-order correction to the global flux must be an even function at $\mathcal{O}(\epsilon^{2})$. To see how this symmetry manifests mathematically due to conservation of flux, consider a Legendre expansion of the general local flux:
\begin{equation}
    \frac{\mathrm{d}F}{\mathrm{d}x} = \left( \frac{\mathrm{d}F}{\mathrm{d}x} \right)^{\text{iso}} + \sum_{l=1}^{\infty} b_{l} P_{l}(x) \, .
\end{equation}
The orthogonality of $P_{l}(x)$ ensures that the entire $l \ge 1$ sum integrates to zero on the interval $x \in [-1, 1]$. The global rate is determined exclusively by the monopole coefficient $(\mathrm{d}F/\mathrm{d}x)^{\text{iso}}$, which only receives geometric corrections to its normalisation. The spin-flip symmetry then enforces that the leading correction must be at $\mathcal{O}(\epsilon^{2})$. Although the higher-order modes $b_{l}$ can have nontrivial corrections of $\epsilon$, they are hidden from the global flux calculation and do not affect the total rate.

Having established a symmetry argument that applies regardless of LC fullness, we now propose explicit interpolation formulae that bridge the two limits.

\textbf{Global flux:} In Section~\ref{sec:bypassing_boundary_layer_analysis}, we observed that the general 1D flux formula is well approximated by the reciprocal sum of the empty and full fluxes. This argument can be used to obtain a simple ansatz for the global flux correction. Consider:
\begin{equation}
\begin{split}
    \frac{1}{F} &\approx \frac{1}{F_{\text{empty}}} + \frac{1}{F_{\text{full}}} \\
    &\approx \frac{1}{F_{\text{empty}}^{\text{iso}}} \left[ 1 - \frac{R_{\text{lc}}^{(0)}}{3\ln{(1/R_{\text{lc}}^{(0)})}} \epsilon^{2} + \mathcal{O}(\epsilon^{4}) \right] + \frac{1}{F_{\text{full}}^{\text{iso}}} \\
    &\approx \frac{1}{F^{\text{iso}}} \left[ 1 - \frac{F^{\text{iso}}}{F_{\text{empty}}^{\text{iso}}} \frac{R_{\text{lc}}^{(0)}}{3\ln{(1/R_{\text{lc}}^{(0)})}} \epsilon^{2} + \mathcal{O}(\epsilon^{4}) \right] \, .
\end{split}
\end{equation}
Using the isotropic expressions (Equations~\ref{eq:final_flux} and~\ref{eq:final_empty_loss_cone_flux}) and the relation between $R_{0}$ and $R_{\text{lc}}^{(0)}$ (Equation~\ref{eq:R0_Rlc_relation}), we obtain a $q$-dependent correction:
\begin{equation}
\begin{split}
    F &\approx F^{\text{iso}} \left[ 1 + \frac{R_{\text{lc}}^{(0)}/3}{q + \ln{(1/R_{\text{lc}}^{(0)})}} \epsilon^{2} + \mathcal{O}(\epsilon^{4}) \right] \, .
\end{split}
\end{equation}
This ansatz correctly recovers the empty and full limits (Equations~\ref{eq:flux_correction} and~\ref{eq:full_lc_flux}), and smoothly interpolates between them. The correction is, however, still negligible since the additional $q$-dependent factor only weakens the already small $\mathcal{O}(\epsilon^{2})$ correction.

\textbf{Local flux:} The local flux $\mathrm{d}F/\mathrm{d}x$ is more subtle because the $x$-dependent boundary induces a ``local fullness parameter'' $q(x,E)=P(E)\bar{D}(E)/R_{\text{lc}}(x)$. Nevertheless, the similarity between the two limits suggests a reciprocal-sum interpolation:
\begin{equation}
    \frac{\mathrm{d}F}{\mathrm{d}x} \approx \left[ \frac{1}{\mathrm{d}F_{\text{empty}}/\mathrm{d}x} + \frac{1}{\mathrm{d}F_{\text{full}}/\mathrm{d}x} \right]^{-1} \, .
\end{equation}
The 2D local fluxes can be expressed as corrections to their 1D counterparts:
\begin{equation}
    \frac{\mathrm{d}F_{\text{empty}}}{\mathrm{d}x} \approx \left( \frac{\mathrm{d}F}{\mathrm{d}x} \right)_{\text{empty}}^{\text{iso}} \left( 1 + \frac{\epsilon}{2} x \right) \, , \quad \frac{\mathrm{d}F_{\text{full}}}{\mathrm{d}x} \approx \left( \frac{\mathrm{d}F}{\mathrm{d}x} \right)_{\text{full}}^{\text{iso}} (1 + \epsilon x) \, ,
\end{equation}
where the isotropic local fluxes are simply a division by two of the global ones, i.e.\ $(\mathrm{d}F/\mathrm{d}x)_{i}^{\text{iso}} = F_{i}^{\text{iso}}/2$ for $i \in \{\text{empty}, \text{full}\}$.

Carrying out the necessary algebra, we find that the local flux in the intermediate regime takes the form:
\begin{equation}
    \frac{\mathrm{d}F}{\mathrm{d}x} \approx \left( \frac{\mathrm{d}F}{\mathrm{d}x} \right)^{\text{iso}} \left[ 1 + A(q) \epsilon x \right] \, ,
\end{equation}
where $(\mathrm{d}F/\mathrm{d}x)^{\text{iso}}$ is Equation~\ref{eq:final_flux} divided by two, and $A(q)$ is:
\begin{equation}
    A(q) = \frac{q + \frac{1}{2}\ln{(1/R_{\text{lc}}^{(0)})}}{q + \ln{(1/R_{\text{lc}}^{(0)})}} \, .
\end{equation}

It must be emphasised that this local reciprocal-sum interpolation is a heuristic ansatz rather than a rigorous solution to the 2D Fokker-Planck equation. By applying the approximation locally at each inclination $x$, this treatment implicitly assumes that inclination slices are decoupled. In reality, the fast $x$-diffusion term in Equation~\ref{eq:enhanced_diffusion_equation} may rapidly smooth out the distinction between inclinations.

Nevertheless, this chapter has determined the analytical behaviours of the loss cone in the limiting regimes; the proposed formulae provide a simple interpolation between the bounded limits, motivated by the structural observation made in the 1D problem (Section~\ref{sec:bypassing_boundary_layer_analysis}). This ansatz provides a first step towards an interpolation framework that ultimately needs to be calibrated against numerical solutions of the 2D diffusion problem.

%% file: chapter4.tex
\chapter{Discussion and Conclusion}

In this chapter, we discuss some of the overarching assumptions and potential implications of the work, as well as potential avenues for future research, before closing with a summary of the dissertation.

\section{Importance of Isotropy} \label{sec:isotropy_discussion}

In Sections~\ref{sec:1d_diffusion_equation} and~\ref{sec:2d_diffusion_equation}, we showed that the isotropy of the field star distribution leads to a suppression of dynamical friction terms in the diffusion coefficients, resulting in a pure divergence form in both 1D and 2D diffusion equations. This can be explained by a simple geometric argument. As long as the field distribution is isotropic, i.e.\ $f_{\star} = f_{\star}(E)$, any dynamical friction due to two-body relaxation must be anti-parallel to the velocity vector (see e.g.\ \cite[Chapter 8]{Binney2008}). This limits the possible back-reaction effects in the chosen coordinates $(R, x)$.

\textbf{Magnitude:} Consider highly eccentric orbits. Since the friction is anti-parallel to the near-radial velocity vector, it barely affects the angular momentum magnitude. Therefore, the friction does not change $R$ to leading order.

\textbf{Orientation:} Consider any orbit. The torque induced by the friction is given by $\mathbf{r} \times \mathbf{F}_{\text{fric}} \propto \mathbf{r} \times \mathbf{v}$, which is parallel to the angular momentum vector. Therefore, the friction cannot change $x$. Note that this result is exact, and it explains why the diffusion coefficients of $x$ do not rely on the limit $R \ll 1$ (see Appendices~\ref{app:diffusion_derivation} and~\ref{app:diffusion_derivation_coordinate_transformation}).

This intuitive geometric argument explains how the clean divergence form of the collision operators (Equations~\ref{eq:boltzmann_intermediate} and~\ref{eq:enhanced_diffusion_equation}) is a direct result of zero back reaction. The discussion of an absence of back reaction dates back to the work of \cite{Binney1988}, which has been cleanly laid out in \cite[Chapter 7]{Binney2008}. Although the same conclusion is drawn, the routes towards it are different. Binney and Lacey's argument applies in any suitable action-angle coordinates, but relies on Maxwellian field stars and zero-mass tracers; the argument here relies on choosing $(R, x)$, but only requires isotropy of the field star distribution.

Isotropy of the field distribution is the key assumption enabling much of the analytical progress in this dissertation. An immediate relaxation of this assumption would be an anisotropic (but still spherical) background (i.e.\ $f_{\star} = f_{\star}(E, L)$). We retain $(E, R, x)$ as the phase space coordinates, but the diffusion problem becomes much richer. The canonical local diffusion coefficients $\langle (\Delta v_{\perp})^{2} \rangle$, $\langle \Delta v_{\parallel} \rangle$ and $\langle (\Delta v_{\parallel})^{2} \rangle$ become vastly more complicated, and indeed one cannot assume a single $\langle (\Delta v_{\perp})^{2} \rangle$ for all perpendicular directions. This may cause the divergence form to vanish, and the resulting diffusion equation may no longer be separable. Nevertheless, we should still be able to write down a Fokker--Planck equation in $(R, x)$, which can always be solved numerically to understand diffusion at the Kerr loss cone boundary.

On the other hand, if spherical symmetry is discarded altogether, the framework breaks down completely as $(R, x)$ no longer qualify as phase space coordinates. We discuss this further in Section~\ref{sec:stellar_dynamical_extensions}.

\section{Discussion and Future Work} \label{sec:future_work}

\subsection{Anisotropic capture fraction}

Accounting for direct captures by a spinning BH presents a unique challenge. Chapter 3 presented an expression for the 1D capture fraction $\mathcal{C}_{\text{cap}}(q, \varphi_c)$ by integrating over the nested capture region. To account for 2D direct capture, the most straightforward, albeit heuristic, method is to assume the 1D physics holds locally. By defining the local parameters $q(x, E)$ and $\varphi_{c}(x)$, the local flux could be modulated by applying $\mathcal{C}_{\text{cap}}(q(x, E), \varphi_{c}(x))$ at each inclination before integrating over $x$.

While intuitive, this prescription may be fundamentally incomplete. Since diffusion in $x$ is rapid, a star that scatters into the observable TDE phase space at one inclination may subsequently diffuse into the direct-capture region before reaching pericentre. Furthermore, the 2D boundary ratio $\varphi_{c}(x) = L_{\text{cap}}(x)/L_{\text{tde}}(x)$ depends strongly on $x$ due to the non-linearity in $L_{\text{cap}}(x)$ (see Figure~\ref{fig:l_cap_vs_chi}), implying that diffusion in $x$ can lead to significant coupling between adjacent inclinations.

Nevertheless, we still expect direct capture to be negligible in the empty regime and to approach a finite fraction in the full regime, suggesting that some interpolation between the two limits should exist. Determining the form of such an interpolation, however, likely requires numerical treatment of the full 2D problem.

\subsection{Towards a numerical solution}

The most direct path to overcoming the limitations of the perturbative treatment is to solve Equation~\ref{eq:enhanced_diffusion_equation} numerically. Future work should implement a partial differential equation (PDE) solver tailored to the 2D diffusion equation. The numerical setup may proceed as follows.

\textbf{Domain and grid:} The computational domain spans $x \in [-1, 1]$ and $R \in [0, 1]$. Because the action happens at $R \ll 1$, a logarithmic grid $s = \ln R$ is likely necessary to properly resolve the boundary layer.

\textbf{Boundary conditions:} At the outer boundary $R=1$, the distribution is pinned to the isotropic field star distribution $f_{\star}(E)$. At the loss cone boundary $R = R_{\text{lc}}(x)$, the solution must be a $\tau$-independent value $f_{\text{lc}}(x)$ that is determined by matching the inner and outer fluxes.

With a robust numerical solution, the exact differential flux $\mathrm{d}F/\mathrm{d}x$ at both the disruption and capture boundaries could be extracted. A full numerical treatment lies beyond the intended scope of this dissertation, whose primary objective is the formulation and theoretical analysis of the 2D problem. Since the limiting regimes are already analytically understood, a numerical solution becomes primarily a technical step towards resolving the intermediate regime.

\subsection{Stellar dynamical extensions} \label{sec:stellar_dynamical_extensions}

This dissertation has focused on classical two-body relaxation in spherical potentials. Realistic galactic nuclei are often subject to additional dynamical processes such as resonant relaxation (RR) and non-sphericality of the gravitational potential.

\textbf{Resonant relaxation:}
RR can strongly enhance angular-momentum diffusion near the BH through long-lived coherent torques \cite{Rauch1996}, although its effects can be quenched by relativistic precession of strongly radial orbits \cite{BarOr2016,Merritt2011}. From a modelling perspective, RR modifies the effective diffusion coefficients in $R$ and $x$. The particularly simple form of Equation~\ref{eq:enhanced_diffusion_equation} relies on a common diffusion coefficient shared by both directions. This form is unlikely to survive once RR is included, since scalar RR (acting on $R$) and vector RR (acting on $x$) operate on different timescales and experience different quenching mechanisms \cite{Kocsis2011}. Nevertheless, the general 2D Fokker--Planck framework should remain applicable, although the separable nature of the resulting diffusion equation may be lost. In any case, numerical solutions offer a last resort.

\textbf{Non-spherical potential:}
The key to enabling the 2D framework is the isotropy of the background distribution, and it is natural to ask how the derivation may break if this assumption is relaxed. Non-spherical potentials can induce torques that change the magnitude and orientation of the angular momentum on timescales much shorter than two-body relaxation \cite{Merritt2004}. They can also support ``centrophilic'' orbits that plunge directly into the loss cone without requiring diffusion \cite{Vasiliev2013}. Concurrently, the breakdown of isotropy means that the zero back-reaction argument in Section~\ref{sec:isotropy_discussion} no longer holds as the wake of a test star will be asymmetric, causing the 2D collision operator to lose its pure divergence structure.

\textbf{Action-angle reformulation:}
More fundamentally, in a non-spherical galaxy, $R$ and $x$ can no longer be treated as global integrals of motion, necessitating a reformulation of the loss cone problem in terms of action variables. Recall that the connection between relativistic and stellar dynamics requires an asymptotic mapping between the Kerr conserved quantities and the classical integrals of motion (see Section~\ref{sec:relativistic_dynamics}). Once spherical symmetry is broken and we are forced to use action variables, it is not clear how a new mapping can be constructed to express the Kerr loss cone boundary in action coordinates. The problem of constructing action variables in Kerr spacetime has been studied in the context of extreme mass ratio inspirals \cite{Kerachian2023}, but the formalism is exceedingly complex \cite{Schmidt2002}. It remains to be seen whether a similar analysis can be done for TDEs.

\subsection{Astrophysical implications} \label{sec:astro_implications}

Before concluding this primarily analytical dissertation, we briefly discuss some of the astrophysical implications of the present work.

\textbf{Preferential disruption and spin evolution:}
The preferential disruption of retrograde stars has two immediate consequences. First, the surviving stellar population will gradually develop a net prograde bias even if the initial distribution was perfectly isotropic. Second, the preferential consumption of retrograde stars deposits a net negative angular momentum into the central BH, damping spin growth and acting as a potential secular spin-down mechanism. This supports the idea that TDEs are unlikely to provide an efficient channel for sustained BH spin growth \cite{Kesden2012,Stone2020}.

\textbf{Debris circularisation and flare dynamics:}
Since the observational signatures of TDEs depend on orbital inclination, the retrograde bias also has direct implications for the observed event population. The orbital inclination of the disrupted star sets the misalignment angle between the resulting debris disc and the BH spin axis, which controls the rate and character of Lense–Thirring precession \cite{Hayasaki2016} and the resulting flare. A retrograde-biased disruption population will therefore preferentially produce misaligned discs, with observational consequences for the quasi-periodic modulation of TDE light curves \cite{Stone2012}.

\textbf{Spin demographics:}
The direct capture fraction discussion in Chapter 3 highlights the importance of properly accounting for direct capture in the intermediate loss cone regime. Since capture by horizon is predominantly a relativistic effect, its modulating effect on the observed TDE rate carries an imprint of the BH spin. Existing efforts \cite{Mummery2023,Kesden2012} have attempted to use the Hills mass cutoff to constrain the spin distribution of the most massive BHs. This approach benefits from the fact that the cutoff is completely determined by the relativistic dynamics and mostly independent of the stellar distribution. However, the results are currently inconclusive. A more statistically robust approach may be to compute the entire $\Gamma(M)$ distribution parametrised by spin, rather than relying on the cutoff alone. This would require a careful treatment of the capture fraction across all regimes, as well as a detailed understanding of the underlying galaxy model.

\textbf{Alternative spacetimes:}
An important advantage of the present framework is that the relativistic boundary calculation in Chapter 2 is modular. Since the local tidal tensor is constructed directly from the metric and geodesic motion, one could in principle replace Kerr by another stationary, axisymmetric spacetime and recompute the loss cone boundaries without altering the stellar dynamics formalism. This may provide a route towards studying TDE rates in charged spinning (Kerr--Newman) BH spacetime, parameterised deviations from Kerr (such as the bumpy \cite{Collins2004} and Johannsen \cite{Johannsen2011} metrics), or even exotic geometries such as a wormhole \cite{Teo1998, Xin2025}.

\newpage
\section{Conclusions} \label{sec:conclusions}

This dissertation develops what is, to our knowledge, the first semi-analytical 2D loss cone framework for predicting tidal disruption event rates in isotropic galaxies hosting rotating BHs. Core methodological developments and key physical findings are summarised as follows:

\begin{itemize}\setlength{\itemsep}{0pt}\setlength{\parskip}{0pt}\setlength{\parsep}{0pt}
    \item \textbf{Relativistic loss cone boundaries:} Using a novel tidal tensor formalism, the tidal disruption and direct capture boundaries in Kerr spacetime are computed as functions of angular momentum magnitude and orientation. The tidal disruption boundary is found to be approximately linear in inclination, while the capture boundary exhibits a strong non-linearity.
    \item \textbf{Analytical 1D direct capture fraction:} The 1D boundary layer problem is solved over nested loss cones to derive a closed-form, energy-dependent fraction that accounts for the proportion of stars directly captured by the event horizon. The analysis demonstrates that direct captures can be ignored in the empty loss cone regime, constitute a fixed proportion of events in the full loss cone regime, and must be accurately accounted for in the intermediate regime.
    \item \textbf{2D diffusion framework:} A 2D Fokker--Planck equation governing stellar diffusion in both angular momentum magnitude ($R = L^{2}/L_{c}^{2}$) and orientation ($x = L_{z}/L$) is derived. By demonstrating that dynamical friction exerts no net back reaction on either $x$ or $R$ for highly eccentric orbits, we show that the 2D collision operator reduces to a flux-conserving, pure divergence form.
    \item \textbf{Preferential disruption:} Through a perturbative expansion of the 2D problem in both the full and empty loss cone limits, we uncover a strong, first-order $\mathcal{O}(\epsilon)$ local inclination bias favouring the disruption of retrograde stars.
    \item \textbf{Robustness of the global TDE rate:} Despite the strong local inclination bias, the global $x$-integrated TDE rate is shown to be extraordinarily robust. In both the empty and full limits, the 2D global flux retains the same form as the classical 1D case for essentially any sufficiently regular loss-cone boundary profile. Anisotropy only enters via a geometric correction to the flux normalisation, which is typically a small $\mathcal{O}(\epsilon^{2})$ effect.
\end{itemize}

Ultimately, this work establishes that while BH spin does not significantly alter the \textit{total} number of TDEs in an isotropic galaxy, it fundamentally affects the \textit{inclination distribution} of the disrupted stars. As next-generation surveys transition TDE astronomy into an era of population-level statistics, accounting for this inherent retrograde bias will be essential for accurately decoding the observed TDE demographics.

%% file: appendix1.tex
\chapter{Solution to Diffusion Equations}

\label{app:diffusion_solution}

This appendix details the derivation of the general solution to the 2D diffusion equation (see Equation~\ref{eq:enhanced_diffusion_equation}) governing the test star distribution function $f(\tau, R, x)$:
\begin{equation}
    \frac{\partial f}{\partial \tau} = P(E) \bar{D}(E) \left[ \frac{\partial }{\partial R} \left( R \frac{\partial f}{\partial R} \right) +  \frac{\partial }{\partial x} \left( \frac{1 - x^{2}}{8 R} \frac{\partial f}{\partial x} \right) \right] \, .
\end{equation}
Once the full 2D solution is obtained, we may specialise to the 1D case by removing all $x$-dependence.

Consider the following separable ansatz:
\begin{equation}
    f(\tau, R, x) = \mathcal{R}(R) \mathcal{X}(x) T(\tau) \, .
\end{equation}
Substituting this into the diffusion equation yields:
\begin{equation}
\frac{1}{P\bar{D} T} \frac{\mathrm{d}T}{\mathrm{d}\tau} = \frac{1}{\mathcal{R}} \frac{\mathrm{d}}{\mathrm{d}R} \left( R \frac{\mathrm{d}\mathcal{R}}{\mathrm{d}R} \right) + \frac{1}{8R \mathcal{X}} \frac{\mathrm{d}}{\mathrm{d}x} \left[ (1 - x^{2}) \frac{\mathrm{d}\mathcal{X}}{\mathrm{d}x} \right] = -\lambda \, .
\end{equation}
We immediately see that the time dependence is just an exponential decay:
\begin{equation}
    T(\tau) \propto \exp{(-\lambda P \bar{D} \tau)} \, .
\end{equation}
Rearranging the spatial components of the equation gives:
\begin{equation}
    \frac{R}{\mathcal{R}} \frac{\mathrm{d}}{\mathrm{d}R} \left( R \frac{\mathrm{d}\mathcal{R}}{\mathrm{d}R} \right) + \lambda R = -\frac{1}{8\mathcal{X}} \frac{\mathrm{d}}{\mathrm{d}x} \left[ (1 - x^{2}) \frac{\mathrm{d}\mathcal{X}}{\mathrm{d}x} \right] = k \, .
\end{equation}
The equation for $\mathcal{X}$ takes the form of a standard Legendre differential equation:
\begin{equation}
    \frac{\mathrm{d}}{\mathrm{d}x} \left[ (1 - x^{2}) \frac{\mathrm{d}\mathcal{X}}{\mathrm{d}x} \right] + 8 k \mathcal{X} = 0 \, .
\end{equation}
For the solution to be regular at $x = \pm 1$, we need $8k = l(l + 1)$ for some integer $l$. The angular solution is then given by the Legendre polynomials:
\begin{equation}
    \mathcal{X}(x) = P_{l}(x) \, .
\end{equation}
For the radial equation governing $\mathcal{R}$:
\begin{equation}
    R \frac{\mathrm{d}}{\mathrm{d}R} \left( R \frac{\mathrm{d}\mathcal{R}}{\mathrm{d}R} \right) + \left[ \lambda R - \frac{l(l + 1)}{8} \right] \mathcal{R} = 0 \, ,
\end{equation}
applying the substitution $z = 2 \sqrt{\lambda R}$ yields:
\begin{equation}
    z^{2} \frac{\mathrm{d}^{2}\mathcal{R}}{\mathrm{d}z^{2}} + z \frac{\mathrm{d}\mathcal{R}}{\mathrm{d}z} + (z^{2} - \nu^{2}) \mathcal{R} = 0 \, .
\end{equation}
This is a Bessel differential equation of order $\nu = \sqrt{l(l + 1)/2}$. The general solution is a linear combination of Bessel functions of the first and second kind:
\begin{equation}
    \mathcal{R}(R) = A J_{\nu}(2 \sqrt{\lambda R}) + B Y_{\nu}(2 \sqrt{\lambda R}) \, ,
\end{equation}
where $B = 0$ since we require the solution to be regular at $R = 0$.

We conclude that the general solution to Equation~\ref{eq:enhanced_diffusion_equation} has the form:
\begin{equation}
    f(\tau, R, x) = \sum_{l = 0}^{\infty} \sum_{m = 1}^{\infty} A_{lm} J_{\nu}(2 \sqrt{\lambda_{lm} R}) P_{l}(x) \exp{(-\lambda_{lm} P \bar{D} \tau)} \, .
\end{equation}
The 2D diffusion equation separates elegantly into special functions. However, determining the exact eigenvalues $\lambda_{lm}$ and coefficients $A_{lm}$ remains analytically intractable, as the anisotropic loss cone boundary condition couples the various modes together.

We now specialise to the isotropic case. Since the loss cone boundary is independent of $x$, we remove any $x$-dependence by retaining only the $l = 0$ mode. Following \cite{Merritt2013}, it is convenient to introduce the scaled variable $y = R/P\bar{D}$. The diffusion equation inside the loss cone is subject to the boundary condition $f(\tau, y_{\text{lc}}) = f(y_{\text{lc}})$ and the initial condition $f(0, y) = 0$ for $0 < y < y_{\text{lc}}$.

Because the boundary condition at $y_{\text{lc}}$ is inhomogeneous, we isolate the transient behaviour by defining a new function $u(\tau, y)$:
\begin{equation}
    u(\tau, y) \equiv f(y_{\text{lc}}) - f(\tau, y) \, .
\end{equation}
The function $u$ satisfies the identical diffusion equation, but is now subject to a homogeneous boundary condition at the loss cone boundary:
\begin{equation}
    u(\tau, y_{\text{lc}}) = 0 \, ,
\end{equation}
along with the modified initial condition $u(0,y) = f(y_{\text{lc}})$.

$u$ satisfies the same diffusion equation as $f$ and has the same separable form of solution:
\begin{equation}
    u(\tau, y) = \sum_{m=1}^{\infty} C_{m} \, J_{0}(\beta_{m} \sqrt{y}) \, \exp{\left(-\beta_{m}^{2}\tau/4\right)} \, .
\end{equation}

We define the eigenvalues $\beta_{m} = 2 \sqrt{\lambda_{m} P \bar{D}}$ to maintain consistency with the notation in \cite{Merritt2013}. The boundary condition $u(\tau, y_{\text{lc}}) = 0$ dictates that the terms $\beta_{m} \sqrt{y_{\text{lc}}}$ must be the consecutive roots of the Bessel function, i.e.\ $J_{0}(\beta_{m} \sqrt{y_{\text{lc}}}) = 0$.

The coefficients $C_m$ are obtained via Bessel function orthogonality. Applying the change of variables $r = \sqrt{y}$ (such that $\mathrm{d}y = 2r\,\mathrm{d}r$) yields\footnote{It is amusing that the substitution $r = \sqrt{y}$ returns the integral variable to the (scaled) angular momentum magnitude $\tilde{L} = L/L_c$.}:
\begin{equation}
    \int_{0}^{y_{\text{lc}}} J_{0}(\beta_{m} \sqrt{y}) J_{0}(\beta_{n} \sqrt{y}) \, \mathrm{d}y
    = 2\int_{0}^{\sqrt{y_{\text{lc}}}} r\, J_{0}(\beta_{m} r) J_{0}(\beta_{n} r) \, \mathrm{d}r
    = y_{\text{lc}}\,\left[J_{1}(\beta_{m}\sqrt{y_{\text{lc}}})\right]^{2} \delta_{mn} \, .
\end{equation}

Multiplying $u(\tau = 0,y)=\sum_{m} C_{m} J_{0}(\beta_{m}\sqrt{y})$ by $J_{0}(\beta_{n}\sqrt{y})$ and integrating over $y \in [0,y_{\text{lc}}]$ gives:
\begin{equation}
    C_{n} = \frac{\displaystyle \int_{0}^{y_{\text{lc}}} u(0,y)\, J_{0}(\beta_{n} \sqrt{y})\,\mathrm{d}y}{\displaystyle y_{\text{lc}}\,\left[J_{1}(\beta_{n}\sqrt{y_{\text{lc}}})\right]^{2}} \, .
\end{equation}
The integral in the numerator evaluates to:
\begin{equation}
    \begin{split}
        \int_{0}^{y_{\text{lc}}} u(0,y)\, J_{0}(\beta_{n} \sqrt{y})\,\mathrm{d}y &= 2 f(y_{\text{lc}})\int_{0}^{\sqrt{y_{\text{lc}}}} r\,J_{0}(\beta_{n} r)\,\mathrm{d}r \\
        &= \frac{2 f(y_{\text{lc}})\sqrt{y_{\text{lc}}}}{\beta_{n}}\,J_{1}(\beta_{n}\sqrt{y_{\text{lc}}}) \, ,
    \end{split}
\end{equation}
where the identity $\int r J_{0}(\beta r)\,\mathrm{d}r = r J_{1}(\beta r)/\beta$ has been applied. 

Substituting this result into the expression for $C_{m}$ yields:
\begin{equation}
    C_{m} = \frac{2 f(y_{\text{lc}})}{\sqrt{y_{\text{lc}}}\,\beta_{m}\,J_{1}(\beta_{m}\sqrt{y_{\text{lc}}})} \, .
\end{equation}

Since $f(\tau,y)=f(y_{\text{lc}})-u(\tau,y)$, we can finally write down the 1D solution for $f(\tau, y)$ presented in Equation~\ref{eq:inside_loss_cone_solution}:
\begin{equation}
    f(\tau, y) = f(y_{\text{lc}}) \left[ 1 - \frac{2}{\sqrt{y_{\text{lc}}}} \sum_{m = 1}^{\infty} \frac{\exp{( -\beta_{m}^{2}\tau/4 )}}{\beta_{m}} \frac{J_{0}(\beta_{m} \sqrt{y})}{J_{1}(\beta_{m} \sqrt{y_{\text{lc}}})} \right] \, .
\end{equation}

%% file: appendix2_v2.tex
\chapter[Diffusion Coefficients via Vector Perturbation]{Diffusion Coefficients via \\Vector Perturbation}
\label{app:diffusion_derivation}

In this appendix, we derive the diffusion coefficients in $R$ and $x$ using an intuitive approach based on vector perturbations. Consider a star located at position $\mathbf{r}$ with velocity $\mathbf{v}$. The angular momentum is $\mathbf{L} = \mathbf{r} \times \mathbf{v}$, its squared magnitude $L^{2} = |\mathbf{r} \times \mathbf{v}|^{2}$, and its projection onto the BH spin axis $L_{z} = \hat{\mathbf{z}} \cdot (\mathbf{r} \times \mathbf{v})$. A local velocity perturbation $\Delta \mathbf{v}$ has its moments defined as:
\begin{equation}
\begin{split}
    \langle \Delta \mathbf{v} \rangle &= \langle \Delta v_{\parallel} \rangle \hat{\mathbf{v}} \, , \\
    \langle \Delta \mathbf{v} \otimes \Delta \mathbf{v} \rangle &= \langle (\Delta v_{\parallel})^{2} \rangle \hat{\mathbf{v}} \otimes \hat{\mathbf{v}} + \frac{1}{2} \langle (\Delta v_{\perp})^{2} \rangle \left( \mathbf{I} - \hat{\mathbf{v}} \otimes \hat{\mathbf{v}} \right) \, ,
\end{split}
\label{eq:app_velocity_moments}
\end{equation}
where $\langle \Delta v_{\parallel} \rangle$, $\langle (\Delta v_{\parallel})^{2} \rangle$, and $\langle (\Delta v_{\perp})^{2} \rangle$ denote the standard diffusion coefficients in the parallel and perpendicular directions, and $\mathbf{I}$ is the identity matrix.

In index notation, the above can be written as:
\begin{equation}
\begin{split}
    \langle \Delta v_{i} \rangle &= \langle \Delta v_{\parallel} \rangle \hat{v}_{i} \, , \\
    \langle \Delta v_{i} \Delta v_{j} \rangle &= \langle (\Delta v_{\parallel})^{2} \rangle \hat{v}_{i} \hat{v}_{j} + \frac{1}{2} \langle (\Delta v_{\perp})^{2} \rangle (\delta_{ij} - \hat{v}_{i} \hat{v}_{j}) \, .
\end{split}
\end{equation}

A velocity perturbation $\Delta \mathbf{v}$ induces a change in $L^{2}$ given by:
\begin{equation}
    \Delta (L^{2}) = 2(\mathbf{L} \times \mathbf{r}) \cdot \Delta \mathbf{v} + |\mathbf{r} \times \Delta \mathbf{v}|^{2} \, .
\end{equation}
Define the auxiliary vector $\mathbf{K} \equiv \mathbf{L} \times \mathbf{r} = r^{2} \mathbf{v} - (\mathbf{r} \cdot \mathbf{v})\mathbf{r}$. Noting that $\mathbf{K} \cdot \hat{\mathbf{v}} = L^{2}/v$ and using the identity $|\mathbf{r} \times \Delta \mathbf{v}|^{2} = r^{2} |\Delta \mathbf{v}|^{2} - (\mathbf{r} \cdot \Delta \mathbf{v})^{2}$, the ensemble average evaluates to:
\begin{equation}
    \langle \Delta (L^{2}) \rangle = r^{2} \langle (\Delta v_{\perp})^{2} \rangle + 2\frac{L^{2}}{v} \langle \Delta v_{\parallel} \rangle + \frac{L^{2}}{2v^{2}}\left( 2\langle (\Delta v_{\parallel})^{2} \rangle - \langle (\Delta v_{\perp})^{2} \rangle \right) \, .
\end{equation}
Since tidal disruption predominantly involves highly eccentric orbits, we drop terms that are $\mathcal{O}(L^{2})$ and obtain:
\begin{equation}
    \langle \Delta (L^{2}) \rangle \approx r^{2} \langle (\Delta v_{\perp})^{2} \rangle + \mathcal{O}(L^{2}) \, .
\end{equation}
Dividing by $L_c^{2}$ yields the drift in $R = L^{2}/L_c^{2}$:
\begin{equation}
    \langle \Delta R \rangle = \frac{r^{2} \langle (\Delta v_{\perp})^{2} \rangle}{L_c^{2}} + \mathcal{O}(R) \, .
\end{equation}
To evaluate the variance $\langle (\Delta R)^{2} \rangle$, we square the leading term of $\Delta (L^{2})$ and take the highly eccentric limit:
\begin{equation}
    \langle (\Delta L^{2})^{2} \rangle \approx 4 \langle (\mathbf{K} \cdot \Delta \mathbf{v})^{2} \rangle \approx 2 r^{2} L^{2} \langle (\Delta v_{\perp})^{2} \rangle \, ,
\end{equation}
which provides the corresponding variance in $R$ once divided by $L_c^4$:
\begin{equation}
    \langle (\Delta R)^{2} \rangle = \frac{2 r^{2} \langle (\Delta v_{\perp})^{2} \rangle}{L_c^{2}} R + \mathcal{O}(R^{2}) \, .
\end{equation}

To evaluate the inclination variable $x = L_z / L$, introduce $\mathbf{W} \equiv \hat{\mathbf{z}} \times \mathbf{r}$ such that $L_z = \mathbf{W} \cdot \mathbf{v}$. A velocity perturbation yields $\Delta L_z = \mathbf{W} \cdot \Delta \mathbf{v}$. Expanding $\Delta x = \Delta(L_z (L^{2})^{-1/2})$ to second order gives:
\begin{equation}
    \Delta x = \frac{\Delta L_z}{L} - \frac{x}{2 L^{2}} \Delta(L^{2}) - \frac{\Delta L_z \Delta(L^{2})}{2 L^3} + \frac{3 x}{8 L^4} (\Delta(L^{2}))^{2} \, .
\label{eq:app_delta_x_expansion}
\end{equation}
Unlike the magnitude $R$, the orientation variables remarkably do not require the highly eccentric limit. The exact expectations for the moments can be evaluated directly:
\begin{equation}
\begin{split}
    \langle \Delta L_z \rangle &= \frac{L_z}{v} \langle \Delta v_{\parallel} \rangle \, , \\
    \langle \Delta L_z \Delta (L^{2}) \rangle &= \frac{L_z}{v^{2}} \left[ 2 L^{2} \langle (\Delta v_{\parallel})^{2} \rangle + (r^{2} v^{2} - L^{2})\langle (\Delta v_{\perp})^{2} \rangle \right] \, , \\
    \langle (\Delta L_z)^{2} \rangle &= \frac{L_z^{2}}{v^{2}} \langle (\Delta v_{\parallel})^{2} \rangle + \frac{1}{2} \left( r^{2} \sin^{2}\theta_r - \frac{L_z^{2}}{v^{2}} \right) \langle (\Delta v_{\perp})^{2} \rangle \, .
\end{split}
\end{equation}
When these expressions are substituted into the Taylor expansion for $\Delta x$ (Equation~\ref{eq:app_delta_x_expansion}), a geometric cancellation occurs. All parallel diffusion terms and all longitudinal corrections sum identically to zero, such that:
\begin{equation}
    \langle \Delta x \rangle = -\frac{1}{4} \frac{r^{2} \langle (\Delta v_{\perp})^{2} \rangle}{L^{2}} x \, ,
\end{equation}
and:
\begin{equation}
    \langle (\Delta x)^{2} \rangle = \frac{r^{2} \langle (\Delta v_{\perp})^{2} \rangle}{2 L^{2}} \left( \sin^{2}\theta_r - x^{2} \right) \, .
\end{equation}
Rewriting $L^{2} = R L_c^{2}$ leads precisely to:
\begin{equation}
\begin{split}
    \langle \Delta x \rangle &= - \frac{r^{2} \langle (\Delta v_{\perp})^{2} \rangle}{4 L_c^{2}} \frac{x}{R} \, , \\
    \langle (\Delta x)^{2} \rangle &= \frac{r^{2} \langle (\Delta v_{\perp})^{2} \rangle}{2 L_c^{2}} \frac{-x^{2} + \sin^{2}\theta_r}{R} \, .
\end{split}
\end{equation}
Finally, to determine the cross-diffusion coefficient $\langle \Delta R \Delta x \rangle$, the second-order cross-expansion is evaluated:
\begin{equation}
    \langle \Delta (L^{2}) \Delta x \rangle \approx \frac{\langle \Delta L_z \Delta(L^{2}) \rangle}{L} - \frac{x}{2 L^{2}} \langle (\Delta L^{2})^{2} \rangle \, .
\end{equation}
Upon substituting the expectations, the terms once again cancel exactly, yielding:
\begin{equation}
    \langle \Delta R \Delta x \rangle = 0 \, .
\end{equation}

Collecting terms, we recover the results in the main text (Equations~\ref{eq:diffusion_coefficients}):
\begin{equation}
\begin{split}
    \langle \Delta R \rangle &= \frac{r^{2} \langle (\Delta v_{\perp})^{2} \rangle}{L_{c}^{2}} + \mathcal{O}(R) \, , \\
    \langle \Delta x \rangle &= - \frac{r^{2} \langle (\Delta v_{\perp})^{2} \rangle}{4 L_{c}^{2}} \frac{x}{R} \, , \\
    \langle (\Delta R)^{2} \rangle &= \frac{2 r^{2} \langle (\Delta v_{\perp})^{2} \rangle}{L_{c}^{2}} R + \mathcal{O}(R^{2}) \, , \\
    \langle \Delta R \Delta x \rangle &= 0 \, , \\
    \langle (\Delta x)^{2} \rangle &= \frac{r^{2} \langle (\Delta v_{\perp})^{2} \rangle}{2 L_{c}^{2}} \frac{-x^{2} + \sin^{2}{\theta_{r}}}{R} \, .
\end{split}
\end{equation}

As a further note, the cross-diffusion coefficient between energy and orientation vanishes by a similar cancellation. Keeping the terms that contribute to the second moment and using the vectors $\mathbf{W}$ and $\mathbf{K}$ defined above gives:
\begin{equation}
\begin{split}
    \langle\Delta E\,\Delta x\rangle
    &=
    \frac{1}{L}
    \left\langle
        (\mathbf{v}\cdot\Delta\mathbf{v})
        (\mathbf{W}\cdot\Delta\mathbf{v})
    \right\rangle
    -\frac{x}{L^2}
    \left\langle
        (\mathbf{v}\cdot\Delta\mathbf{v})
        (\mathbf{K}\cdot\Delta\mathbf{v})
    \right\rangle\\
    &=
    \frac{L_z}{L}\langle(\Delta v_\parallel)^2\rangle
    -x\langle(\Delta v_\parallel)^2\rangle \\
    &=0 \, ,
\end{split}
\end{equation}
where $\mathbf{W}\cdot\mathbf{v}=L_z$ and $\mathbf{K}\cdot\mathbf{v}=L^2$ are used in the second line.

We show additional diffusion coefficients involving energy in Appendix~\ref{app:diffusion_derivation_coordinate_transformation}, where a coordinate transformation approach is used. The complete symbolic verification, including the fixed-energy reduction to the coefficients derived here, is provided in the supporting code repository\footnote{\url{https://github.com/Walker-Xin/tidal_rate}}.

%% file: appendix3_v2.tex
\chapter[Diffusion Coefficients via Coordinate Transformation]{Diffusion Coefficients via \\Coordinate Transformation}
\label{app:diffusion_derivation_coordinate_transformation}

In Appendix~\ref{app:diffusion_derivation}, the diffusion coefficients were derived by perturbing the angular-momentum vector directly. Here we recover them independently using the general coordinate-transformation formalism of \cite{Merritt2013a}. It is useful to perform the transformation first in $(E,Y,x)$, where:
\begin{equation}
\begin{split}
    E &= \frac{v^{2}}{2}+\Phi(r) \, ,\\
    Y &= L^{2}=r^{2}v^{2}(1-\mu^{2}) \, ,\\
    x &= \cos\iota=\frac{L_z}{L}
       =\sin\theta_r\sin\phi_v \, .
\end{split}
\label{eq:app_EYx_coordinates}
\end{equation}
The original velocity coordinates are $(v,\mu=\cos\theta_v,\phi_v)$. Their invariant line element is:
\begin{equation}
    \mathrm{d}s^{2}
    =\mathrm{d}v^{2}
    +\frac{v^{2}}{1-\mu^{2}}\mathrm{d}\mu^{2}
    +v^{2}(1-\mu^{2})\mathrm{d}\phi_v^{2} \, .
\end{equation}
For a general set of velocity coordinates $V^\lambda$, the first and second moments transform according to \cite[Equation 5.120]{Merritt2013a}:
\begin{equation}
\begin{split}
    \langle\Delta V^\lambda\rangle
    &=
    \partial_i V^\lambda T^i
    +\frac{1}{2}
    \left(
        \partial_i\partial_j V^\lambda
        -\Gamma^k_{ij}\partial_k V^\lambda
    \right)S^{ij} \, ,\\
    \langle\Delta V^\lambda\Delta V^\nu\rangle
    &=\partial_iV^\lambda\partial_jV^\nu S^{ij} \, ,
\end{split}
\label{eq:diffusion_coefficient_transformation}
\end{equation}
where $\Gamma^k_{ij}$ is the Christoffel symbol of the velocity-space metric, and for an isotropic field-star distribution:
\begin{equation}
\begin{split}
    T^i
    &=\frac{v^i}{v}\langle\Delta v_\parallel\rangle \, ,\\
    S^{ij}
    &=\frac{v^iv^j}{v^2}
      \langle(\Delta v_\parallel)^2\rangle
    +\left(\delta^{ij}-\frac{v^iv^j}{v^2}\right)
      \frac{\langle(\Delta v_\perp)^2\rangle}{2} \, .
\end{split}
\end{equation}

For compactness, define:
\begin{equation}
    A_\parallel\equiv\langle\Delta v_\parallel\rangle \, ,
    \qquad
    B_\parallel\equiv\langle(\Delta v_\parallel)^2\rangle \, ,
    \qquad
    B_\perp\equiv\langle(\Delta v_\perp)^2\rangle \, .
\end{equation}
Applying Equation~\ref{eq:diffusion_coefficient_transformation} to Equation~\ref{eq:app_EYx_coordinates} gives the three drift coefficients:
\begin{equation}
\begin{aligned}
    \langle\Delta E\rangle
        &=
        vA_\parallel+\frac{1}{2}(B_\parallel+B_\perp) \, ,\\
    \langle\Delta Y\rangle
        &=
        \frac{2Y}{v}A_\parallel
        +\frac{Y}{v^2}B_\parallel
        +r^2B_\perp
        -\frac{Y}{2v^2}B_\perp \, ,\\
    \langle\Delta x\rangle
        &=
        -\frac{r^2B_\perp}{4Y}x \, .
\end{aligned}
\label{eq:app_EYx_drifts}
\end{equation}
The six independent second moments are:
\begin{equation}
\begin{aligned}
    \langle(\Delta E)^2\rangle
        &=v^2B_\parallel \, ,\\
    \langle\Delta E\,\Delta Y\rangle
        &=2YB_\parallel \, ,\\
    \langle\Delta E\,\Delta x\rangle
        &=0 \, ,\\
    \langle(\Delta Y)^2\rangle
        &=
        \frac{4Y^2}{v^2}B_\parallel
        +2r^2YB_\perp
        -\frac{2Y^2}{v^2}B_\perp \, ,\\
    \langle\Delta Y\,\Delta x\rangle
        &=0 \, ,\\
    \langle(\Delta x)^2\rangle
        &=
        \frac{r^2B_\perp}{2Y}
        \left(\sin^2\theta_r-x^2\right) \, .
\end{aligned}
\label{eq:app_EYx_diffusion_coefficients}
\end{equation}
Thus the local diffusion tensor has the block-diagonal form:
\begin{equation}
    \mathbf{D}_{(E,Y,x)}
    =
    \begin{pmatrix}
        v^2B_\parallel
        &2YB_\parallel
        &0\\
        2YB_\parallel
        &\displaystyle
         \frac{4Y^2}{v^2}B_\parallel
         +2r^2YB_\perp
         -\frac{2Y^2}{v^2}B_\perp
        &0\\
        0
        &0
        &\displaystyle
         \frac{r^2B_\perp}{2Y}
         (\sin^2\theta_r-x^2)
    \end{pmatrix}.
\label{eq:app_EYx_block_diagonal}
\end{equation}
At the level of the diffusion tensor, the orientation variable $x$ is decoupled from $E$ and $Y$. This result does not require the highly eccentric limit.

A secondary transformation from $(E,Y,x)$ to $(E,R,x)$ can be carried out, with $R=Y/L_c^2(E)$. This results in cumbersome expressions involving the derivatives of $L_c(E)$ due to the energy dependence of the circular angular momentum.

However, for the application in this work, energy diffusion can be neglected (see Section~\ref{sec:1d_diffusion_equation}). We may therefore hold $E$ fixed and regard $L_c(E)$ as a fixed normalisation. The transformation from $Y$ to $R$ then reduces to a simple rescaling, yielding the desired diffusion coefficients in $(R,x)$ at fixed $E$:
\begin{equation}
\begin{aligned}
    \left.\langle\Delta R\rangle\right|_E
        &=
        \frac{2R}{v}A_\parallel
        +\frac{R}{v^2}B_\parallel
        +\frac{r^2}{L_c^2}B_\perp
        -\frac{R}{2v^2}B_\perp \, ,\\
    \left.\langle\Delta x\rangle\right|_E
        &=
        -\frac{r^2B_\perp}{4L_c^2}\frac{x}{R} \, ,\\
    \left.\langle(\Delta R)^2\rangle\right|_E
        &=
        \frac{4R^2}{v^2}B_\parallel
        +\frac{2r^2R}{L_c^2}B_\perp
        -\frac{2R^2}{v^2}B_\perp \, ,\\
    \left.\langle\Delta R\,\Delta x\rangle\right|_E
        &=0 \, ,\\
    \left.\langle(\Delta x)^2\rangle\right|_E
        &=
        \frac{r^2B_\perp}{2L_c^2}
        \frac{\sin^2\theta_r-x^2}{R} \, .
\end{aligned}
\label{eq:app_fixed_E_Rx_coefficients}
\end{equation}
Finally, taking $R\ll1$ yields
\begin{equation}
\begin{aligned}
    \left.\langle\Delta R\rangle\right|_E
        &=\frac{r^2B_\perp}{L_c^2}+\mathcal{O}(R) \, ,\\
    \left.\langle(\Delta R)^2\rangle\right|_E
        &=\frac{2r^2B_\perp}{L_c^2}R+\mathcal{O}(R^2) \, ,
\end{aligned}
\end{equation}
in agreement with the results in Appendix~\ref{app:diffusion_derivation}.

The complete diffusion structure therefore has a useful interpretation. Energy and angular-momentum magnitude form a coupled block since $\langle\Delta E\,\Delta Y\rangle\neq0$. In contrast, orientation has no mixed diffusion with either variable: $\langle\Delta E\,\Delta x\rangle=\langle\Delta Y\,\Delta x\rangle=0$. After the averaging $\langle\sin^2\theta_r\rangle=(1+x^2)/2$ described in Section~\ref{sec:angular_diffusion_coefficients}, the collison operator can be decomposed as:
\begin{equation}
    \left(\frac{\partial f}{\partial t}\right)_{c}
    =
    \left(\frac{\partial f}{\partial t}\right)_{c,E,Y}
    +
    \frac{r^{2} B_\perp}{8Y}
    \mathcal{L}_x f \, ,
    \qquad
    \mathcal{L}_x
    \equiv
    \frac{\partial}{\partial x}
    \left[
        (1-x^2)\frac{\partial}{\partial x}
    \right].
\end{equation}
Here $\mathcal{L}_x$ is the Legendre operator: its regular eigenfunctions on $-1\leq x\leq1$ are the Legendre polynomials $P_l(x)$, with eigenvalues $-l(l+1)$. Thus the angular part of the diffusion operator reduces naturally to a familiar eigenfunction problem. There remains a parametric coupling to $(E, Y)$ because the angular diffusion rate depends on them, but no mixed derivatives involving $x$ occur.

The coordinate transformations in this appendix are implemented in the supporting code repository\footnote{\url{https://github.com/Walker-Xin/tidal_rate}}.